\documentclass[fleqn,usenatbib]{mnras}

\usepackage[T1]{fontenc}

\usepackage{graphicx}	
\usepackage{amsmath}	
\usepackage{amssymb}	
\usepackage{multirow}
\usepackage{pdflscape}
\usepackage{lmodern}

\title[Tracing Chemical Depletion in Evolved Binaries Hosting Transition Discs]{Tracing Chemical Depletion in Evolved Binaries \\Hosting Second-Generation Transition Discs}

\author[Mohorian et al. (2025)]{Maksym Mohorian,$^{1,2}$\thanks{E-mail: maksym.mohorian@students.mq.edu.au} Devika Kamath,$^{1,2,3}$ Meghna Menon,$^{1,2}$ Anish M. Amarsi,$^{4}$ 
\newauthor Hans Van Winckel,$^{5}$ Claudia Fava,$^{1,2}$ and Kateryna Andrych$^{1,2}$
\\
$^{1}$School of Mathematical and Physical Sciences, Macquarie University, Balaclava Road, Sydney, NSW 2109, Australia\\
$^{2}$Astrophysics and Space Technologies Research Centre, Macquarie University, Balaclava Road, Sydney, NSW 2109, Australia\\
$^{3}$INAF, Osservatorio Astronomico di Roma, Via Frascati 33, I-00077 Monte Porzio Catone, Italy\\
$^{4}$Theoretical Astrophysics, Department of Physics and Astronomy, Uppsala University, Box 516, SE-751 20 Uppsala, Sweden\\
$^{5}$Institute of Astronomy, KU Leuven, Celestijnenlaan 200D, 3001 Leuven, Belgium
}

\date{Accepted XXX. Received YYY; in original form ZZZ}

\pubyear{2025}

\begin{document}\label{firstpage}
\pagerange{\pageref{firstpage}--\pageref{lastpage}}
\maketitle
\begin{abstract}
The mechanisms responsible for chemical depletion across diverse astrophysical environments are not yet fully understood. In this paper, we investigate chemical depletion in post-AGB/post-RGB binary stars hosting second-generation transition discs using high-resolution optical spectra from HERMES/Mercator and UVES/VLT. We performed a detailed chemical abundance analysis of 6 post-AGB/post-RGB stars and 6 post-AGB/post-RGB candidates with transition discs in the Galaxy and in the Large Magellanic Cloud. The atmospheric parameters and elemental abundances were obtained through 1D LTE analysis of chemical elements from C to Eu, and 1D NLTE corrections were incorporated for elements from C to Fe. Our results confirmed that depletion efficiency, traced by the [S/Ti] abundance ratio, is higher in post-AGB/post-RGB binaries with transition discs compared to the overall sample of post-AGB/post-RGB binaries. We also examined correlations between derived abundances and binary system parameters (astrometric, photometric, orbital, pulsational). Additionally, we compared the depletion patterns in our sample to those observed in young stars with transition discs and in the interstellar medium. We confirmed that the depletion is significantly stronger in post-AGB/post-RGB binaries with transition discs than in young stars with transition discs. Furthermore, we found that [X/Zn] abundance ratio trends of volatile and refractory elements in post-AGB/post-RGB binaries with transition discs generally resemble similar trends in the interstellar medium (except for trends of [Si/Zn] and [Mg/Zn] ratios). These findings, although based on a limited sample, provide indirect constraints for depletion mechanism in circumbinary discs around post-AGB/post-RGB stars.
\end{abstract}


\section{Introduction}\label{sec:int}
One the most complex aspects of binary star evolution is the interaction between binaries and circumbinary discs \citep{heath2020DiscBinaryInteraction, penzlin2022DiscBinaryInteraction, coleman2022DiscBinaryInteractionPlanets, zagaria2023DiscBinaryInteractionPlanets}. These discs, characterised by intricate dynamics and varying dust content, significantly influence observable properties of the host binary, such as infrared excess \citep{itoh2015DiscBinaryInteractionIRexcess}, accretion rates \citep{izzard2023DiscBinaryInteractionAccretion}, orbital eccentricity \citep{heath2020DiscBinaryInteraction}, and jet activity \citep{bollen2022Jets}. The diversity of dust content in circumbinary discs results from dust condensation being dependent on the chemical composition of the condensing mixture and the physical conditions of the environment, such as pressure and temperature \citep{lodders2003CondensationTemperatures, wood2019CondensationTemperatures}. While the process of dust condensation is well-studied for protoplanetary discs around young single stars \citep{lagage1994DustDepletionPlanetIndicator, birnstiel2016DustEvolutionAndPlanetesimals}, the effect of binarity on dust condensation in circumbinary discs around evolved binary stars remains poorly explored \citep{oomen2020MESAdepletion, miguel2024DiscBinaryInteractionDustFormation}.

During the evolution along the asymptotic giant branch (AGB) or red giant branch (RGB), low- and intermediate-mass stars  ($M\sim0.8-8M_\odot$) in a binary system can fill their Roche lobe, leading to mass loss that terminates AGB/RGB evolution. The primary star then transitions to the post-asymptotic giant branch (post-AGB) or post-red giant branch (post-RGB) stage \citep{vanwinckel2003Review, kamath2016PostRGBDiscovery, vanwinckel2018Binaries, kluska2022GalacticBinaries}. While the binary interactions between the post-AGB/post-RGB primary star and the secondary component \citep[often a main sequence star;][]{oomen2018OrbitalParameters} remain poorly understood, observational studies reveal that these interactions often result in the formation of a stable disc of circumstellar gas and dust \citep[with radii <\,1\,000\,AU;][]{deruyter2006discs, deroo2006CompactDisc, bujarrabal2015KeplerianRotation, hillen2016IRAS08, bujarrabal2018IRAS08, kluska2022GalacticBinaries, corporaal2023DiscParameters, gallardocava2023thesis}.

The presence of the disc around post-AGB binaries was observationally established by the distinct pattern in spectral energy distribution \citep[SED;][]{deruyter2005discs, deruyter2006discs, kamath2014SMC, gezer2015WISERVTau, kamath2015LMC, kluska2022GalacticBinaries}. This pattern includes a near-infrared (near-IR) dust excess, indicative of hot dust in the system \citep{vanwinckel2003Review, oomen2018OrbitalParameters}. Interferometric imaging studies resolved the inner rim of circumbinary discs in several post-AGB/post-RGB binary systems, which in many cases is close to the dust sublimation radius \citep[typically, $\sim$5--30 AU;][]{kluska2019DiscSurvey, corporaal2023DiscParameters}. In addition, high-resolution polarimetric imaging studies with SPHERE revealed complex substructures in circumbinary discs around post-AGB/post-RGB stars, including rings, spirals, and arc-like features \citep{ertel2019Imaging, andrych2023Polarimetry}. Moreover, the outer regions of these discs are known to display crystallisation \citep{gielen2011silicates, hillen2015ACHerMinerals} and grain growth \citep{scicluna2020GrainGrowth}. Additionally, circumbinary discs around post-AGB/post-RGB inaries display Keplerian rotation based on position–velocity maps of $^{12}$CO \citep{bujarrabal2015KeplerianRotation, cava202389HerCOmaps}.

Recently, \citet{kluska2022GalacticBinaries} compiled a comprehensive sample of 85 Galactic post-AGB binaries and their low-luminosity analogues, dusty post-RGB stars. Using the near-IR and mid-IR colours (2MASS $H-K$ and WISE $W_1-W_3$, respectively), \citet{kluska2022GalacticBinaries} categorised the discs around Galactic post-AGB/post-RGB binaries in three groups: i) full discs, where the dust in the disc extends from the dust sublimation radius outward, ii) transition discs, where IR colours suggest the presence of large dust cavities in the inner discs, and iii) discs with significant lack of IR excess, which points to virtual absence of circumbinary dust (we refer to these discs as dust-poor, though this group includes gas-poor debris discs). A subsequent mid-IR interferometric study of full and transition disc candidates in the Galaxy by \citet{corporaal2023DiscParameters} confirmed the presence of this gap (the dust inner rims in transition disc systems are 2.5--7.5 times larger than the corresponding dust sublimation radii). This subset of transition disc targets is the centrepiece of this study (see Section~\ref{sec:tar}).

The interaction between the binary star and the circumbinary disc significantly affects the surface composition of the post-AGB/post-RGB star. In particular, the post-AGB/post-RGB binaries with circumbinary discs exhibit photospheric chemical depletion (hereafter referred to as depletion) with a notable underabundance of refractory elements (i.e. those with condensation temperatures $T_{\rm cond}>1250$ K, such as Al, Fe, Ti, and the majority of the slow neutron capture process elements) relative to volatile elements \citep[i.e. those with condensation temperatures $T_{\rm cond}<1250$ K, like S, Zn, Na, and K;][]{gielen2009Depletion, gezer2015WISERVTau, kamath2019depletionLMC}. The exact mechanism behind chemical depletion in post-AGB/post-RGB stars is not yet fully understood. However, it is believed to result from the chemical fractionation of gas and dust in the circumbinary disc and subsequent accretion of a small portion of this pure gas onto the photosphere of the primary component \citep{waters1992GasDustFractionation, oomen2020MESAdepletion}. The efficiency of the gas-dust fractionation is specific to each chemical element and depends on its condensation temperature,  $T_{\rm cond}$ \citep{lodders2003CondensationTemperatures, wood2019CondensationTemperatures}. Consequently, mostly refractory dust particles are settling in the mid-plane of the disc, while mostly volatile gas is partially re-accreted onto the binary \citep{mosta2019ReaccretionInnerRim, munoz2019ReaccretionInnerRim}.

Observational studies of depletion in post-AGB/post-RGB binaries in the Galaxy and the Magellanic Clouds \citep{giridhar1998RVTauVars, maas2002RUCenSXCen, deruyter2005discs, giridhar2005rvtau, deruyter2006discs, maas2007t2cep, vanwinckel2012AFCrt, rao2014RVTauAbundances, desmedt2012J004441, desmedt2014LeadMCs, desmedt2015LMC2sEnrichedPAGBs, desmedt2016LeadMW} showed that the relative abundances [X/H]\footnote{$[$X/H$]\,=\,\log\frac{N(X)}{N(H)}-\log\frac{N(X)_\odot}{N(H)_\odot}+12$, where N(X) and N(H) are the number abundances of element X and hydrogen, respectively. Symbol $\odot$ denotes the corresponding solar abundances.} are prominently decreased for those elements, which have high condensation temperatures $T_{\rm cond}$. This leads to a prominent break in the plots of condensation temperature $T_{\rm cond}$ vs. relative abundance [X/H] for post-AGB/post-RGB binaries \citep{oomen2019depletion}. The condensation temperature at the break -- the turn-off temperature $T_{\rm turn-off}$ -- has a wide range of values within the Galactic subsample \citep[from 800 to 1500 K;][]{kluska2022GalacticBinaries}.

The rate (or efficiency) of the chemical depletion in post-AGB/post-RGB binary stars is traced by a volatile-to-refractory abundance ratio, usually [Zn/Ti] ratio \citep{gezer2015WISERVTau, oomen2019depletion}. Depletion efficiency may be categorised into the following groups: mild ([Zn/Ti] < 0.5 dex), moderate (0.5 dex < [Zn/Ti] < 1.5 dex), or strong ([Zn/Ti] > 1.5 dex). \citet{kluska2022GalacticBinaries} showed that the observed depletion efficiency ([Zn/Ti] ratio) is generally the highest in the subsample of transition disc candidates. Additionally, the high-temperature end of depletion profile in the $T_{\rm cond}$-[X/H] plots (i.e. for elements with $T_{\rm cond}>T_{\rm turn-off}$) may follow a linear trend (``saturated'' profile) or a two-piece linear fit with a horizontal plateau at higher condensation temperatures \citep[``plateau'' profile;][]{waelkens1991depletion, oomen2019depletion, oomen2020MESAdepletion}. Theoretical studies using the detailed MESA stellar evolution models confirmed that the observed depletion profiles in post-AGB binaries (including the breaks at $T_{\rm turn-off}$ and ``plateau'' start) may be qualitatively reproduced by dilution of re-accreted metal-poor gas from the disc with the pristine composition of the stellar surface \citep{oomen2019depletion, oomen2020MESAdepletion}.

C, N, and O (CNO elements) are generally excluded from the depletion profiles, because it is difficult to separate their 
depletion from the effects of nucleosynthetic and mixing processes that occur during AGB/RGB phase \citep{mohorian2024EiSpec, menon2024EvolvedBinaries}. Current stellar evolution models predict that surface abundances of C and N are significantly modified by mixing processes on AGB/RGB, while surface abundance of O is unaffected by mixing processes in low-mass ($M<2M_\odot$) AGB stars and in RGB stars \citep{ventura2008aton3, karakas2014dawes, ventura2020CNOinAGB, kobayashi2020OriginOfElements, kamath2023models}. This highlights the complexity of disentangling chemical impacts on CNO abundances from evolution and re-accretion.

In this study, we further explore the disc-binary interaction by systematically investigating chemical depletion in post-AGB/post-RGB binary stars. Our focus is on the subset of post-AGB/post-RGB binaries hosting second-generation transition discs, as these systems share key similarities with young stellar objects (YSOs) that host planet-forming discs. The goals of this study are: i) to homogeneously derive the depletion profiles of transition disc targets, ii) to examine the connection between depletion and other observational parameters, and iii) to establish a comparative study between depletion patterns observed in post-AGB/post-RGB binary stars, YSOs, and the interstellar medium (ISM). To achieve our goals, we derived precise atmospheric parameters and elemental abundances using 1D local thermodynamic equilibrium (LTE) models for chemical elements from C to Eu. In addition, we accounted for 1D non-LTE (NLTE) effects for a representative set of chemical elements (C, N, $\alpha$-elements, and Fe). In Section~\ref{sec:tar}, we provide an overview of our target sample. In Section~\ref{sec:dob}, we introduce the photometric and spectroscopic data used in our research. In Section~\ref{sec:san}, we present the methodology of deriving atmospheric parameters and elemental abundances. In Section~\ref{sec:res}, we present results of our detailed abundance analysis of transition disc targets. In Section~\ref{sec:dpl}, we correlate the obtained chemical depletion profiles with other parameters of the studied binaries (astrometric, photometric, spectroscopic, pulsational, and orbital parameters) and compare with chemical depletion in young stars and ISM. Finally, in Section~\ref{sec:con}, we present our conclusions.

\section{Target sample}\label{sec:tar}
\begin{figure*}
    \centering
    \includegraphics[width=.99\linewidth]{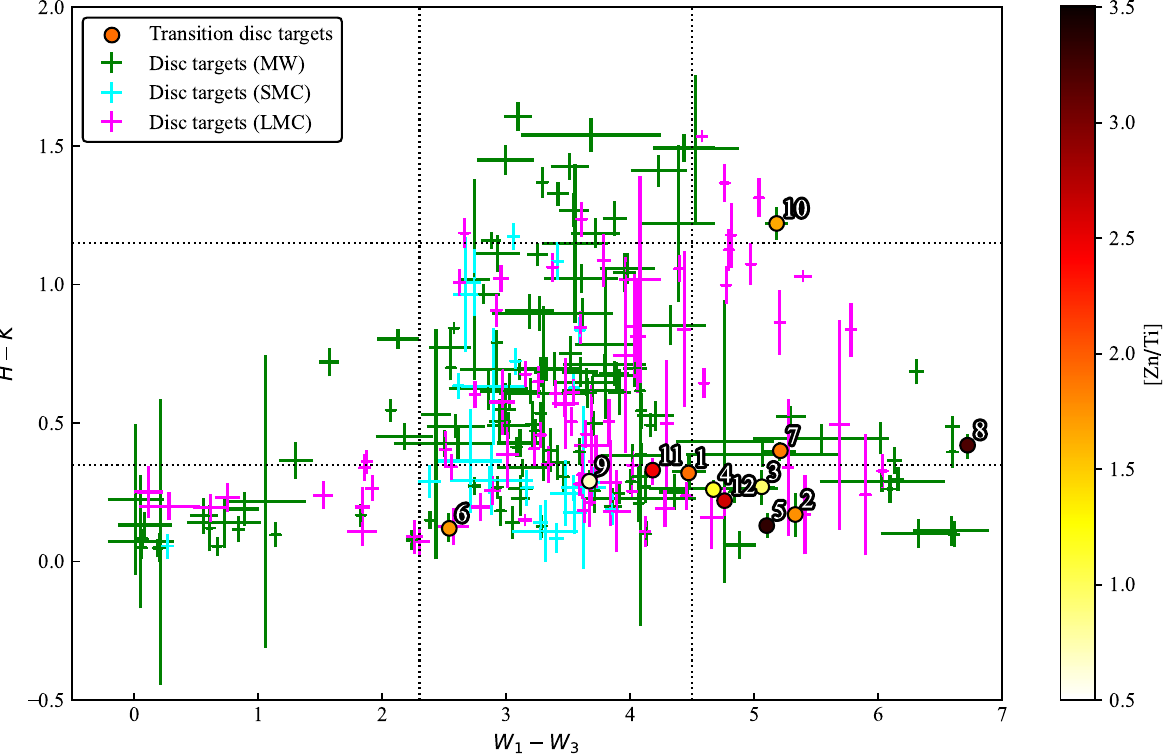}
    \caption{Updated IR colour-colour plot of post-AGB/post-RGB binary stars in the Galaxy and in the Magellanic Clouds. NIR magnitudes ($H$ and $K$) are adopted from 2MASS 6X, while MIR magnitudes ($W_1$ and $W_3$) are adopted from AllWISE (for more details, see Section~\ref{sec:tar}). Grey error bars represent disc targets in the Galaxy \citep{kluska2022GalacticBinaries}, green error bars represent disc targets in the SMC \citep{kamath2014SMC}, blue error bars represent disc targets in the LMC \citep{kamath2015LMC}. Our sample of transition disc targets is marked with circles coloured by [Zn/Ti] abundance ratios from this study (see Table~\ref{tab:fnlabu}). Adopted contours represent the rough demarcation between different disc categories \citep{kluska2022GalacticBinaries}.}\label{fig:colplt}
\end{figure*}

In this study, we focus on post-AGB/post-RGB binary stars with disc-type SED\footnote{The disc-type SED is characteristic of the binarity of the host star \citep{gezer2015WISERVTau, kamath2016PostRGBDiscovery, oomen2018OrbitalParameters, kluska2022GalacticBinaries}.} of the transition type (see Fig.~\ref{fig:colplt} and Table~\ref{tab:tarlst}). \citet{kluska2022GalacticBinaries} presented seven high-probability ($W_1-W_3>4.5$; category 2) and three moderate-probability ($2.3<W_1-W_3<4.5$, $H-K<0.3$, [Zn/Ti]>0.7 dex; category 3) transition disc candidates in the Galaxy. A mid-IR interferometric study by \citet{corporaal2023DiscParameters} confirmed the inner gaps in six of these Galactic candidates -- hence, they are referred to as transition disc stars (see targets \#1--\#6 in Table~\ref{tab:tarlst}). The remaining subsample of transition disc targets (targets \#7--\#10) are referred to as transition disc candidates for the rest of the paper.

We complemented the Galactic subsample of transition disc targets with two moderate-probability transition disc candidates (targets \#11 and \#12) in the Large Magellanic Cloud (LMC) following the procedure outlined in \citet{kluska2022GalacticBinaries}. In brief, we used the near-IR colours \citep[$H-K$ from 2MASS;][]{skrutskie20062MASS} and mid-IR colours \citep[$W_1-W_3$ from WISE;][]{wright2010WISE} to select new targets from the overall sample of post-AGB/post-RGB binary stars with disc-type SEDs in the Small Magellanic Cloud (SMC) \citep{kamath2014SMC} and in the LMC \citep{kamath2015LMC}. From all SMC and LMC targets in categories 2 and 3, we selected those for which high-resolution optical spectra were available (see Fig.~\ref{fig:colplt}).

We note that we excluded SS Lep and V777 Mon (Red Rectangle) from our preliminary sample despite their high mid-IR excesses because of the large corresponding uncertainties ($(W_1-W_3)_{\rm SS~Lep}\,=\,4.76\pm0.39$ mag, $(W_1-W_3)_{\rm V777~Mon}\,=\,4.53\pm0.35$ mag) as compared to the other 12 targets, which had the uncertainty in $W_1-W_3$ below 0.08 mag. We also note that transition disc candidates AF Crt (\#8) and V1504 Sco (\#10) are observed edge-on \citep{vanwinckel2012AFCrt, kluska2022GalacticBinaries}, which makes their luminosity estimates (see Section~\ref{ssec:doblum}) less reliable. Nevertheless, despite high inclinations, the depletion profiles of AF Crt (\#8) and V1504 Sco (\#10) are similar to those of other transition disc targets.

Our final sample consists of 12 transition disc targets: six confirmed transition disc stars (all in the Galaxy) and six transition disc candidates (four in the Galaxy and two in the LMC). We note that the surface composition of all 12 targets in our sample was previously studied by different groups using different methods (see Table~\ref{tab:litpar}). In this study, we perform a homogeneous analysis of our targets within the context of disc-binary interaction, accounting for NLTE effects, and re-define depletion efficiency in post-AGB/post-RGB binaries with transition discs. We also note that all targets are Type II Cepheid variables, which allowed us to calculate their luminosities using period-luminosity-colour (PLC) relation (see Section~\ref{ssec:doblum}). We present the target details individually in Appendix~\ref{app:lit}.

\begin{table*}
    \caption{Names, coordinates, and selection criteria of the target sample (see Section~\ref{sec:tar}). $H$ and $K$ magnitudes were adopted from the 2MASS 6X catalogue; $W_1$ and $W_3$ magnitudes were adopted from the AllWISE catalogue. [Zn/Ti] abundance ratio is a proxy for the efficiency of depletion process (see Section~\ref{sec:int}).} \label{tab:tarlst}
    \begin{tabular}{|c@{\hspace{0.001cm}}|c@{\hspace{0.001cm}}|c@{\hspace{0.001cm}}|c@{\hspace{0.001cm}}|c@{\hspace{0.001cm}}|c@{\hspace{0.001cm}}|c@{\hspace{0.001cm}}|c@{\hspace{0.001cm}}|c@{\hspace{0.001cm}}|c@{\hspace{0.001cm}}|}
    \hline
        ~ & \multicolumn{3}{c}{Names} & \multicolumn{2}{c}{Coordinates} & \multicolumn{3}{c}{Selection criteria} \\
        ID & IRAS/OGLE & 2MASS & Variable & R.A. & Dec. & $H-K$ & $W_1-W_3$ & [Zn/Ti] \\
        ~ & ~ & ~ & ~ & (deg) & (deg) & (mag) & (mag) & (dex) \\\hline
        \multicolumn{9}{c}{Transition disc stars \citep{corporaal2023DiscParameters}} \\\hline
        1 & 06072+0953 & J06095798+0952318 & CT Ori & 092.491630 & +09.875519 & 0.324 & 4.473 & 1.9$^a$ \\
        2 & 06472--3713 & J06485640--3716332 & ST Pup & 102.235100 & --37.275900 & 0.168 & 5.329 & 2.1$^b$ \\
        3 & 12067--4508 & J12092381--4525346 & RU Cen & 182.348800 & --45.426400 & 0.268 & 5.059 & 1.0$^c$ \\
        4 & 18281+2149 & J18301623+2152007 & AC Her & 277.567500 & +21.866670 & 0.263 & 4.672 & 0.7$^d$ \\
        5 & 18564--0814 & J18590869--0810140 & AD Aql & 284.786283 & --08.170671 & 0.132 & 5.103 & 2.5$^d$ \\
        6 & 19163+2745 & J19181955+2751031 & EP Lyr & 289.581300 & +27.850890 & 0.123 & 2.537 & 1.3$^e$ \\ \hline
        \multicolumn{9}{c}{Transition disc candidates \citep[categories 2 and 3 in][]{kluska2022GalacticBinaries}} \\\hline
        7 & 06034+1354 & J06061490+1354191 & DY Ori & 091.562130 & +13.905310 & 0.398 & 5.206 & 2.1$^e$ \\
        8 & 11472--0800 & J11494804--0817204 & AF Crt & 177.450151 & --08.289059 & 0.417 & 6.721 & 3.4$^f$ \\
        9 & 16278--5526 & J16315414--5533074 & GZ Nor & 247.975624 & --55.552108 & 0.288 & 3.669 & 0.8$^g$ \\
        10 & 17233--4330 & J17265864--4333135 & V1504 Sco & 261.744331 & --43.553756 & 1.221 & 5.179 & 1.4$^h$ \\
        11 & LMC--029$^1$ & J05030498--6840247 & LMC V0770 & 075.770630 & --68.673500 & 0.326 & 4.181 & 2.3$^i$ \\
        12 & LMC--147$^1$ & J05315099--6911463 & LMC V3156 & 082.962495 & --69.196213 & 0.218 & 4.758 & 2.5$^j$ \\ \hline
    \end{tabular}\\
    \textbf{Notes:} $^1$OGLE names of two LMC targets were shortened from OGLE LMC-T2CEP-029 and OGLE LMC-T2CEP-147 to LMC-029 and LMC-147, respectively. The superscripts of the [Zn/Ti] values indicate the individual chemical abundance studies: $^a$\cite{gonzalez1997CTOri}; $^b$\cite{gonzalez1996STPup}; $^c$\cite{maas2002RUCenSXCen}; $^d$\cite{giridhar1998RVTauVars}; $^e$\cite{gonzalez1997EPLyrDYOriARPupRSgt}; $^f$\cite{vanwinckel2012AFCrt}; $^g$\cite{gezer2019GKCarGZNor}; $^h$\cite{maas2005DiscPAGBs}; $^i$\cite{kamath2019depletionLMC}; $^j$\cite{reyniers2007LMC147}.
\end{table*}
\begin{table*}
    \caption{Literature data on photometric parameters, orbital parameters, luminosity estimates, and depletion parameters of transition disc targets (see Section~\ref{sec:tar}).} \label{tab:litpar}
    \begin{tabular}{|c@{\hspace{0.05cm}}|c@{\hspace{0.05cm}}|c@{\hspace{0.05cm}}|c@{\hspace{0.05cm}}|c@{\hspace{0.05cm}}|c@{\hspace{0.05cm}}|c@{\hspace{0.05cm}}|c@{\hspace{0.05cm}}|c@{\hspace{0.05cm}}|c@{\hspace{0.05cm}}|c@{\hspace{0.05cm}}|}
    \hline
        && \multicolumn{2}{c}{Photometric parameters} & \multicolumn{2}{c}{Orbital parameters} & \multicolumn{3}{c}{Previous luminosity estimates} & \multicolumn{2}{c}{Depletion parameters} \\
        ID & Name & RVb & $P_{\rm puls}$ & $P_{\rm orb}$ & $e$ & $L_{\rm SED}$ & $L_{\rm PLC}$ & $L_{\rm IR}/L_*$ & $T_{\rm turn-off}$ & Profile \\
        ~ & ~ & ~ & (d) & (d) & ~ & ($L_\odot$) & ($L_\odot$) & ~ & (K) & ~ \\\hline
        1 & CT Ori & no & 33.65$^f$ & - & - & 15100$^c$ & - & 0.55$^d$ & 1200 & S \\
        2 & ST Pup & no & 18.73$^e$ & 406$^b$ & 0.00$^b$ & 690$^c$ & - & 0.72$^d$ & 800 & S \\
        3 & RU Cen & no & 32.37$^g$ & 1489$^b$ & 0.62$^b$ & 1100$^c$& - & 0.40$^d$ & 800 & P \\
        4 & AC Her & no & 37.73$^h$ & 1189$^b$ & 0.00$^b$ & 2400$^c$ & 3600$^i$ & 0.21$^d$ & 1200 & U \\
        5 & AD Aql & no & 32.7$^h$ & - & - & 11500$^c$ & - & 0.51$^d$ & 1000 & S \\
        6 & EP Lyr & no & 41.59$^g$ & 1151$^b$ & 0.39$^b$ & 5500$^c$ & 7100$^i$ & 0.02$^d$ & 800 & P \\ \hline
        7 & DY Ori & no & 30.155$^a$ & 1248$^b$ & 0.22$^b$ & 21500$^c$ & - & 0.55$^d$ & 1000 & U \\
        8 & AF Crt & no & 31.5$^f$ & - & - & 280$^c$ & - & 1.83$^d$ & 1000 & S \\
        9 & GZ Nor & no & 36.2$^l$ & - & - & 1400$^c$ & - & 0.22$^d$ & 800 & P \\
        10 & V1504 Sco & yes & 22.0$^f$ & 735$^f$ & - & 1100$^c$ & - & 4.69$^d$ & 1000 & S \\
        11 & LMC V0770 & no & 31.245$^j$ & - & - & 3300$^m$ & 2629$^j$ & 0.63$^k$ & -- & -- \\
        12 & LMC V3156 & no & 46.795$^j$ & - & - & 5900$^j$ & 6989$^j$ & 0.84$^k$ & -- & -- \\ \hline
    \end{tabular}\\
    \textbf{Notes:} Literature data on depletion profiles is adopted from \cite{oomen2019depletion}: `S' means `saturated', `P' means `plateau', `U' means `uncertain'. The source list: $^a$\cite{pawlak2019ASAS}, $^b$\cite{oomen2018OrbitalParameters}, $^c$\cite{oomen2019depletion}, $^d$\cite{kluska2022GalacticBinaries}, $^e$\cite{walker2015STPup}, $^f$\cite{kiss2007T2Cepheids}, $^g$\cite{bodi2019RVTauVars}, $^h$\cite{giridhar1998RVTauVars}, $^i$\cite{bollen2022Jets}, $^j$\cite{manick2018PLC}, $^k$\cite{vanaarle2011PAGBsInLMC}, $^l$\cite{gezer2019GKCarGZNor}, $^m$\cite{kamath2015LMC}.\\
\end{table*}

\section{Data and observations}\label{sec:dob}
In this section, we present the photometric data (see Section~\ref{ssec:dobpht}) used to derive luminosities of transition disc targets. The luminosities were derived using SED fitting and PLC relation for Type II Cepheids (see Section~\ref{ssec:doblum}). We also present the spectroscopic data used to calculate atmospheric parameters and elemental abundances (see Section~\ref{ssec:dobspc}).

\subsection{Photometric data}\label{ssec:dobpht}
To obtain the SEDs of our target sample (see Appendix~\ref{app:sed}), we followed the procedure originally developed by \citet{degroote2013SEDfitting} and recently presented in \citet{mohorian2024EiSpec}. In brief, we collected the photometric magnitudes across various wavelength bands, which span from optical to far-infrared (far-IR; see Table~\ref{tabA:phomag}), including data from Johnson-Cousins system \citep{johnson1953Filters,cousins1976Filters}, Tycho-2 catalogue \citep{hog2000Tycho2}, Sloan Digital Sky Survey \citep[SDSS, ][]{york2000SDSSphotometry}, Two Micron All Sky Survey \citep[2MASS;][]{skrutskie20062MASS}, WISE \citep{wright2010WISE}, AKARI \citep{ishihara2010AKARI}, Infrared Astronomical Satellite \citep[IRAS;][]{neugebauer1984IRAS}, Photodetector Array Camera and Spectrometer \citep[PACS;][]{poglitsch2010PACS}, and Spectral and Photometric Imaging REceiver \citep[SPIRE;][]{griffin2010SPIRE}. In Appendix~\ref{app:sed}, we present the SEDs of transition disc targets, fitted with updated \textit{Gaia} DR3 distances (see Section~\ref{ssec:doblum}).

\subsection{Determination of luminosities from SED fitting and PLC relation}\label{ssec:doblum}
In this study, we determined the luminosities of the target sample using two methods: i) through SED fitting (referred to as SED luminosity, $L_{\rm SED}$) following the methodology outlined in \citet{mohorian2024EiSpec}; and ii) employing the PLC relation (referred to as PLC luminosity, $L_{\rm PLC}$) following the procedure outlined in \citet{menon2024EvolvedBinaries}. In this subsection, we provide a brief overview of these methods. In Table~\ref{tab:fnlvls}, we present the estimated SED and PLC luminosities.

To determine the SED luminosities, we selected appropriate Kurucz model atmospheres \citep{castelli2003ATLAS9} to fit the initial photometric data points (the bolometric IR luminosity $L_{\rm IR}$ was obtained through integration of star-subtracted IR excess) and computed the de-reddened model atmospheres for each target through an extensive parameter grid search \citep{mohorian2024EiSpec}. The search was performed by the minimisation of the $\chi^2$ value in the parameter space of four variables: effective temperature $T_{\rm eff}$, surface gravity $\log g$, total reddening (extinction parameter) $E(B-V)$, and angular diameter of the star $\theta$. The total reddening comprised both interstellar and circumstellar contributions. For interstellar reddening, we adopted the wavelength-dependent extinction law \citep{cardelli1989SEDextinction}, assuming an $R_V$\,=\,3.1. Additionally, we used the Bailer-Jones geometric distances, denoted as $z_{\rm BJ}$, along with their corresponding lower and upper limits, $z_{\rm BJL}$ and $z_{\rm BJU}$ \citep{bailerjones2021distances}. These more precise geometric distances were computed based on \textit{Gaia} DR3 parallaxes, incorporating a direction-dependent prior distance. We note that throughout our computations, we assumed isotropic radiation emission from the stars. We also note that stellar variability was not accounted for, resulting in an increased $\chi^2$ value for high-amplitude variables. Once the solution for the de-reddened model atmosphere was found, we integrated it to obtain the SED luminosity $L_{\rm SED}$ and the relative bolometric IR luminosity $L_{\rm IR}/L_\ast\,=\,L_{\rm IR}/L_{\rm SED}$. The luminosity uncertainties were derived for each respective target by computing the standard deviation of the lower and upper luminosity bounds, caused by uncertainties of the geometric distances and the photometric data points.

To derive the PLC luminosity, we used the calibrated relation \citep{menon2024EvolvedBinaries} given by
\begin{equation}
    M_{bol} = m\times\log P_0 + c - \mu + BC + 2.55\times(V-I)_0,
\end{equation}
where $M_{bol}$ represents the absolute bolometric magnitude obtained using Wesenheit (colour-corrected) V-band magnitude \citep[$WI_V\,=\,V-2.55(V-I)_0$; see][]{ngeow2005wesenheit}, while the parameters $m\,=\,-3.59$ and $c\,=\,18.79$ correspond to the calibrated slope and intercept of the linear fit, respectively. $P_0$ represents the observed fundamental pulsation period in days, $\mu\,=\,18.49$ denotes the distance modulus to the LMC, $BC$ signifies the bolometric correction derived from the effective temperature \citep{flower1996BoloCorr, torres2010BoloCorrErrata}, and $(V-I)_0$ denotes the intrinsic (de-reddened) colour of each star (reddening value is adopted from SED fits). The uncertainties of PLC luminosity are primarily influenced by the uncertainties of reddening.

We note that SED luminosities require \textit{Gaia} parallax measurements (underlying the derivation of geometric distances), which are plagued by orbital motion in case of binary systems \citep{kluska2022GalacticBinaries}. Additionally, since the targets in this study are Type II Cepheid pulsating variables with periods ranging from 18.73 to 46.7 days (see Table~\ref{tab:litpar}), their atmospheric parameters, particularly effective temperature $T_{\rm eff}$, surface gravity $\log g$, and microturbulent velocity $\xi_t$, undergo significant variations throughout the pulsation cycle \citep{mohorian2024EiSpec}. This causes a considerable scatter in the photometric data points leading to increased uncertainties of SED luminosity (see Appendix~\ref{app:sed}). Hence, we regard the PLC luminosity to be more precise and reliable compared to the SED luminosity in our targets.

\begin{table*}
    \caption{Derived luminosities and atmospheric parameters of transition disc targets. The columns are as follows: col. 1: Target ID; col. 2: Target name; col. 3: SED luminosity (see Section~\ref{ssec:doblum}); col. 4: infrared luminosity; col. 5: PLC luminosity (adopted; see Section~\ref{ssec:doblum}); col. 6, 7, 8, 9: derived atmospheric parameters (see Section~\ref{ssec:respro}).}\label{tab:fnlvls}
    \begin{tabular}{|c|c|c|c|c|c|c|c|c|}\hline
        \multirow{2}{*}{ID} & \multirow{2}{*}{Name} & \multirow{2}{*}{$\log\dfrac{L_{\rm SED}}{L_\odot}$} & \multirow{2}{*}{$\log\dfrac{L_{\rm IR}}{L_{\rm SED}}$} & \multirow{2}{*}{$\log\dfrac{L_{\rm PLC}}{L_\odot}$} & $T_{\rm eff}$ & $\log g$ & [Fe/H] & $\xi_{\rm t}$ \\
        &&&&& (K) & (dex) & (dex) & (km/s) \\ \hline
        1 & CT Ori & 3.41$\pm$0.13 & --0.26$\pm$0.13 & 3.26$\pm$0.16 & 5940$\pm$120 & 1.01$\pm$0.18 & --1.89$\pm$0.11 & 3.37$\pm$0.10 \\
        2 & $^\ast$ST Pup & 2.87$\pm$0.08 & --0.14$\pm$0.08 & 2.96$\pm$0.17 & 5340$\pm$80 & 0.20$\pm$0.10 & --1.92$\pm$0.08 & 2.83$\pm$0.03 \\
        3 & RU Cen & 4.00$\pm$0.22 & --0.40$\pm$0.22 & 3.50$\pm$0.27 & 6120$\pm$80 & 1.46$\pm$0.15 & --1.93$\pm$0.08 & 3.26$\pm$0.10 \\
        4 & AC Her & 3.79$\pm$0.11 & --0.68$\pm$0.11 & 3.71$\pm$0.21 & 6140$\pm$100 & 1.27$\pm$0.16 & --1.47$\pm$0.08 & 3.92$\pm$0.12 \\
        5 & AD Aql & 3.41$\pm$0.19 & --0.29$\pm$0.19 & 2.84$\pm$0.29 & 6200$\pm$170 & 1.67$\pm$0.45 & --2.20$\pm$0.09 & 2.98$\pm$0.36 \\
        6 & EP Lyr & 3.74$\pm$0.16 & --1.70$\pm$0.16 & 3.96$\pm$0.23 & 6270$\pm$160 & 1.24$\pm$0.18 & --2.03$\pm$0.17 & 2.48$\pm$0.10 \\ \hline
        7 & $^\ast$DY Ori & 2.81$\pm$0.10 & --0.26$\pm$0.13 & 3.50$\pm$0.17 & 6160$\pm$70 & 0.88$\pm$0.14 & --2.03$\pm$0.04 & 2.48$\pm$0.09 \\
        8 & $^\ast$AF Crt & -- & 0.26$\pm$0.08 & 2.51$\pm$0.17 & 6110$\pm$110 & 0.96$\pm$0.21 & --2.47$\pm$0.05 & 4.87$\pm$0.16 \\ 
        9 & $^\ast$GZ Nor & 3.24$\pm$0.15 & --0.66$\pm$0.22 & 2.71$\pm$0.29 & 4830$\pm$20 & 0.00$\pm$0.18 & --1.89$\pm$0.11 & 5.95$\pm$0.18 \\
        10 & V1504 Sco & -- & 0.67$\pm$0.11 & 3.71$\pm$0.35 & 5980$\pm$90 & 0.98$\pm$0.17 & --1.05$\pm$0.07 & 4.29$\pm$0.05 \\ 
        11 & $^\ast$LMC V0770 & 3.42$\pm$0.11 & --0.20$\pm$0.19 & 3.46$\pm$0.20 & 5750$\pm$100 & 0.00$\pm$0.18 & --2.61$\pm$0.05 & 2.20$\pm$0.01 \\
        12 & $^\ast$LMC V3156 & 3.75$\pm$0.08 & --0.08$\pm$0.16 & 3.99$\pm$0.19 & 6160$\pm$130 & 1.38$\pm$0.20 & --2.48$\pm$0.04 & 4.28$\pm$0.12 \\ \hline
    \end{tabular}\\
    \textbf{Notes:} We highlight with asterisks ($^\ast$) those targets, for which we used the ATLAS9 model atmospheres (see Section~\ref{ssec:sanlte}). For the uncertainty of infrared luminosity, we assume that the error of SED luminosity, primarily due to reddening uncertainty, dominates over the error from fitting the infrared bump, which is influenced by the uncertainty in photometric data points. SED luminosities of edge-on targets AF Crt (\#8) and V1504 Sco (\#10) were removed from the table as SED fitting is not applicable when the optical light is dominated by scattering.
\end{table*}

\subsection{Spectroscopic data}\label{ssec:dobspc}
In this subsection, we present the optical high-resolution spectra used in this study, which were obtained from the High Efficiency and Resolution Mercator Echelle Spectrograph mounted on Mercator telescope (HERMES/Mercator; see Section~\ref{sssec:dobspcher}) and Ultraviolet and Visual Echelle Spectrograph mounted on Very Large Telescope (UVES/VLT; see Section~\ref{sssec:dobspcuvs}). In Table~\ref{tab:obslog}, we present the final selection of spectral visits (for all considered visits, see Appendix~\ref{app:vis}). In Fig.~\ref{fig:spcmon}, we show exemplary spectral regions for all targets, focusing on the S and Zn lines, which are crucial for deriving the initial metallicity [M/H]$_{\rm 0,min}$, because Fe is depleted in transition disc targets (see Section~\ref{sec:san}).

To ensure the accuracy of the abundance analysis of the pulsating transition disc targets, our selection of optical visits depended on the time span of observational dataset for each target. Our selection strategy is as follows (see Table~\ref{tab:obslog}):
\begin{enumerate}
    \item ST Pup (\#2), RU Cen (\#3), GZ Nor (\#9), V1504 Sco (\#10): We used the only available spectrum for each of these targets.
    \item LMC V3156 (\#11) and LMC V0770 (\#12): Multiple spectra of these targets were taken in less than two-day span (1.08 days and 0.07 days, respectively). Since the pulsation periods of these targets are greater than 30 days, we merged the RV-corrected spectra of each target to increase the S/N ratio.
    \item CT Ori (\#1), AC Her (\#4), AD Aql (\#5), EP Lyr (\#6), DY Ori (\#7), and AF Crt (\#8): These targets are part of long-term spectral monitoring program using HERMES (see Section~\ref{sssec:dobspcher}). Since the observations of these targets were generally taken with a gap of at least few days, we selected the individual visits of these targets with the highest S/N values.
\end{enumerate}

\begin{table}
    \caption{Observation log of the target sample. For more information on the selection criteria for the spectral visits, see Section~\ref{sec:tar}. In Appendix~\ref{app:vis}, we provide the full observational log of transition disc targets.} \label{tab:obslog}
    \begin{tabular}{|c@{\hspace{0.01cm}}|c@{\hspace{0.01cm}}|c@{\hspace{0.01cm}}|c@{\hspace{0.01cm}}|c@{\hspace{0.01cm}}|c@{\hspace{0.01cm}}|}
    \hline
        ID & Name & Facility & MJD & RV (km/s) & S/N \\ \hline
        1 & CT Ori & H/M & 56254.1414 & 47.12$\pm$0.96 & 45 \\
        2 & ST Pup & H/M & 51085.2822 & 23.03$\pm$0.38 & 50 \\
        3 & RU Cen & H/M & 51627.1188 & -0.40$\pm$0.28 & 55 \\
        4 & AC Her & H/M & 57876.1807 & -54.18$\pm$0.57 & 45 \\
        5 & AD Aql & H/M & 56870.9410 & 54.36$\pm$0.96 & 40 \\
        6 & EP Lyr & H/M & 55005.1064 & 29.37$\pm$1.29 & 45 \\ \hline
        7 & DY Ori & H/M & 58452.1440 & 11.67$\pm$0.90 & 40 \\
        8 & AF Crt & H/M & 56298.2332 & 27.72$\pm$1.08 & 40 \\
        9 & GZ Nor & U/V & 56819.4194 & -121.61$\pm$0.26 & 50 \\
        10 & V1504 Sco & H/M & 51626.3207 & 28.43$\pm$0.25 & 50 \\
        11 & LMC V0770 & U/V & 56589.1761 & 288.96$\pm$3.37 & 45 \\
        12 & LMC V3156 & U/V & 53409.9187 & 280.06$\pm$1.26 & 45 \\ \hline
    \end{tabular}\\
    \textbf{Notes:} H/M means HERMES/Mercator, U/V means UVES/VLT. RV is the radial velocity of the spectrum used in the analyses. S/N is the average signal-to-noise ratio of the spectrum.
\end{table}
\begin{figure*}
    \centering
    \includegraphics[width=.59\linewidth]{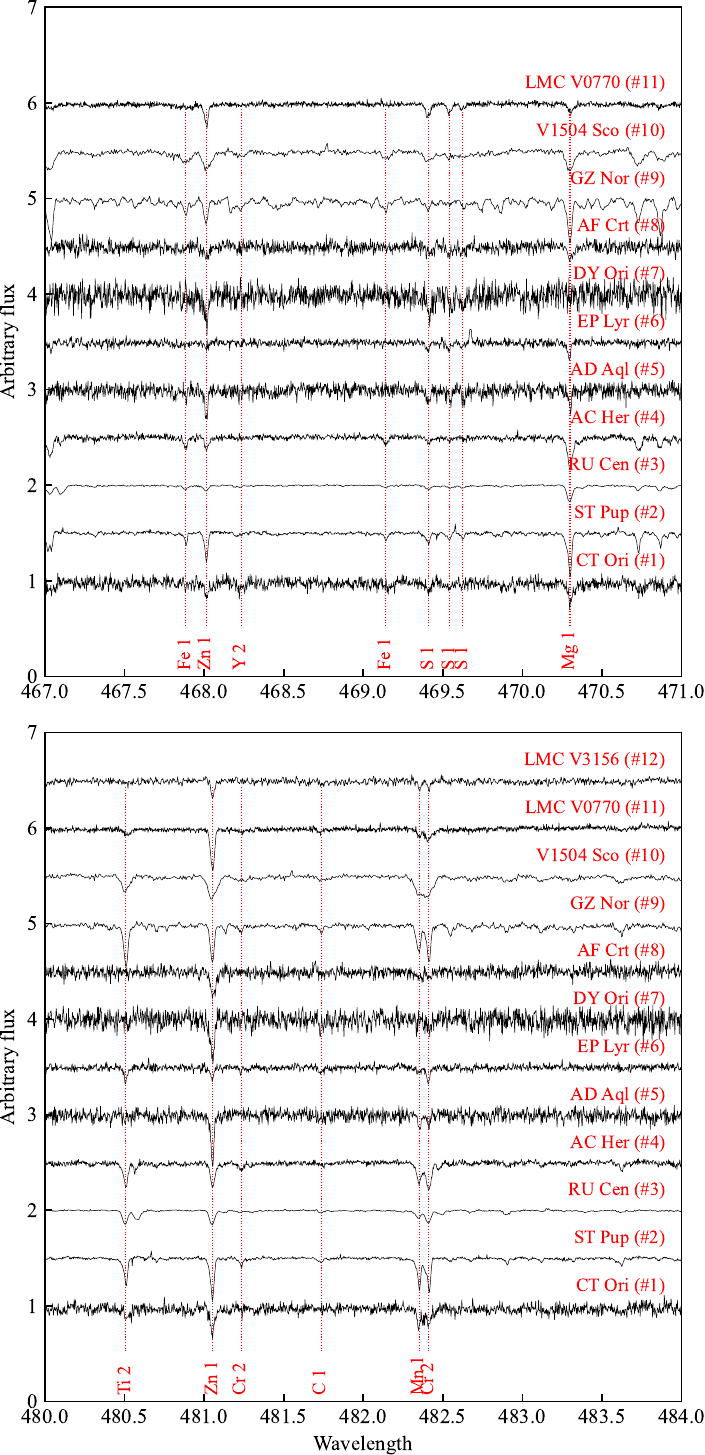}
    \caption{Comparison of the normalised spectra of all sample stars in two spectral regions containing lines of volatile elements S and Zn. All spectra are normalised, corrected for radial velocity, and shifted in flux for clarity. The dashed red vertical lines indicate positions of line peaks. The names of the stars are provided in the plot. The observed spectrum of LMC V3156 does not contain the region between 467.5 and 470.5 nm (for more details, see Section~\ref{ssec:dobspc}).}\label{fig:spcmon}
\end{figure*}

\subsubsection{HERMES spectra}\label{sssec:dobspcher}
For CT Ori (\#1), ST Pup (\#2), RU Cen (\#3), AC Her (\#4), AD Aql (\#5), EP Lyr (\#6), DY Ori (\#7), AF Crt (\#8), and V1504 Sco (\#10), we used the high-resolution (R\,=\,$\lambda/\Delta\lambda\sim$ 85 000) optical spectra obtained within an extensive monitoring initiative (June 2009--ongoing). This initiative involved HERMES \citep{raskin2011hermes} installed on the 1.2-m Mercator telescope at the Roque de los Muchachos Observatory, La Palma. 

This monitoring program resulted in the collection of a substantial dataset of high-resolution optical spectra for post-AGB systems, thoroughly detailed in \citet{vanwinckel2018Binaries}. The HERMES spectra of transition disc targets were reduced using the standard pipeline, as outlined in \citet{raskin2011hermes}. The complete log of HERMES observations can be found in Appendix~\ref{app:vis}, with the selected visits listed in Table~\ref{tab:obslog}.

\subsubsection{UVES spectra}\label{sssec:dobspcuvs}
For GZ Nor (\#9), LMC V0770 (\#11), and LMC V3156 (\#12) we employed high-resolution optical spectra (R\,=\,$\lambda/\Delta\lambda\sim$ 80 000 in the Blue arm; R\,=\,$\lambda/\Delta\lambda\sim$ 110 000 in the Red arm) obtained with the UVES \citep{dekker2000UVES}. UVES is mounted on the 8-m UT2 Kueyen Telescope at the VLT located at the Paranal Observatory of ESO in Chile.

To reduce the UVES spectra, we followed the standard steps for UVES reduction pipeline \citep[frame extraction, flat-field correction, wavelength calibration, cosmic clipping; see][]{dekker2000UVES}. The full observational log of UVES spectra can be found in Appendix~\ref{app:vis}, and the selected visits are listed in Table~\ref{tab:obslog}.

\section{Spectral analysis}\label{sec:san}
In this study, we used E-iSpec \citep[explained in detail in][]{mohorian2024EiSpec} -- a modified version of iSpec \citep{blancocuaresma2014, blancocuaresma2019} to investigate the chemical composition of the target sample. Our chemical analysis included precise derivation of atmospheric parameters (effective temperature $T_{\rm eff}$, surface gravity $\log g$, metallicity [Fe/H], and microturbulent velocity $\xi_{\rm t}$) and abundances [X/H] of all elements, which had detectable spectral features in the spectra of transition disc targets. Additionally, we used \texttt{Balder} \citep{amarsi2018Balder} to calculate NLTE corrections for the abundances of a representative set of 11 chemical elements: C, N, O, Na, Mg, Al, Si, S, K, Ca, and Fe. In the following subsections, we present the methodology of abundance analysis of transition disc stars and candidates (for individual depletion profiles, see Appendix~\ref{app:dpl}).

\subsection{LTE analysis using E-iSpec}\label{ssec:sanlte}
To calculate the atmospheric parameters and elemental abundances from atomic spectral lines, we followed the procedure presented in \citet{mohorian2024EiSpec}. In brief, we used the Moog radiative transfer code \citep[equivalent width method;][]{sneden1973moog} with the VALD3 line list \citep{kupka2011vald}, the recently updated solar abundances from \citet{asplund2021solar}, and LTE model atmospheres: spherically-symmetric MARCS models \citep{gustafsson2008MARCS} and plane-parallel ATLAS9 models \citep{castelli2003ATLAS9}\footnote{In general, spherically-symmetric models are known to perform better for giant stars due to their extended atmospheres \citep{meszaros2012ModelsAPOGEE}. However, half of our target sample had atmospheric parameters outside of MARCS parameter grid values; for such targets we used the ATLAS9 model atmospheres instead.}.

\begin{figure*}
    \centering
    \includegraphics[width=.38\linewidth]{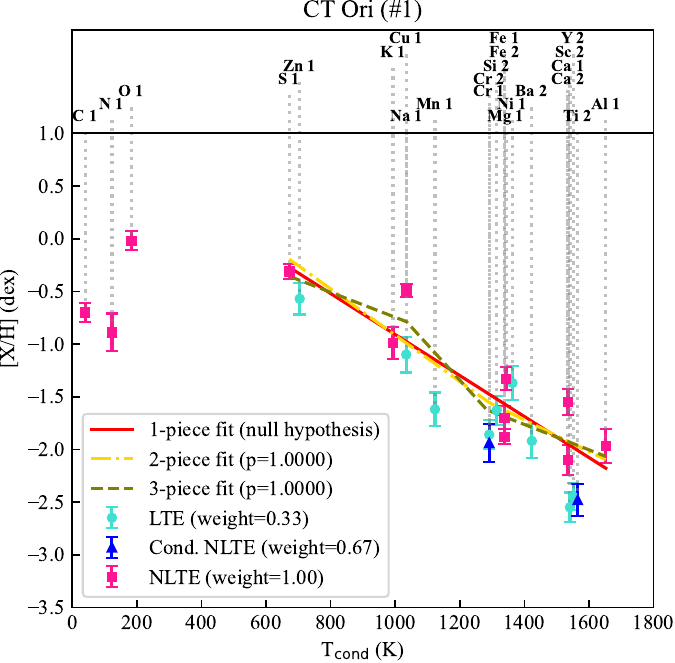}
    \includegraphics[width=.38\linewidth]{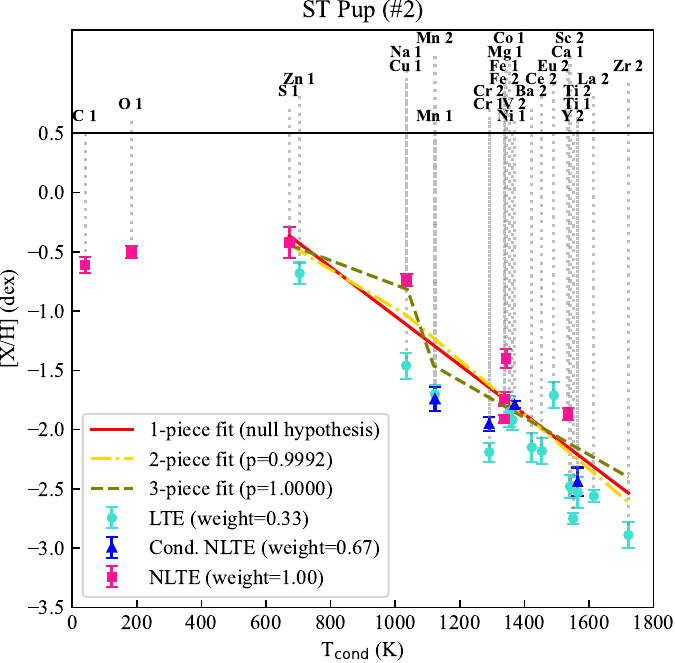}
    \includegraphics[width=.38\linewidth]{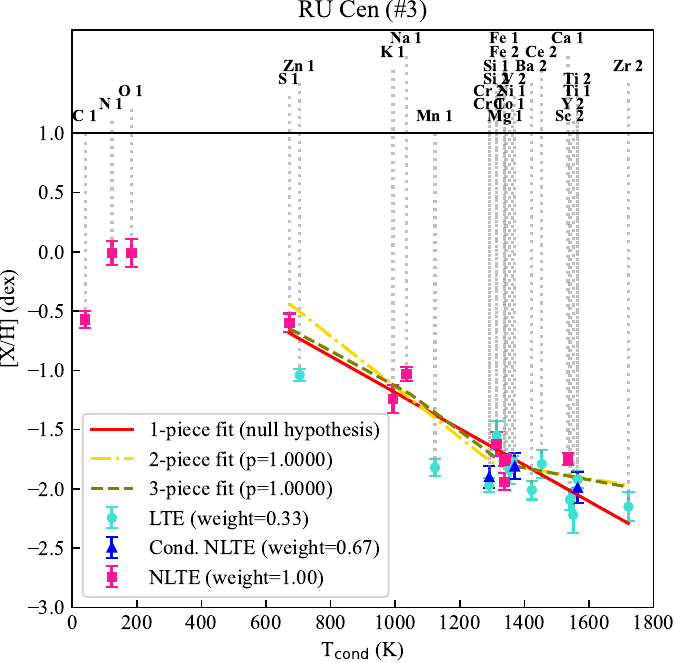}
    \includegraphics[width=.38\linewidth]{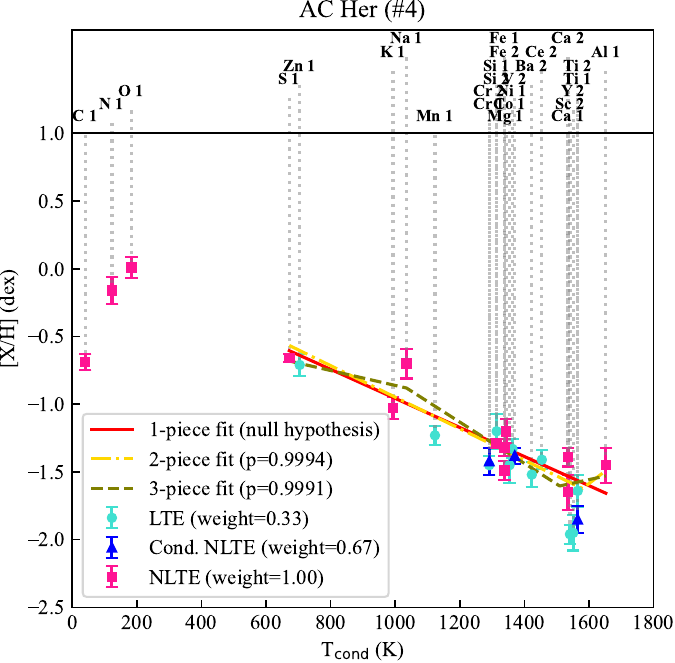}
    \includegraphics[width=.38\linewidth]{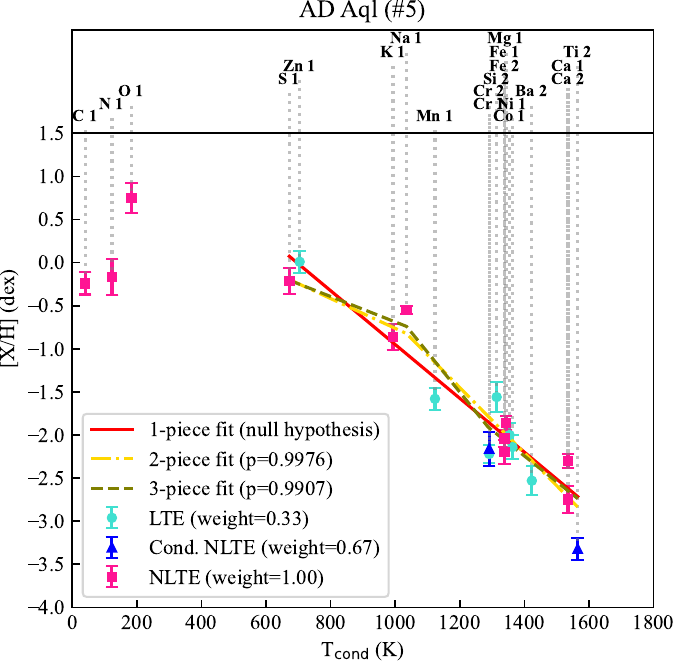}
    \includegraphics[width=.38\linewidth]{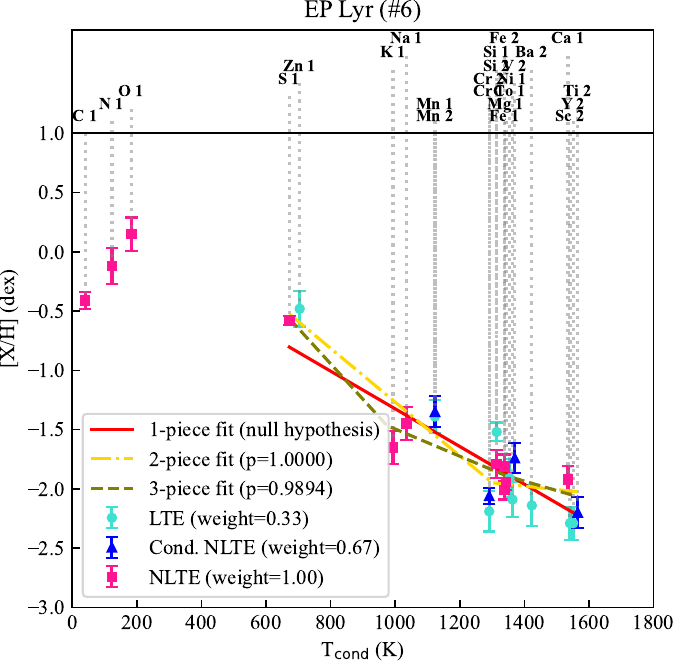}
    \caption{Elemental abundances of transition disc stars (CT Ori, ST Pup, RU Cen, AC Her, AD Aql, and EP Lyr) as functions of condensation temperature \citep{lodders2003CondensationTemperatures, wood2019CondensationTemperatures}. The legend for the symbols and colours used is included within the plot. ``Cond. NLTE'' means conditionally NLTE abundance (derived from spectral lines of \ion{Ti}{ii}, \ion{V}{ii}, \ion{Cr}{ii}, or \ion{Mn}{ii}; for more details, see Section~\ref{ssec:respro}).}\label{fig:dplstr}
\end{figure*}

\begin{figure*}
    \centering
    \includegraphics[width=.38\linewidth]{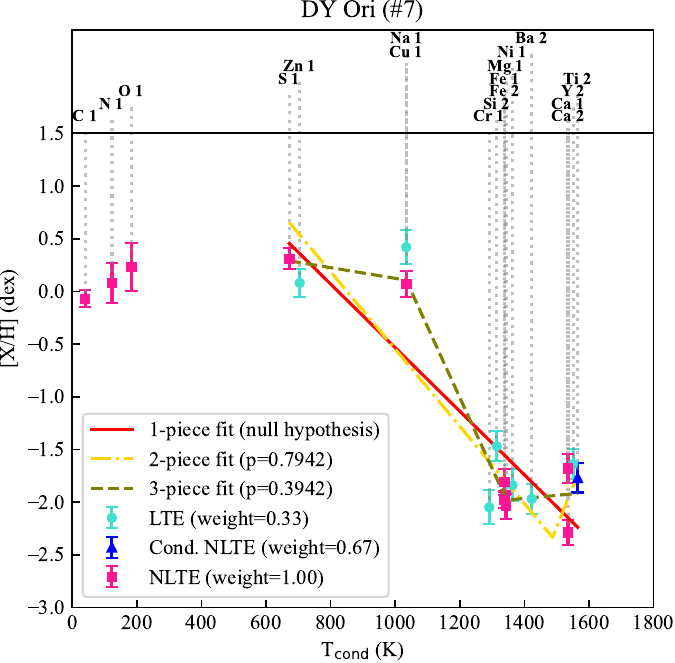}
    \includegraphics[width=.38\linewidth]{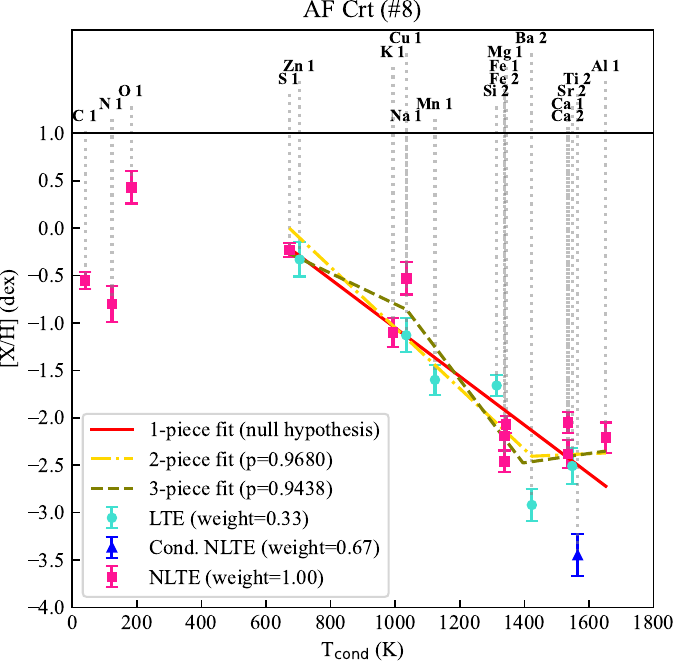}
    \includegraphics[width=.38\linewidth]{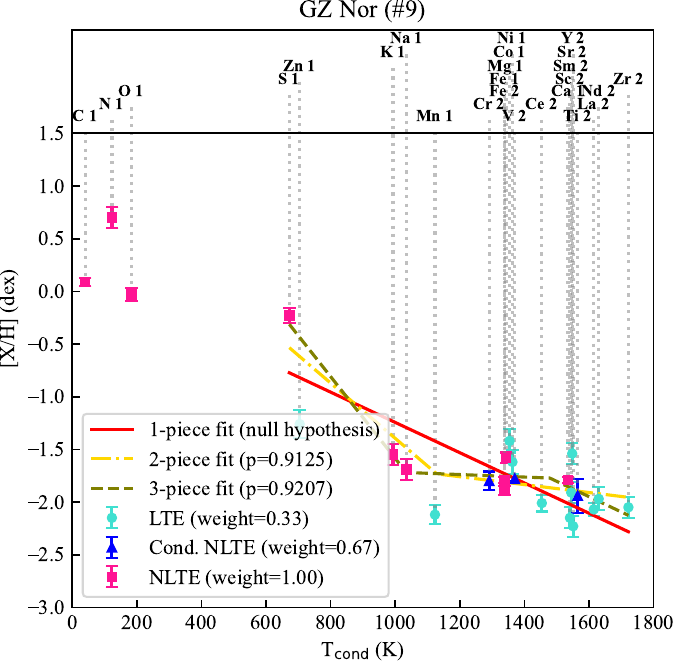}
    \includegraphics[width=.38\linewidth]{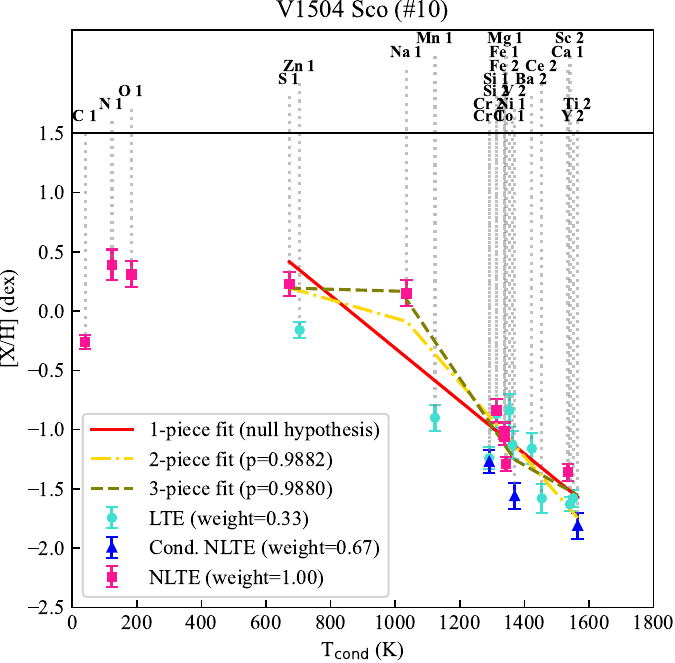}
    \includegraphics[width=.38\linewidth]{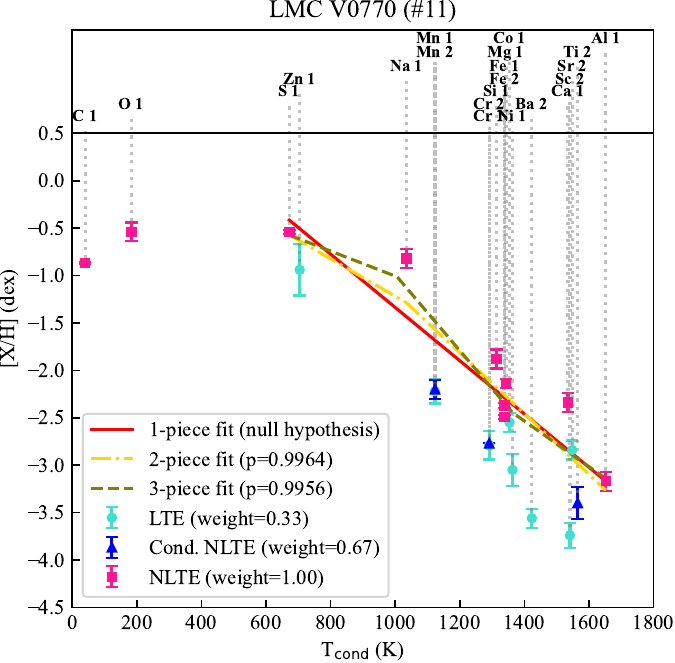}
    \includegraphics[width=.38\linewidth]{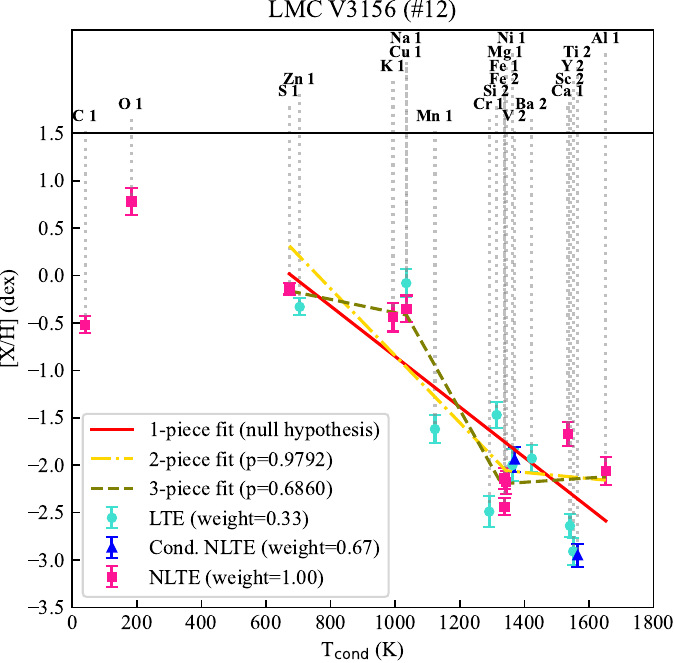}
    \caption{Elemental abundances of transition disc candidates (DY Ori, AF Crt, GZ Nor, V1504 Sco, LMC V0770, and LMC V3156) as functions of condensation temperature \citep{lodders2003CondensationTemperatures, wood2019CondensationTemperatures}. The legend for the symbols and colours used is included within the plot. ``Cond. NLTE'' means conditionally NLTE abundance (derived from spectral lines of \ion{Ti}{ii}, \ion{V}{ii}, \ion{Cr}{ii}, or \ion{Mn}{ii}; for more details, see Section~\ref{ssec:respro}.}\label{fig:dplcnd}
\end{figure*}

In the original version of E-iSpec \citep{mohorian2024EiSpec}, the uncertainties were calculated manually. In this study, we updated E-iSpec by automating the process of uncertainty calculation for elemental abundances, while following the same approach as in manual method. In brief, the total uncertainty is calculated as the sum in quadrature of random and systematic uncertainties. We note that we set the random uncertainty to be 0.1 dex for those ionisations, for which we used only one spectral line to derive the elemental abundance. We assume that the abundance change caused by metallicity affects [X/Fe], but does not affect [X/H]. In Fig.~\ref{fig:dplstr} and \ref{fig:dplcnd}, we show the derived elemental abundances with uncertainties for all transition disc stars and candidates, respectively. Furthermore, we updated E-iSpec with a functionality to recalculate atmospheric parameters and elemental abundances using NLTE abundance corrections from \texttt{Balder} (see Section~\ref{ssec:sannlte}).

In Table~\ref{tab:fnlvls}, we present the estimated SED and PLC luminosities (see Section~\ref{ssec:doblum}) and the derived atmospheric parameters of the target sample (the targets, for which we used ATLAS9 model atmospheres, are highlighted with asterisks). In Tables~\ref{tab:fnlabu} and \ref{tabA:fnlabu}, we provide the results of our abundance analysis using E-iSpec: the selected abundance ratios and the elemental abundances, respectively all 12 targets is provided in Appendix~\ref{app:lst}.

\begin{table*}
    \caption{Selected abundance ratios of transition disc targets (see Table~\ref{tabA:fnlabu} for a full list of [X/H] abundances). The columns are as follows: col. 1: Target ID; col. 2: Target name; col. 3, 4, 5: proxies for depletion efficiency (we adopt the NLTE-corrected [S/Ti] ratio, see Section~\ref{ssec:reseff}); col. 6: NLTE-corrected C/O ratio; col. 7: minimal initial metallicity (see Section~\ref{ssec:reseff}); col. 8: depletion strength (see Section~\ref{ssec:dplyso}).}\label{tab:fnlabu}
    \begin{tabular}{|c|c|c|c|c|c|c|c|}\hline
        ~ & ~ & LTE & LTE & NLTE & NLTE & NLTE & NLTE \\
        ID & Name & [Zn/Ti] & [Zn/Fe] & [S/Ti] & C/O & [M/H]$_{\rm0,min}$ & $\Delta_{\rm g/d}$ \\
        ~ & ~ & (dex) & (dex) & (dex) & ~ & (dex) & ~ \\ \hline
        1 & CT Ori & 1.91$\pm$0.30 & 1.31$\pm$0.23 & 2.17$\pm$0.25 & 0.12$\pm$0.04 & --0.31$\pm$0.10 & 3720 \\
        2 & ST Pup & 1.76$\pm$0.21 & 1.23$\pm$0.15 & 2.02$\pm$0.35 & 0.46$\pm$0.11 & --0.42$\pm$0.23 & 3090 \\
        3 & RU Cen & 0.95$\pm$0.18 & 0.90$\pm$0.14 & 1.39$\pm$0.28 & 0.16$\pm$0.06 & --0.60$\pm$0.15 & 2190 \\
        4 & AC Her & 1.14$\pm$0.18 & 0.78$\pm$0.18 & 1.19$\pm$0.14 & 0.12$\pm$0.03 & --0.66$\pm$0.04 & 680 \\
        5 & AD Aql & 3.33$\pm$0.26 & 2.21$\pm$0.28 & 3.11$\pm$0.34 & 0.06$\pm$0.03 & --0.21$\pm$0.21 & 9770 \\
        6 & EP Lyr & 1.72$\pm$0.28 & 1.53$\pm$0.31 & 1.62$\pm$0.20 & 0.16$\pm$0.06 & --0.58$\pm$0.07 & 2690 \\ \hline
        7 & DY Ori & 1.85$\pm$0.27 & 2.06$\pm$0.22 & 2.08$\pm$0.35 & 0.30$\pm$0.15 & 0.31$\pm$0.21 & 19500 \\
        8 & AF Crt & 3.12$\pm$0.40 & 2.13$\pm$0.30 & 3.22$\pm$0.31 & 0.06$\pm$0.03 & --0.23$\pm$0.09 & 16980 \\
        9 & GZ Nor & 0.68$\pm$0.29 & 0.63$\pm$0.22 & 1.71$\pm$0.28 & 0.78$\pm$0.16 & --0.23$\pm$0.12 & 4570 \\
        10 & V1504 Sco & 1.65$\pm$0.18 & 0.90$\pm$0.20 & 2.04$\pm$0.11 & 0.16$\pm$0.05 & 0.23$\pm$0.00 & 1950 \\
        11 & LMC V0770 & 2.46$\pm$0.44 & 1.55$\pm$0.32 & 2.86$\pm$0.19 & 0.28$\pm$0.06 & --0.54$\pm$0.02 & 8910 \\
        12 & LMC V3156 & 2.62$\pm$0.21 & 2.11$\pm$0.19 & 2.81$\pm$0.19 & 0.03$\pm$0.01 & --0.14$\pm$0.07 & 19950 \\ \hline
    \end{tabular}
\end{table*}

\subsection{NLTE abundance corrections using Balder}\label{ssec:sannlte}
The departures from LTE typically (but not always) grow with increasing temperature $T_{\rm eff}$, decreasing surface gravity $\log g$, or decreasing metallicity [Fe/H] \citep{lind2012NLTE}. These departures primarily stem from the intense radiation field at shorter wavelengths, which is not effectively compensated by thermal collisions in the stellar photosphere. Among the various species, neutral and relatively low-ionisation elements like \ion{Fe}{i} or \ion{Ti}{i} experience overionisation, resulting in weakened spectral lines. Conversely, the dominant ionisation stage like \ion{Fe}{ii} or \ion{Ti}{ii} closely follows the Saha distribution, and remains mostly unaffected by NLTE effects \citep{amarsi2016NLTEiron, amarsi2022NLTEiron}. Thus, we adopted [Fe/H] abundance derived from \ion{Fe}{ii} lines as the metallicity and applied NLTE corrections to elemental abundances \citep[see][and references therein]{amarsi2020NLTEgalah}. 

To study the chemical depletion in post-AGB/post-RGB binary stars with transition discs, we selected C, N, O, Na, Mg, Al, Si, S, K, Ca, and Fe as a representative set of chemical elements. We calculated the NLTE corrections for each individual studied spectral line of these elements using the code \texttt{Balder}, which is based on \texttt{Multi3D} \citep{leenaarts2009Multi3D}.

The method of NLTE correction calculation closely follows that described in \citet{amarsi2020NLTEgalah}. We list the model atoms used in this work in Table~\ref{tab:nonref}. The NLTE calculations were performed on a small number of LTE spherically-symmetric MARCS model atmospheres \citep{gustafsson2008MARCS} using two small grids: the hotter grid spanned 5\,500\,K\,$\leq\,T_{\rm eff}\,\leq$\,6\,250\,K, 1.0\,dex\,$\leq\,\log g\,\leq$\,2.0\,dex, and --2.5\,dex\,$\leq\,[{\rm Fe/H}]\,\leq$\,--1.0\,dex, while the cooler grid spanned 4\,750\,K\,$\leq\,T_{\rm eff}\,\leq$\,5\,250\,K, $\log g\,=\,0.5$\,dex, and --2.0\,dex\,$\leq\,[{\rm Fe/H}]\,\leq$\,--1.5\,dex. Calculations were performed for a range of abundances from $\mathrm{[X/H]\,=\,-4.5}$ to $+2.0$ in steps of $0.5\,\mathrm{dex}$, and microturbulences $\xi_{\rm t}$\,=\,2.0 km/s and 5.0 km/s. The theoretical equivalent widths were used to calculate NLTE abundance corrections for each individual spectral line $i$, denoted as $\Delta_{i}^{\rm diff.}$. These were interpolated onto the stellar parameters of interest, with edge values adopted for stars outside of this theoretical grid. The resulting line-by-line NLTE abundances were given by:
\begin{equation}
    [{\rm X/H}]_{i}^{\rm NLTE} = [{\rm X/H}]_{i}^{\rm LTE} + \Delta_{i}^{\rm diff.},
\end{equation}
where superscript ``diff.'' highlights that we use the relative abundance scale [X/H], which requires $\Delta_{i}^{\rm diff.}$ to also include the NLTE corrections to the solar absolute abundances of the corresponding elements. We note that we assume the differences between the ATLAS9 model atmospheres (used for a subsample of targets) and the MARCS models (used in NLTE calculations) to be of secondary importance compared to the total uncertainties of the LTE abundances as well as the uncertainties of the NLTE models. For the uncertainties of NLTE abundances, we assume the same systematic component as in the LTE abundances, and recalculate the random component in similar way (for one spectral line, $\sigma_{\rm random}\,=\,0.1$\,dex; for more spectral lines $\sigma_{\rm random}$ is the standard deviation).

\section{Elemental abundances of transition disc targets}\label{sec:res}
In this section, we present the results of our LTE and NLTE abundance analysis of optical spectra of transition disc targets (see Section~\ref{ssec:respro}), define depletion indicators used for our sample (see Section~\ref{ssec:reseff}), and compare derived depletion profiles with those from previous studies (see Section~\ref{ssec:rescmp}).

\subsection{Observed depletion profiles}\label{ssec:respro}
In Fig.~\ref{fig:dplstr} and \ref{fig:dplcnd}, we present the depletion patterns together with the corresponding fits for each target. We mark elemental abundances corrected for NLTE effects with pink squares. For Fe-peak elements beyond the $\alpha$-process (from Ti to Fe), we considered the abundances derived from the ionised lines to be likely less sensitive to NLTE effects (i.e., conditionally NLTE; \ion{Ti}{ii}, \ion{V}{ii}, \ion{Cr}{ii}, \ion{Mn}{ii}). We denote these abundances with blue triangles. We mark the LTE abundances of the all remaining ionisations with light blue circles. The depletion profiles were fitted from S to Zr based on the arbitrary reliability of the abundance measurement: NLTE abundances were weighted as 1, conditionally NLTE abundances were weighted as 0.67, and LTE abundances were weighted as 0.33. In this approach, the abundance of each ionisation was treated separately, which provided more weight to those elements, for which the abundances of two ionisations were measured. We note that the chosen weighting is an arbitrary measure used to emphasise the difference between fully-LTE depletion profile and NLTE-corrected depletion profile.

One-piece linear fits (red solid lines) represent the homogeneous depletion profile without any break temperatures caused by the onset of depletion ($T_{\rm turn-off}$) or plateau ($T_{\rm plateau}$). Two-piece and three-piece linear fits (yellow dashed-dotted and green dashed lines, respectively) represent the depletion with any or both of these break temperatures, respectively. Two-piece and three-piece fits were tested against the one-piece linear fit (which was considered null hypothesis) using the likelihood ratio test: if the p-value of the two- or three-piece linear fit is less than 0.05, this fit offers significantly better\footnote{Meaning that the decrease in the $\chi^2$ value for a more elaborate model is large enough to justify the reduction in degrees of freedom by invoking new constraints (parameters).} goodness-of-fit for the depletion profile than the one-piece linear fit.

The major differences between LTE and NLTE-corrected abundances in transition disc targets are summarised below:
\begin{enumerate}
    \item For our targets, NLTE corrections decrease C/O ratio by up to $\sim$30\% (mainly due to NLTE corrections for the high-excitation \ion{C}{i} lines used here\footnote{For all targets except GZ Nor (\#9), the O abundances are based on the low-excitation forbidden [\ion{O}{i}] lines that are insensitive to NLTE effects (e.g. \citep{amarsi2016oxygen}.}). However, for GZ Nor (\#9), NLTE corrections increase C/O ratio by two times.
    \item The average line-to-line scatter of NLTE abundances is generally lower than the average line-to-line scatter of LTE abundances, with the most prominent reduction of the scatter for Na (0.04 dex), Al (0.05 dex), Si (0.07 dex), K (0.04 dex), and Ca (0.04 dex).
    \item The final depletion profiles of transition disc targets are well-fitted by one-piece linear trends. This result highlights that derived depletion profiles of all transition disc targets are saturated.
    \item There are few prominent (but statistically insignificant) deviations from one-piece fits of depletion profiles: i) [Na/H] and [Cu/H] in DY Ori (\#7), ii) [S/H] in GZ Nor (\#9), iii) [Na/H] in V1504 Sco (\#10), and iv) [Mn/H] in RU Cen (\#3) and GZ Nor (\#9). The deviations in [Na/H], [Cu/H], and [S/H] abundances may be caused by the differences in chemical composition and conditions between the transition disc targets and the solar-mixture gas assumed in chemical equilibrium calculations \citet{wood2019CondensationTemperatures} (especially, for S in the least O-rich transition disc target from our sample, C/O$_{\rm GZ~Nor}$\,=\,0.78). We note that RU Cen (\#3) and GZ Nor (\#9) clearly show saturation if we temporarily set aside the [Mn/H] abundance. The explanation for this is that the NLTE correction for Mn lines in metal-poor giants generally are positive \citep[{[Mn/H]}$_{\rm NLTE}$ – {[Mn/H]}$_{\rm LTE}\,\sim\,$+0.6 dex for {[Fe/H]} = –-3 dex;][]{bergemann2008MnNLTE, amarsi2020NLTEgalah}, bringing [Mn/H] abundance closer to the linear decline of the depletion profile. We also note that the behaviour of the condensation temperatures for environments with different C/O ratios is out of the scope of this work, though it is a promising path for the future study.
\end{enumerate}

Finally, we also note that an alternative explanation for the enhancement of [Na/H]\,=\,0.15 dex in V1504 Sco involves the first and second dredge-ups, which may be responsible for [Na/H] enhancement in progenitors with intermediate masses $M_\ast>4.5M_\odot$ \citep{karakas2014dawes}.

\subsection{Definition and rationale for depletion indicators}\label{ssec:reseff}
To characterise the depletion profiles, we defined [S/H]$_{\rm NLTE}$ as the lower limit of the initial metallicity [M/H]$_{\rm 0,min}$, and [S/Ti]$_{\rm NLTE}$ as the NLTE depletion scale (we use [Ti/H]$_{\rm LTE}$ derived from \ion{Ti}{ii} lines as [Ti/H]$_{\rm NLTE}$). Our reasoning for selecting S as volatile indicator and Ti as refractory indicator is as follows:
\begin{enumerate}
    \item C, N, and O in post-AGB/post-RGB stars are modified to an unknown extent by the convective and non-convective mixing processes on AGB/RGB, such as dredge-ups \citep{kobayashi2011IsotopeEvolutionModels, ventura2020CNOinAGB, kamath2023models, mohorian2024EiSpec}. Moreover, CNO elements may be partially depleted, as seen in protoplanetary discs, through a poorly constrained process of CO and N$_2$ molecules (the major carriers of volatile CNO elements in the disc) converting into CO$_2$ and NH$_3$ ice and freezing-out onto dust grains \citep[possible explanations include dispersal of gas disc, interactions between the gas and the dust, and chemical reprocessing; see][and references therein]{reboussin2015Odepletion, krijt2016Odepletion, bai2016Odepletion, xu2017Odepletion, francis2022Odepletion, furuya2022CNdepletion}.
    \item In protoplanetary discs and in the ISM, S is depleted into sulphide minerals to an unknown extent (e.g., FeS or FeS$_2$) \citep{kama2019Sdepletion, konstantopoulou2022ISMdepletion}. However, given that S follows the abundance profiles of our target sample, this element generally is the least depleted after the exclusion of C, N, and O.
    \item S and Ti are $\alpha$-elements with multiple spectral features available in optical range and the most different condensation temperature (Sc is more refractory, yet has significantly less optical spectral lines), but these elements share similar nucleosynthetic history so that the intrinsic [S/Ti] ratio is supposed to be close to zero in the absence of depletion \citep{kobayashi2020OriginOfElements}. Additionally, our approach allows for the first time to use NLTE [S/Ti] ratio instead of LTE [Zn/Ti] ratio.
\end{enumerate}

Finally, the turn-off temperature $T_{\rm turn-off}<1000$ K for all transition disc targets (see Fig.~\ref{fig:dplstr} and \ref{fig:dplcnd}). Since the one-piece fits are statistically preferred for each target, and there are only few derived elemental abundances with condensation temperatures $T_{\rm cond}<1000$ K, setting $T_{\rm turn-off}$ becomes less trivial. Hence, we define the upper limit of turn-off temperatures in the transition disc sample to be located between S and Zn condensation temperatures ($T_{\rm turn-off}\,=\,700$ K).

\subsection{Comparison with literature depletion profiles}\label{ssec:rescmp}
In Appendix~\ref{app:dpl}, we provide a detailed comparison of depletion profiles of transition disc targets from the literature and from our homogeneous analysis. The key findings of this comparison are summarised below:
\begin{enumerate}
    \item Depletion profiles for RU Cen (\#3), EP Lyr (\#6), DY Ori (\#7), and GZ Nor (\#9), previously classified as ``plateau'' profiles based on LTE abundances \citep{oomen2019depletion}, become ``saturated'' profiles when NLTE corrections are applied (as indicated by p-values).
    \item In previous studies, the onset of depletion, characterised by $T_{\rm turn-off}$, was reported at varying temperatures for our sample, ranging from 800 K for four out of 10 Galactic targets to 1200 K for CT Ori (\#1) and AC Her (\#4) \citep{kluska2022GalacticBinaries}. In contrast, our analysis reveals a consistently minimal $T_{\rm turn-off}$ across all transition disc targets ($T_{\rm turn-off}$\,=\,700 K).
\end{enumerate}

Overall, the spectral data collected in this study enabled the determination of a more extensive set of elemental abundances compared to those reported in the literature for transition disc targets (see Table~\ref{tab:litpar}). The derived elemental abundances are discussed in the context of depletion in Section~\ref{sec:dpl}.

\begin{table}
    \centering
    \caption{The references for model atoms used in this study (see Section~\ref{ssec:sannlte}).} \label{tab:nonref}
    \begin{tabular}{|c|c|}
    \hline
        Element & Reference \\ \hline
        C & \cite{amarsi2019NLTEcarbon} \\
        N & \cite{amarsi2020NLTEnitrogen} \\
        O & \cite{amarsi2018NLTEoxygen} \\
        Na & \cite{lind2011NLTEsodium} \\
        Mg & \cite{asplund2021solar} \\
        Al & \cite{nordlander2017NLTEaluminium} \\
        Si & \cite{amarsi2017NLTEsilicon} \\
        S & Amarsi et al. (in prep.) \\
        K & \cite{reggiani2019NLTEpotassium} \\
        Ca & \cite{asplund2021solar} \\
        Fe & \cite{amarsi2022NLTEiron} \\ \hline
    \end{tabular}
\end{table}

\section{Discussion}\label{sec:dpl}
In this section, we analyse the obtained results to gain a deep understanding of the conditions and factors driving the depletion process in post-AGB/post-RGB binaries with transition discs. We do this by comparing the chemical depletion parameters in our targets with: i) other observational parameters of our target sample, ii) chemical depletion parameters in transition disc YSOs, and iii) chemical depletion parameters in the interstellar medium (ISM).

\subsection{Correlation analysis of known parameters in post-AGB/post-RGB binaries}\label{ssec:dplcor}
To comprehensively investigate our diverse sample of the most chemically depleted subclass of post-AGB/post-RGB binaries (transition disc targets) we conducted a correlation analysis on a representative selection of observational parameters to address specific questions, including:
\begin{enumerate}
    \item Photometric parameters (IR colours $H-K$ and $W_1-W_3$, SED luminosity $L_{\rm SED}$, PLC luminosity $L_{\rm PLC}$, dust-to-star luminosity ratio $L_{\rm IR}/L_\ast$): to explore potential connections between the current IR excess and depletion efficiency.
    \item Orbital parameters (orbital period $P_{\rm orb}$, eccentricity $e$): to explore whether certain orbital parameter configurations are more prone to depletion.
    \item Pulsational parameter (fundamental pulsation period $P_{\rm puls}$): to investigate the impact of pulsations on the depletion profile.
    \item Astrometric parameters (coordinates R.A. and Dec.): to study the spatial distribution of transition disc targets.
    \item Spectroscopic parameters (effective temperature $T_{\rm eff}$, surface gravity $\log g$, metallicity [Fe/H], C/O ratio, initial metallicity [M/H]$_{\rm 0,min}$, [Zn/Ti] and [S/Ti] abundance ratios): to analyse the shape and the scale of depletion patterns in our targets.
\end{enumerate}

The resulting correlation matrix is depicted in Fig.~\ref{fig:cormat} (see Appendix~\ref{app:cor} for individual dependence plots). To examine the behaviour of the dependencies, a logarithmic scale was applied to all parameters except for coordinates and eccentricity. In the following discussion, we address families of correlations, ordered from the most expected to the least.

We found several obvious physical correlations and anti-correlations (Spearman's coefficient $|\rho|\geq0.6$), including: i) spectroscopic parameters related to depletion, such as [Zn/Ti], [S/Ti], [S/Ti]$_{\rm NLTE}$, [Zn/Fe], and [Fe/H]; ii) orbital and other parameters (P$_{\rm orb}$ and $e$ are known for six and five targets, respectively), and iii) surface gravity $\log g$, effective temperature $T_{\rm eff}$, luminosity $L_\ast/L_\odot$, and $\log$(C/O) ratio.

The correlation between $\log L_{\rm IR}/L_\ast$ and [S/Ti]$_{\rm NLTE}$ abundance ratio is rather unexpected. By excluding the edge-on targets AF Crt (\#8) and V1504 Sco (\#10), and relatively dust-poor target EP Lyr (\#6), the correlation between the IR luminosity and the depletion scale becomes even more prominent. However, this correlation is likely caused by the sample bias.

Another notable finding was a strong anti-correlation of the fundamental pulsation period with $W_1-W_3$ colour. However, the circumbinary dust around our targets is produced from a previous phase of mass loss rather than pulsations \citep{vanwinckel2003Review}, so this correlation may be caused by sample bias as well.

Finally, we note that the adopted luminosities (see PLC luminosities in Table~\ref{tab:fnlvls}) infer the post-RGB evolutionary status of ST Pup (\#2), AD Aql (\#5), AF Crt (\#8), and GZ Nor (\#9). We found that the surface [O/H] abundance in these four post-RGB binaries generally deviates the most from the corresponding value predicted by the linear fit of the depletion profile. However, the surface [O/H] abundance in post-AGB binaries generally aligns with the corresponding linear fit of the depletion profile (see Fig.~\ref{fig:dplstr} and \ref{fig:dplcnd}). To quantify this deviation, we used a ``de-scaled'' [O/S] abundance ratio using the following equation:
\begin{equation}
    {\rm [O/S]_{NLTE}^{descaled}} = {\rm \frac{[O/S]_{obs}}{[O/S]_{calc}}},
\end{equation}
where [O/S]$_{\rm obs}$ is the observed [O/S] abundance ratio, while [O/S]$_{\rm calc}\,=\,{\rm [S/Ti]\cdot\frac{\textit{T}_{cond,O}-\textit{T}_{cond,S}}{\textit{T}_{cond,S}-\textit{T}_{cond,Ti}}}$ is the [O/S] abundance ratio calculated by scaling the corresponding [S/Ti] abundance ratio. In our sample, the values of ${\rm [O/S]_{NLTE}^{descaled}}$ range from $\sim0$ (when [O/H] and [S/H] abundances are similar) to 1 (when [O/H] abundance follows the linear depletion trend).

The correlation between ``de-scaled'' abundance ratio ${\rm [O/S]_{NLTE}^{descaled}}$ and luminosity $L_{\ast}$ is not significant for the whole sample, but it is prominent for the subsample of confirmed transition disc targets (\#1--\#6; see Fig.~\ref{figA:allcor2}). This connection may hint at O depletion being higher in post-AGB binaries than in post-RGB binaries (in other words, $T_{\rm turn-off}$ is lower for depletion profiles of post-AGB binaries rather than for those of post-RGB binaries). This is consistent with the results from \citet{mohorian2024EiSpec}, where the average turn-off temperature for two post-RGB binaries SZ~Mon and DF~Cyg ($T_{\rm turn-off,~post-RGB}\approx1300$K) was found to be higher than the average value for the post-AGB binary sample \citep[$T_{\rm turn-off,~post-AGB}\approx1100$K;][]{oomen2019depletion}. Our investigation will be extended in future work, as it is beyond the scope of this study.

\begin{figure*}
    \centering
    \includegraphics[width=.99\linewidth]{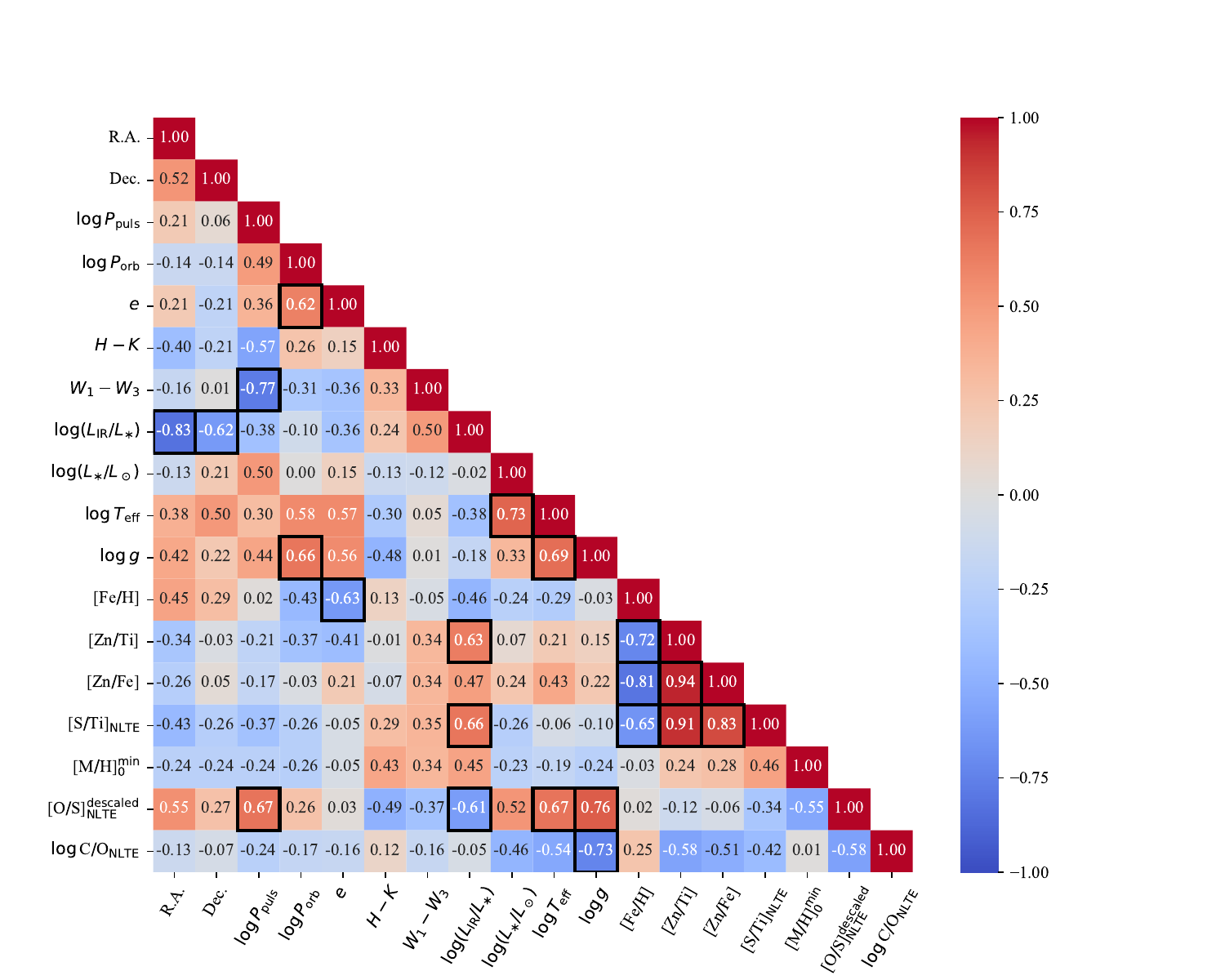}
    \caption{Correlation matrix of various stellar parameters of transition disc targets. The strong correlations and anti-correlations (i.e., with Spearman's correlation coefficients $|\rho|\geq0.6$) are highlighted with black boxes (for more details, see Section~\ref{ssec:dplcor}).}\label{fig:cormat}
\end{figure*}

\subsection{Parallels with chemical depletion in young stars hosting transition discs}\label{ssec:dplyso}
Transition discs around young T~Tauri ($M_{\rm T~Tauri}\,<\,2\,M_\odot$) and Herbig Ae/Be ($2\,M_\odot\,<\,M_{\rm Herbig~Ae/Be}\,<\,10\,M_\odot$) stars are similar to post-AGB/post-RGB transition discs in structure and many physical properties, including a broad near-IR excess indicative of hot dust in the disc, Keplerian rotation, dust disc mass, and dust mineralogy \citep[see, e.g.,][and references therein]{follette2017PPDGrainEvolution, deruyter2005discs, corporaal2023DiscParameters, andrych2023Polarimetry, andrych2024IRAS08}. 

Interestingly, similar to post-AGB/post-RGB binaries, a subclass of young T~Tauri and Herbig Ae/Be stars hosting transition discs also exhibits photospheric depletion of refractory elements, a phenomenon known as the $\lambda$ Boo phenomenon \citep[see, e.g.,][]{andrievsky2002lambdaBooAbundances, jura2015lambdaBoo, kama2015DiscDepletionLinkinYSOs, jermyn2018Depletion, murphy2020lambdaBoo}. While both young T~Tauri and Herbig Ae/Be stars with $\lambda$ Boo-like depletion and post-AGB/post-RGB binaries show similar photospheric underabundances of refractory elements, the chemical depletion process is significantly more efficient in post-AGB/post-RGB systems (as indicated by volatile-to-refractory abundance ratios). For instance, the young stars with $\lambda$ Boo-like depletion show underabundances of Mg, Si, and Fe in the range from 0 to 1 dex \citep{kama2015DiscDepletionLinkinYSOs}, whereas post-AGB/post-RGB binaries display [Zn/Ti] abundance ratios in the range from 0 to 3.5 dex \citep{kluska2022GalacticBinaries}.

In post-AGB/post-RGB binaries, the separation of volatile-rich gas and refractory-rich dust remains poorly understood. In contrast, in young T~Tauri and Herbig Ae/Be stars with transition discs and displaying the $\lambda$ Boo phenomenon, the dust-gas separation is linked to several theoretical mechanisms, including grain growth \citep{dullemond2001GrainGrowth}, photoevaporation \citep{alexander2006Photoevaporation}, dead zones \citep{regaly2012DeadZones}, and embedded giant planets \citep{birnstiel2010EmbeddedPlanets}. Furthermore, \citet{folsom2012LambdaBooFraction} suggested that up to a third of Herbig Ae/Be stars hosting protoplanetary discs show signs of depletion and harbour giant planets. Given the structural and chemical parallels between transition discs in $\lambda$ Boo stars and those in post-AGB/post-RGB binaries, we explore the potential role of giant planets in carving the inner gaps in transition discs around post-AGB/post-RGB binaries.

To study the depletion efficiency in protoplanetary transition discs around young stars, \citet{kama2015DiscDepletionLinkinYSOs} used the photospheric composition of Herbig Ae/Be single stars as a proxy for the chemical composition of the accreted matter assuming that this matter (with accretion rates of $\sim10^{-9}-10^{-6}M_\odot$/yr) quickly dominates the original surface chemistry of a star. Previously, \citet{turcotte2002AccretionDomination} showed that for a young star at an age of 10$^6$ years, the domination of accreted matter may be achieved with the accretion rates of as low as $\sim10^{-11}M_\odot$/yr.

To investigate the depletion efficiency in circumbinary discs around post-AGB/post-RGB stars, \citet{oomen2019depletion} modelled the accretion rate onto the binary from a viscously evolving disc for a range of accretion rates and disc masses. They showed that to fit the observed parameters, re-accretion in post-AGB/post-RGB stars require significantly larger initial accretion rates than in young stars ($>3\times10^{-7}M_\odot$/yr). Following the approach from \citet{turcotte2002AccretionDomination}, we estimate that the re-accreted matter should dominate the original surface material in post-AGB/post-RGB binaries within $\sim100$ years. However, this approach is limited by the assumptions of the negligible impact of binary interaction and the simplified chemical composition of the accreted matter \citep[see Fig.~2 in][]{oomen2019depletion}.

To compare the depletion scales in our sample and in YSOs, we calculated the depletion strength $\Delta_{\rm g/d}$ \citep{kama2015DiscDepletionLinkinYSOs} given by
\begin{equation}
    \Delta_{\rm g/d} = 100\times10^{\rm [V/H]-[R/H]} = 100\times10^{\rm [V/R]},
\end{equation}
where [V/H] and [R/H] are the abundances of volatile and refractory tracing elements, respectively. By definition, the solar composition corresponds to depletion strength $\Delta_{\rm g/d,\ \odot}\,=\,100$.

For the [V/H] and [R/H] abundances, \citet{kama2015DiscDepletionLinkinYSOs} combined volatile [C/H] and [O/H], and refractory [Fe/H], [Mg/H], and [Si/H], respectively. As mentioned in Section~\ref{sec:res}, we considered volatile [S/H] and refractory [Ti/H] to be a more reliable scale of dust depletion. However, for comparison consistency, we used the NLTE-corrected abundances [S/H]$_{\rm NLTE}$ and [Fe/H]$_{\rm NLTE}$, as the depletion tracers in this subsection. Therefore, the expression for the depletion strength $\Delta_{\rm g/d}$ in transition disc targets is given by
\begin{equation}
    \Delta_{\rm g/d} = 100\times10^{\rm [S/H]_{NLTE}-[Fe/H]_{NLTE}} = 100\times10^{\rm [S/Fe]_{NLTE}}.
\end{equation}

For young stars hosting transition discs, the depletion strength $\Delta_{\rm g/d}$ was found to be below $\approx10^3$ \citep[see Fig.~2 in][]{kama2015DiscDepletionLinkinYSOs}. However, for post-AGB/post-RGB targets osting transition discs, our calculated values of $\Delta_{\rm g/d}$ lie in the range of higher values (700--20\,000; see Table~\ref{tab:fnlabu}). This notable increase of the depletion strength $\Delta_{\rm g/d}$ in our transition disc targets points at an increased dust depletion efficiency, which may hint at a more effective dust fractionation in the inner circumbinary disc. An alternative explanation could be the dilution efficiency being higher in post-AGB/post-RGB stars due to their smaller atmospheres \citep{vanwinckel2003Review}. However, to solidify our qualitative comparison, there is a clear need to model discs around post-AGB/post-RGB binaries incorporating more sophisticated disc dynamics and more accurate stellar luminosities. Moreover, the assumed chemical composition of the re-accreted matter should be revised taking into account observed patterns, as demonstrated in the present study.

\subsection{Parallels with chemical depletion in ISM}\label{ssec:dplism}
The ISM is enriched by various sources, including stellar mass-loss, star formation, and supernovae  \citep{zhukovska2008ISMCompositionFromStars, bierbaum2011ISMcomposition, hofner2018MassLossAGB, saintonge2022ISMSourcesOfEnrichment}. The gas and dust in the ISM have different chemical compositions, and this difference, known as depletion, is studied by measuring ion column densities in the gas phase \citep{jenkins2009ISMdepletion}. The gas-phase ISM abundances can vary due to factors like differences in star formation, in nucleosynthetic history, or in the condensation of metals into dust grains \citep{decia2016ISMdepletion, konstantopoulou2022ISMdepletion}. Distinguishing between these factors, especially at low metallicities, is crucial for studying depletion in the ISM \citep{jenkins2014DustDepletionInISM}.

\begin{figure*}
    \centering
    \includegraphics[width=.45\linewidth]{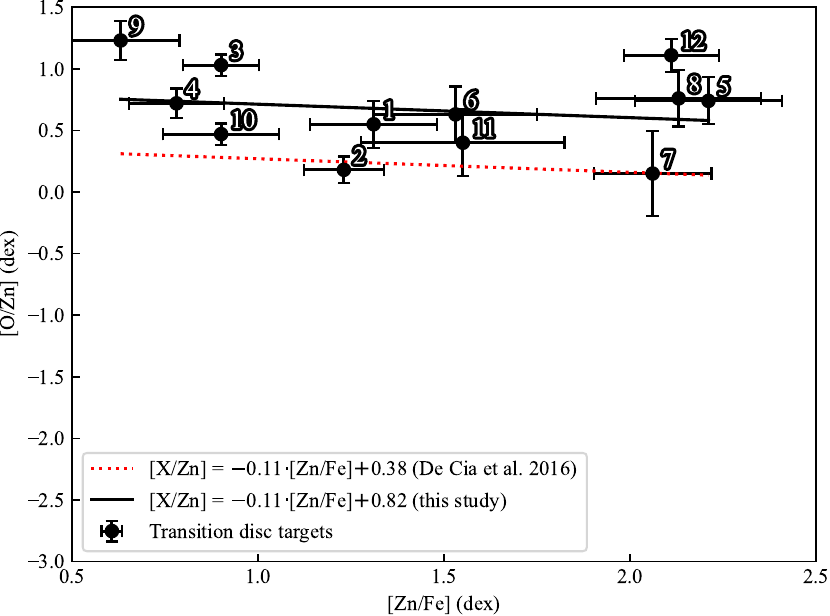}
    \includegraphics[width=.45\linewidth]{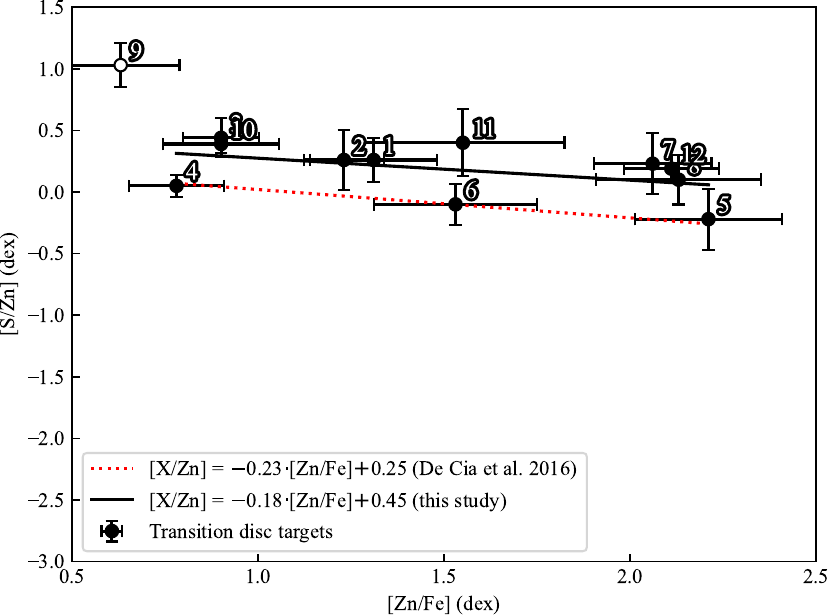}
    \includegraphics[width=.45\linewidth]{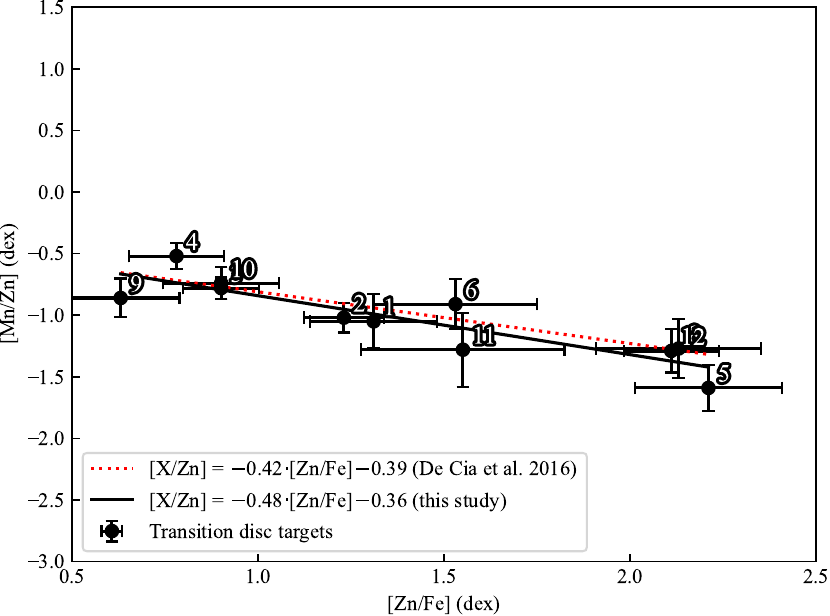}
    \includegraphics[width=.45\linewidth]{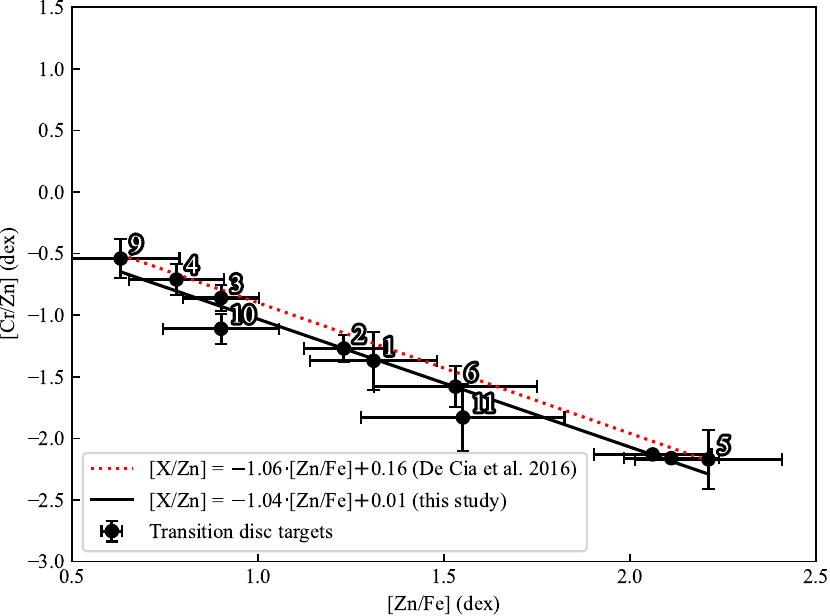}
    \includegraphics[width=.45\linewidth]{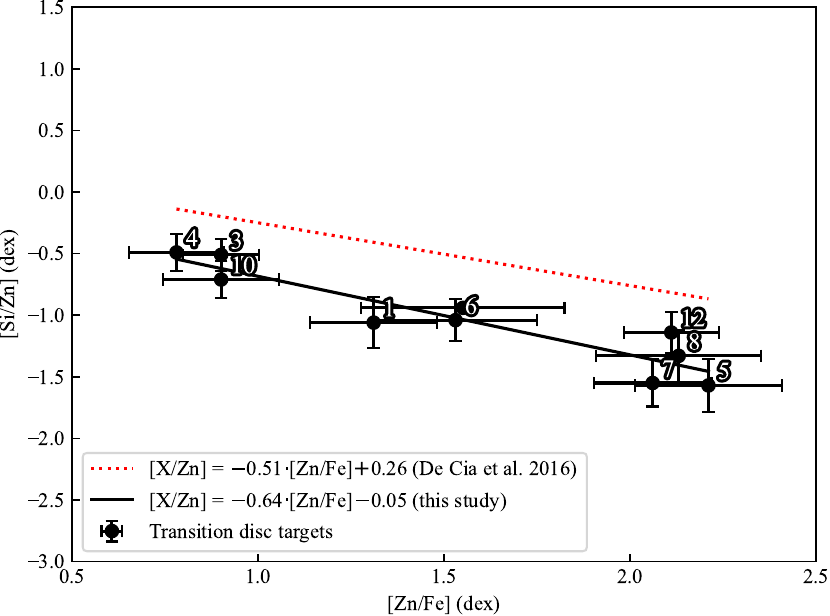}
    \includegraphics[width=.45\linewidth]{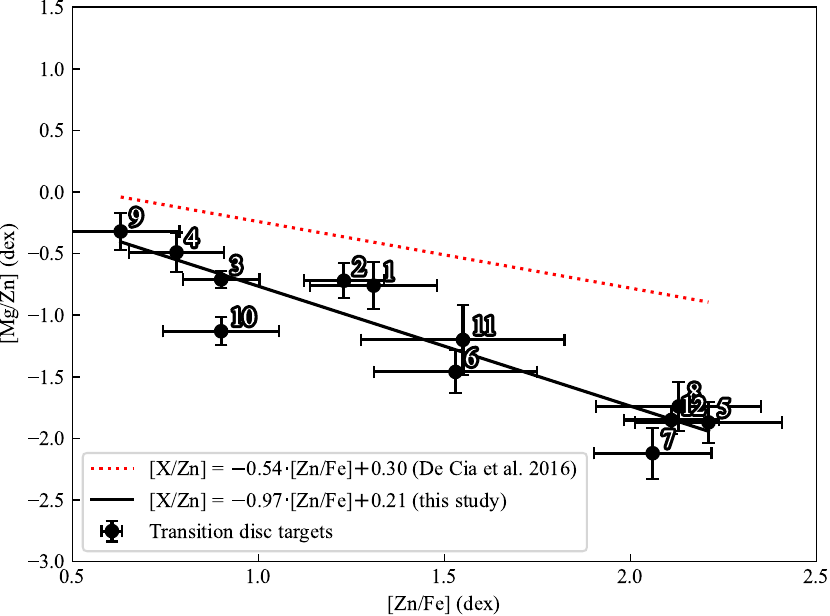}
    \caption{Comparison of [X/Zn] ratio trends for \ion{O}{i}, \ion{S}{i}, \ion{Mn}{i}, \ion{Cr}{i}, \ion{Si}{ii}, and \ion{Mg}{i} between transition disc targets and the ISM \citep{decia2016ISMdepletion}. [Zn/Fe] was calculated based on abundances from \ion{Zn}{i} and \ion{Fe}{ii} spectral lines. The legend for the symbols and colours used is included within the plot. We note that for [S/Zn] subplot we excluded GZ Nor (\#9) from fitting because of the unique depletion profile of this target (see Fig.~\ref{fig:dplcnd}). We also note that for [Cr/Zn] ratio of GZ Nor (\#9) we used the abundance derived from \ion{Cr}{ii} lines, while for [Si/Zn] ratio of LMC V0770 (\#11) we used the abundance derived from \ion{Si}{i} lines (for more details, see Section~\ref{ssec:dplism}).}\label{fig:cmpism}
\end{figure*}

A homogeneous research on depletion across various ISM environments, from the Galaxy to damped Ly-$\alpha$ absorbers, was conducted by \citet{decia2016ISMdepletion}. In their study, the ISM sites were distinguished not by the location, but by [Zn/Fe] abundance ratio: their pointings in the Galaxy and in the Magellanic Clouds occupied the region of [Zn/Fe]>0.5 dex, while their pointings towards damped Ly-$\alpha$ absorbers covered the region of [Zn/Fe]<1 dex. In Fig.~\ref{fig:cmpism}, the ISM trends of [X/Zn] abundance ratios for O, S, Mn, Cr, Si, and Mg are denoted with dotted red lines and the corresponding abundance ratios in transition disc targets are shown in black circles and are fitted with black lines. We note that we used NLTE abundances of O, S, and Mg, conditionally NLTE abundances of Cr and Fe (derived from spectral lines of \ion{Cr}{ii} and \ion{Fe}{ii}, respectively), and LTE abundances of Mn, Si, and Zn.

We found that the variation in depletion efficiencies of different chemical elements between our transition disc targets and the ISM depends on the volatility of these elements (traced by the corresponding condensation temperatures):

\begin{enumerate}
    \item O ($T_{\rm cond}$\,=\,183 K; highly volatile): The [O/Zn] ratio in transition disc targets shows a similar slope as in the ISM \citep{decia2016ISMdepletion}, but is enhanced by $\sim0.4$ dex. However, when compared with the updated trends of the extended ISM sample \citep[see Appendix~B2 in][]{konstantopoulou2022ISMdepletion}, the agreement becomes satisfactory.
    \item S ($T_{\rm cond}$\,=\,672 K; moderately volatile): The [S/Zn] trend in our data is similar to the one observed in the ISM with a slight enhancement by $\sim$0.2 dex. However, when distant Ly-$\alpha$ absorbers are excluded (leaving only lines of sight in the Galaxy and the LMC), the agreement becomes satisfactory \citep[see Fig.~1 in][]{konstantopoulou2022ISMdepletion}. We highlight that despite being highly volatile, S may be depleted into dust grains to unknown extent, similarly to interstellar S in the Galaxy \citep{jenkins2009ISMdepletion} and towards the damped Ly-$\alpha$ absorbers \citep{decia2016ISMdepletion}.
    \item Mn ($T_{\rm cond}$\,=\,1123 K; moderately volatile): The patterns of [Mn/Zn] ratios match in our target sample and in the ISM. After accounting for the LTE underabundance of Mn in F and G stars within the solar neighbourhood \citep{battistini2015MnNucleosynthesis}, the slope of [Mn/Zn] vs [Fe/Zn] approaches unity \citep[see Table~3 in][]{decia2016ISMdepletion}.
    \item Cr ($T_{\rm cond}$\,=\,1291 K; moderately refractory): The [Cr/Zn] trends also match in our target sample and in the ISM. The Cr and Fe abundance ratios are identical within the error bars ([Cr/Zn]$\sim$1.04$\times$[Fe/Zn]) in the transition disc targets and in the ISM.
    \item Si ($T_{\rm cond}\,=\,1314$ K; moderately refractory): In evolved binaries, the slope of [Si/Zn] trend is --0.64, which is lower than the corresponding slope in the ISM. Since Si and Fe have similar condensation temperatures ($\Delta T_{\rm cond}\,=\,24$ K) and solar abundances ($\sim7.5$ dex), their absolute abundances $\log\varepsilon$(Si) and $\log\varepsilon$(Fe)\footnote{Absolute abundance of an element X is given by: $\log\varepsilon(X)\,=\,\log\frac{N(X)}{N(H)}\,=\,[{\rm X/H}] + \log\varepsilon_\odot$(X).} in transition disc targets show a 2:3 ratio \citep[Si and Fe abundances in the ISM show a 1:2 ratio;][]{decia2016ISMdepletion}.
    \item Mg ($T_{\rm cond}\,=\,1343$ K; moderately refractory): The [Mg/Zn] trend in our targets shows a steeper slope than in the ISM (--0.97 and --0.54, respectively). Given the close proximity of condensation temperatures ($\Delta T_{\rm cond}\,=\,5$ K) and solar abundances of Mg and Fe ($\sim7.5$ dex), their absolute abundances $\log\varepsilon$(Mg) and $\log\varepsilon$(Fe) in transition disc targets show 1:1 abundance ratio.
\end{enumerate}

To summarise, the observed abundance ratios in the transition disc post-AGB/post-RGB targets and in the ISM generally display similar trends for both volatile and refractory elements (O, S, Mn, and Cr). This may be attributed to the fact that all our sample targets are O-rich (C/O<1, see Table~\ref{tab:fnlabu}), similar to the O-rich ISM environments studied by \citet{decia2016ISMdepletion}. However, we detected lower slopes in the abundance trends of Si and Mg in the transition disc targets relative to the ISM.

As mentioned above, the [X/Zn] trends represent both the pure depletion and the nucleosynthetic over- or underabundance of an element X. To remove the nucleosynthetic effects, \citet{decia2016ISMdepletion} converted the Galactic abundance patterns from \citet{mcwilliam1997GalacticChemicalEvolution} to [Zn/Fe] scale and provided corrections for the slopes of [Cr/Zn], [Si/Zn], and [Mg/Zn]. Applying these corrections, we obtained [Si/Zn]$_{\rm corr}$ $\sim$ 0.76 $\times$ [Fe/Zn]$_{\rm corr}$, and [Cr/Zn]$_{\rm corr}$ $\sim$ [Mg/Zn]$_{\rm corr}$ $\sim$ 1.04 $\times$ [Fe/Zn]$_{\rm corr}$. Since the solar abundances of Cr and Fe differ by $\sim2$ dex, the slope of Cr trend points at independence of the depletion efficiency of an element from its absolute abundance in post-AGB/post-RGB binaries with transition discs. The corrected slopes of Si and Mg trends in post-AGB/post-RGB binaries with transition discs point to Si, Mg, and Fe being depleted in the stellar surface with a number abundance ratio of 0.76:1.04:1, respectively.

Previous studies of circumbinary discs around post-AGB/post-RGB binaries detected Mg-rich end members of olivines (forsterite) and pyroxenes (enstatite) and did not detect Fe-rich dust grains \citep{gielen2008SPITZERsurvey, gielen2009Depletion, gielen2010corrigendum, gielen2011silicates, hillen2015ACHerMinerals}. Despite this, our measurements of Fe depletion in transition disc targets suggest the existence of Fe grains, such as Fe alloys, Fe oxides, fayalite, or ferrosilite. This highlights the uniqueness of the depletion profile analysis in offering an independent insight into the dust composition in the transition discs around post-AGB/post-RGB binary stars. We note that precise and consistent modelling of infrared spectral features in transition disc targets is beyond the scope of this study, though combining our results with mid-infrared observations from current mid-infrared facilities like MIRI/JWST is a promising avenue for the future research.

\section{Conclusions}\label{sec:con}
In this study, we aimed to investigate the mechanisms responsible for depletion, properties of the depletion patterns, and correlations to other observational parameters in post-AGB/post-RGB binary stars with transition discs. In our analysis, we used high-resolution optical spectra from HERMES/Mercator and UVES/VLT.

Using E-iSpec, we performed a detailed chemical abundance analysis of six confirmed transition disc stars and six transition disc candidates in the Galaxy and in the LMC. In addition to the derived LTE abundances, we used \texttt{Balder} software to calculate the NLTE corrections for a representative subset of chemical elements from C to Fe. We found that these corrections significantly affect the surface C/O ratios of transition disc targets and modify the depletion patterns. The resulting NLTE-corrected depletion profiles are saturated, meaning that the current surface abundances of all transition disc targets trace the composition of the re-accreted gas, rather than the original photospheric material. Moreover, we confirmed that depletion efficiency in transition disc systems, as traced by [S/Ti] abundance ratio, is higher than in other post-AGB/post-RGB binary stars.

In addition, we explored correlations between the derived abundances and various observational parameters associated with the binary system (e.g., astrometric, photometric, orbital, pulsational). For transition disc targets, the detected correlations generally align with those reported in the literature for the broader sample of post-AGB/post-RGB binary stars. Notably, we identified a moderate correlation between stellar luminosity and the normalised [O/S] abundance ratio. This hints at the turn-off temperature $T_{\rm turn-off}$ in transition disc targets being typically lower for depletion profiles in post-AGB binaries than in post-RGB binaries.

We also investigated the links between chemical depletion in our target sample and other environments (around YSOs with transition discs and in the ISM). We confirmed that the homogeneously derived depletion strength ($\Delta_{\rm g/d,~pAGB/pRGB}\,=\,700-20000$) in post-AGB/post-RGB binaries with transition discs is significantly higher than the values observed in young stars with transition discs ($\Delta_{\rm g/d,~YSO} < 1\,000$). However, the depletion patterns in post-AGB/post-RGB binaries with transition discs, as traced by the [X/Zn] vs. [Zn/Fe] slopes, resemble those seen in the ISM for both volatile elements (O, S, Mn) and refractory elements (Cr, Fe). We also found that the refractory elements Si and Mg deviate from this trend, indirectly offering a rough estimate of the amount of unobservable Fe dust grains in the transition discs around post-AGB/post-RGB binaries. It is important to note that these findings are based on a limited sample size. Our ongoing analysis of optical spectra of post-AGB/post-RGB binaries in the Galaxy, SMC and LMC aims to expand the target sample to enable a more statistically robust investigation, allowing for further validation and confirmation of current results.

\section*{Acknowledgements}\label{sec:ack}
The spectroscopic results presented in this paper are based on observations made with the Mercator Telescope, operated on the island of La Palma by the Flemish Community, at the Spanish Observatorio del Roque de los Muchachos of the Instituto de Astrofisica de Canarias. This research is based on observations collected at the European Organisation for Astronomical Research in the Southern Hemisphere under ESO programmes 074.D-0619 and 092.D-0485. This research was supported by computational resources provided by the Australian Government through the National Computational Infrastructure (NCI) under the National Computational Merit Allocation Scheme and the ANU Merit Allocation Scheme (project y89).

MM1 acknowledges the International Macquarie Research Excellence Scholarship (iMQRES) program for the financial support during the research. MM1, DK, and MM2 acknowledge the ARC Centre of Excellence for All Sky Astrophysics in 3 Dimensions (ASTRO 3D), through project CE170100013. AMA acknowledges support from the Swedish Research Council (VR 2020-03940) and from the Crafoord Foundation via the Royal Swedish Academy of Sciences (CR 2024-0015). HVW acknowledges support from the Research Council, KU Leuven under grant number C14/17/082.

\section*{Data availability}
The data underlying this article are available in the article and in its online supplementary material.

\bibliographystyle{mnras}
\bibliography{BiblioList}
\label{lastpage}
\appendix
\section{Individual target details}\label{app:lit}
In this Appendix, we discuss the previous studies of our target sample, from which we adopted pulsational and orbital parameters, together with luminosity estimates (see Table~\ref{tab:litpar}). In Fig.~\ref{figA:allmap}, we show the astrometric distribution of transition disc targets.

\subsection{Transition disc stars}\label{ssec:tarhst}
In this subsection, we present the targets with circumbinary discs, for which the reported dust inner rims $R_{\rm in}$ are at least 2.5 times larger than the expected dust sublimation radius  $R_{\rm sub}$ \citep{corporaal2023DiscParameters}. The presence of large, dust-free cavities in the discs around CT Ori, ST Pup, RU Cen, AC Her, AD Aql, and EP Lyr indicates these systems have transition discs, similar to those found around YSOs.

\subsubsection{CT Ori (\#1)} 
CT Ori, characterised as an RV Tau variable (spectroscopic class B\footnote{RV Tau variables of spectroscopic classes A, B, and C are metal-rich, metal-poor with enhanced carbon, and metal-poor without carbon enhancement, respectively \citep{preston1963RVTauSpecGroups}.}), exhibits a fundamental\footnote{The fundamental pulsation period for Type II Cepheids is twice shorter than the double period which usually is the best-fit value for phase-folded light curves \citep{stobie1970PeriodInconsistency}.} pulsation period of 33.65 days with no RVb phenomenon (slow variation in mean magnitude with a long secondary period $P\sim600-2600$ days), as observed by \citet{kiss2007T2Cepheids}. The orbital parameters of CT Ori are not constrained yet. Based on SED fitting, the luminosity of CT Ori was estimated to be $L_{\rm SED}\,=\,15100L_\odot$ by \citet{oomen2019depletion}. \citet{kluska2022GalacticBinaries} obtained a disc-star luminosity ratio for CT Ori $L_{\rm IR}/L_*\,=\,0.55$. \citet{corporaal2023DiscParameters} confirmed the transition disc nature of CT Ori with $R_{\rm in}/R_{\rm sub}\sim4.5$. \citet{gonzalez1997CTOri} conducted a comprehensive abundance analysis of CT Ori and identified significant depletion with [Fe/H]\,=\,--2.0 dex and [Zn/Ti]\,=\,1.9 dex.

\subsubsection{ST Pup (\#2)} 
ST Pup is classified as a W Vir pulsating variable. \citet{walker2015STPup} has determined the fundamental pulsation period of ST Pup (18.73 days). Despite the absence of the RVb phenomenon, \citet{oomen2018OrbitalParameters} calculated the orbital parameters of ST Pup using data from the long-term radial-velocity monitoring campaign ($P_{\rm orb}\,=\,406\pm2$ d, $e\,=\,0.00+0.04$). \citet{oomen2019depletion} reported the luminosity estimate for ST Pup through SED fitting ($L_{\rm SED}\,=\,690L_\odot$). \citet{kluska2022GalacticBinaries} obtained a moderate disc-star luminosity ratio of $L_{\rm IR}/L_*\,=\,0.72$. According to \citet{corporaal2023DiscParameters}, the ratio of dust inner rim to dust sublimation radius for ST Pup is $R_{\rm in}/R_{\rm sub}\sim3$. \citet{gonzalez1996STPup} analysed the chemical composition of ST Pup, revealing significant depletion ([Fe/H]\,=\,--1.5 dex, [Zn/Ti]\,=\,2.1 dex).

\subsubsection{RU Cen (\#3)} 
RU Cen, classified as an RV Tau variable (spectroscopic class B), has a fundamental pulsation period of 32.37 days, as documented by \citet{bodi2019RVTauVars} (RVb phenomenon was not detected). By using radial velocities from various spectral observations, \citet{oomen2018OrbitalParameters} determined the orbital parameters of RU Cen to be $P_{\rm orb}\,=\,1489\pm10$ days and $e\,=\,0.62\pm0.07$. The luminosity of RU Cen was estimated to be $L_{\rm SED}\,=\,1100L_\odot$ based on SED fitting by \citet{oomen2019depletion}. \citet{kluska2022GalacticBinaries} provided a moderate disc-star luminosity ratio of $L_{\rm IR}/L_*\,=\,0.40$. \citet{corporaal2023DiscParameters} reported $R_{\rm in}/R_{\rm sub}\sim3.5$ for RU Cen. \citet{maas2002RUCenSXCen} conducted a detailed abundance analysis of RU Cen and reported significant depletion with [Fe/H]\,=\,--1.9 dex and [Zn/Ti]\,=\,1.0 dex.

\subsubsection{AC Her (\#4)} 
AC Her, an RV Tau pulsating variable of spectroscopic class B, has a fundamental pulsation period of 37.73 days as determined by \citet{giridhar1998RVTauVars}. There is no apparent RVb phenomenon in this star. Using radial velocities from long-term spectral observations, \citet{oomen2018OrbitalParameters} derived the orbital parameters of AC Her ($P_{\rm orb}\,=\,1189\pm1$ days and $e\,=\,0.0+0.05$). The luminosity of AC Her was estimated to be $L_{\rm SED}\,=\,2\,400\,L_\odot$ based on SED fitting by \citet{oomen2019depletion}. \citet{bollen2022Jets} modelled jets in this system and obtained an independent estimate of luminosity for this target $L_{\rm jet\,model}\,=\,3\,600\,L_\odot$. \citet{kluska2022GalacticBinaries} reported a low disc-star luminosity ratio $L_{\rm IR}/L_*\,=\,0.21$. In the study by \citet{corporaal2023DiscParameters}, the ratio of inner rim to sublimation radius for AC Her was found to be $R_{\rm in}/R_{\rm sub}\sim7.5$. \citet{giridhar1998RVTauVars} performed a detailed abundance analysis of AC Her and found moderate depletion with [Fe/H]\,=\,--1.4 dex and [Zn/Ti]\,=\,0.7 dex.

\subsubsection{AD Aql (\#5)} 
AD Aql, classified as an RV Tau variable (spectroscopic class B), has a fundamental pulsation period of 32.7 days, as documented by \citet{giridhar1998RVTauVars} (RVb phenomenon was not detected for this target). The orbital parameters of AD Aql are not studied yet. The luminosity of AD Aql was estimated to be $L_{\rm SED}\,=\,11\,500\,L_\odot$ based on SED fitting by \citet{oomen2019depletion}. \citet{kluska2022GalacticBinaries} obtained a rather moderate disc-star luminosity ratio of $L_{\rm IR}/L_*\,=\,0.51$. \citet{corporaal2023DiscParameters} reported $R_{\rm in}/R_{\rm sub}\sim6$ for AD Aql. \citet{giridhar1998RVTauVars} analysed the chemical composition of AD Aql and reported high depletion with [Fe/H]\,=\,--2.1 dex and [Zn/Ti]\,=\,2.5 dex.

\subsubsection{EP Lyr (\#6)} 
EP Lyr is an RV Tau pulsating variable of spectroscopic class B. This target possesses a fundamental pulsation period of 41.59 days \citep{bodi2019RVTauVars} with no evident long-period variation. \citet{oomen2018OrbitalParameters} employed radial velocities from long-term
radial-velocity monitoring campaign to derive the orbital parameters of EP Lyr ($P_{\rm orb}\,=\,1\,151\pm14$ days and $e\,=\,0.39\pm0.09$). \citet{oomen2019depletion} estimated the luminosity of EP Lyr through SED fitting ($L_{\rm SED}\,=\,5\,500\,L_\odot$). Furthermore, \citet{bollen2022Jets} modelled jets in this system and obtained an independent estimate of luminosity for EP Lyr $L_{\rm jet\,model}\,=\,7\,100\,L_\odot$. The disc-star luminosity ratio of $L_{\rm IR}/L_*\,=\,0.02$ \citep{kluska2022GalacticBinaries} is the lowest in the target sample. \citet{corporaal2023DiscParameters} derived the ratio of inner rim to sublimation radius $R_{\rm in}/R_{\rm sub}\sim3.5$ for EP Lyr. \citet{gonzalez1997EPLyrDYOriARPupRSgt} performed a detailed abundance analysis of EP Lyr, revealing moderate depletion with [Fe/H]\,=\,--1.8 dex and [Zn/Ti]\,=\,1.3 dex.

\subsection{Transition disc candidates}\label{ssec:tarcnd}
In this subsection, we present the targets classified by \citet{kluska2022GalacticBinaries} as category 2 ($W_1-W_3>4.5$) and category 3 ($2.3<W_1-W_3<4.5$, $H-K<0.3$). Even though category 3 of colour-colour plot is populated by both full disc and transition disc candidates \citep{kluska2022GalacticBinaries}, these subclasses may be distinguished by the depletion scale: [Zn/Ti]$_{\rm full-disc}\sim0$ dex, [Zn/Ti]$_{\rm transition-disc}>0.7$ dex.

\subsubsection{DY Ori (\#7)} 
DY Ori is an RV Tau pulsator of spectroscopic class B. ASAS \citep{pawlak2019ASAS} provides the fundamental pulsation period of DY Ori (30.155 days) with no obvious RVb phenomenon. However, \citet{oomen2018OrbitalParameters} used radial velocities of multiple spectral observations of DY Ori to derive the orbital parameters of this target ($P_{\rm orb}\,=\,1\,248\pm36$ d, $e\,=\,0.22\pm0.08$). \citet{oomen2019depletion} estimated the luminosity of DY Ori based on the SED fitting ($L_{\rm SED}\,=\,21\,500\,L_\odot$). \citet{kluska2022GalacticBinaries} reported a moderate disc-star luminosity ratio $L_{\rm IR}/L_*\,=\,0.55$. \citet{gonzalez1997EPLyrDYOriARPupRSgt} conducted a detailed abundance analysis of DY Ori and detected high depletion with [Fe/H]\,=\,--2.0 dex and [Zn/Ti]\,=\,2.1 dex.

\subsubsection{AF Crt (\#8)} 
AF Crt, a W Vir pulsator, has a double pulsation period of 31.5 days as determined by \citet{kiss2007T2Cepheids} without any apparent variation in mean magnitude on the order of hundreds of days. The orbital parameters of AF Crt are not constrained yet. The luminosity of AF Crt was estimated to be $L_{\rm SED}\,=\,280\,L_\odot$ based on SED fitting by \citet{oomen2019depletion}. \citet{kluska2022GalacticBinaries} derived a rather high disc-star luminosity ratio $L_{\rm IR}/L_*\,=\,1.83$. \citet{vanwinckel2012AFCrt} performed a detailed abundance analysis of AF Crt and reported high depletion of this target with [Fe/H]\,=\,--2.7 dex and [Zn/Sc]\,=\,3.4 dex (a slightly less refractory Sc was used for volatile-to-refractory ratio, since Ti abundance was not calculated).

\subsubsection{GZ Nor (\#9)} 
GZ Nor, classified as an RV Tau pulsator, demonstrates a fundamental pulsation period of 36.2 days \citep{gezer2019GKCarGZNor}. No evident RVb phenomenon was observed and no orbital parameters were derived for GZ Nor. \citet{oomen2019depletion} determined the luminosity of GZ Nor through SED fitting $L_{\rm SED}\,=\,1\,400\,L_\odot$. \citet{kluska2022GalacticBinaries} provided a mild disc-star luminosity ratio of $L_{\rm IR}/L_*\,=\,0.22$. \citet{gezer2019GKCarGZNor} studied the chemical composition of GZ Nor and detected moderate depletion with [Fe/H]\,=\,--2.1 dex and [Zn/Ti]\,=\,0.8 dex.

\subsubsection{V1504 Sco (\#10)} 
V1504 Sco is an RV Tau variable star of spectroscopic class B. \citet{kiss2007T2Cepheids} provided the fundamental pulsation period of this star (22.0 d) with a detected RVb phenomenon. Studying the mean magnitude variation, \citet{kiss2007T2Cepheids} derived the orbital period of V1504 Sco ($P_{\rm orb}\,=\,735\pm230$ d). \citet{oomen2019depletion} estimated the luminosity of V1504 Sco based on the SED fitting ($L_{\rm SED}\,=\,1\,100\,L_\odot$). \citet{kluska2022GalacticBinaries} reported an extreme disc-star luminosity ratio $L_{\rm IR}/L_*\,=\,4.69$. \citet{maas2005DiscPAGBs} conducted a detailed abundance analysis of V1504 Sco and detected high depletion with [Fe/H]\,=\,--1.0 dex and [Zn/Ti]\,=\,1.4 dex.

\subsubsection{LMC V0770 (\#11)} 
LMC V0770, an RV Tau pulsator, exhibits a fundamental pulsation period of 31.2 days, as determined by \citet{manick2018PLC}. No variation in the mean magnitude was detected and no orbital parameters were constrained for LMC V0770. \citet{oomen2019depletion} used SED fitting to calculate the luminosity of LMC V0770 ($L_{\rm SED}\,=\,3\,300\,L_\odot$). Moreover, \citet{manick2018PLC} used PLC relation to obtain an independent estimate of luminosity ($L_{\rm PLC}\,=\,2\,629\,L_\odot$). According to \citet{kluska2022GalacticBinaries}, the disc-star luminosity ratio for LMC V0770 is moderate $L_{\rm IR}/L_*\,=\,0.63$. \citet{kamath2019depletionLMC} analysed the chemical composition of LMC V0770 and reported a high depletion with [Fe/H]\,=\,--2.6 dex and [Zn/Ti]\,=\,2.3 dex.

\subsubsection{LMC V3156 (\#12)} 
LMC V3156, an RV Tau pulsator, has a fundamental pulsation period of 46.7 days, as provided by \citet{manick2018PLC}. There is no apparent RVb phenomenon in the light curves of this target and no known orbital parameters. The luminosity of LMC V3156 was estimated by \citet{manick2018PLC} using two methods: SED fitting and PLC relation ($L_{\rm SED}\,=\,5\,900\,L_\odot$, $L_{\rm PLC}=6\,989\,L_\odot$). \citet{vanaarle2011PAGBsInLMC} reported a disc-star luminosity ratio of $L_{\rm IR}/L_*\,=\,0.84$. \citet{reyniers2007LMC147} conducted a detailed abundance analysis of LMC V3156 and found significantly high depletion with [Fe/H]\,=\,--2.4 dex and [Zn/Ti]\,=\,2.5 dex.

\section{SED plots of the target sample}\label{app:sed}
\begin{figure*}
    \centering
    \includegraphics[width=.49\linewidth]{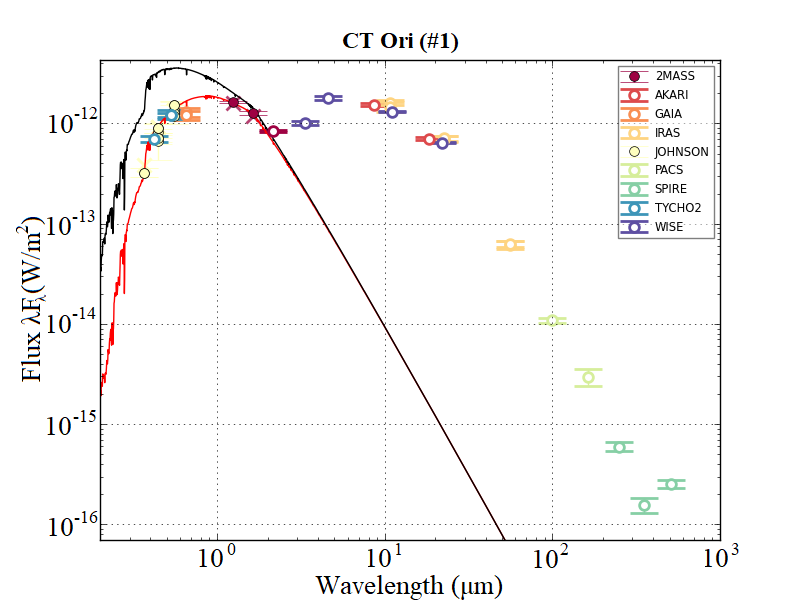}
    \includegraphics[width=.49\linewidth]{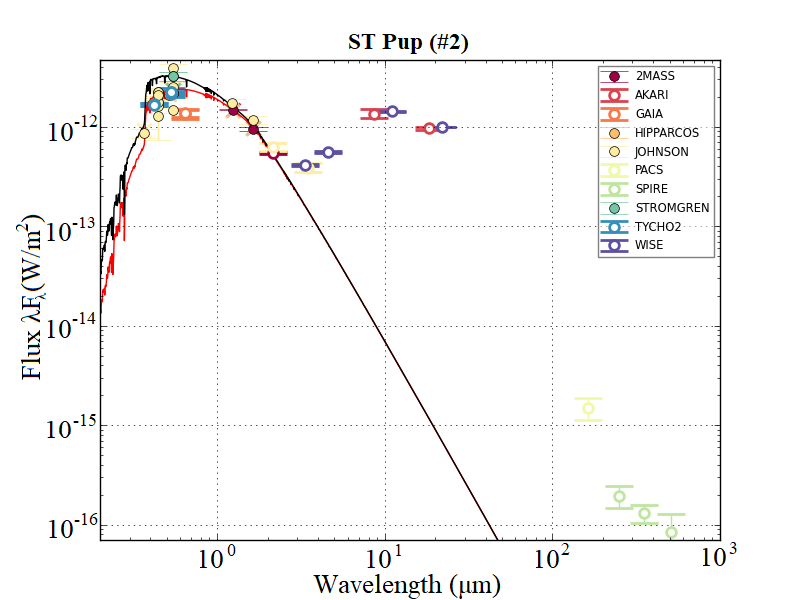}
    \includegraphics[width=.49\linewidth]{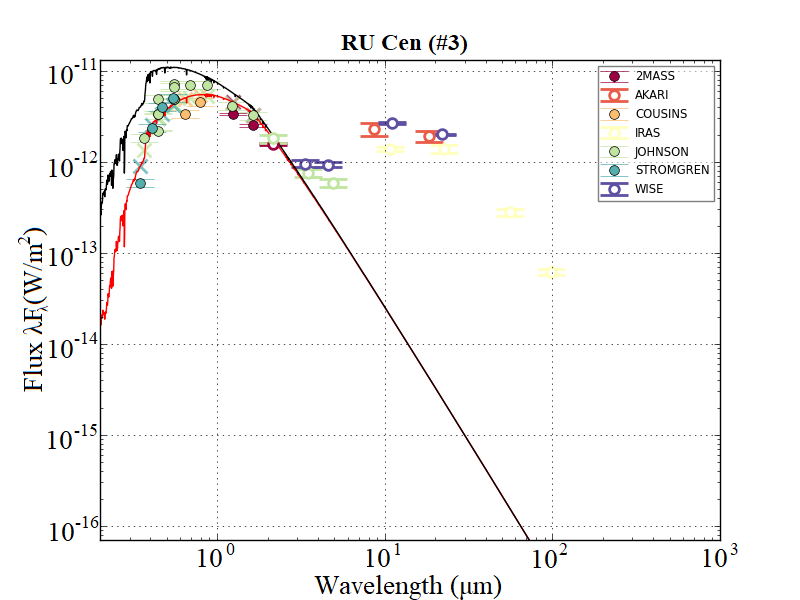}
    \includegraphics[width=.49\linewidth]{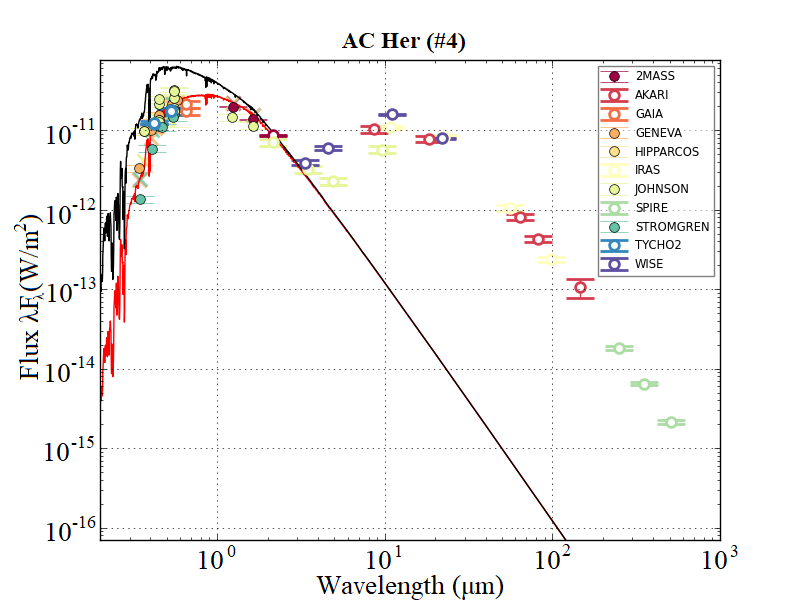}
    \includegraphics[width=.49\linewidth]{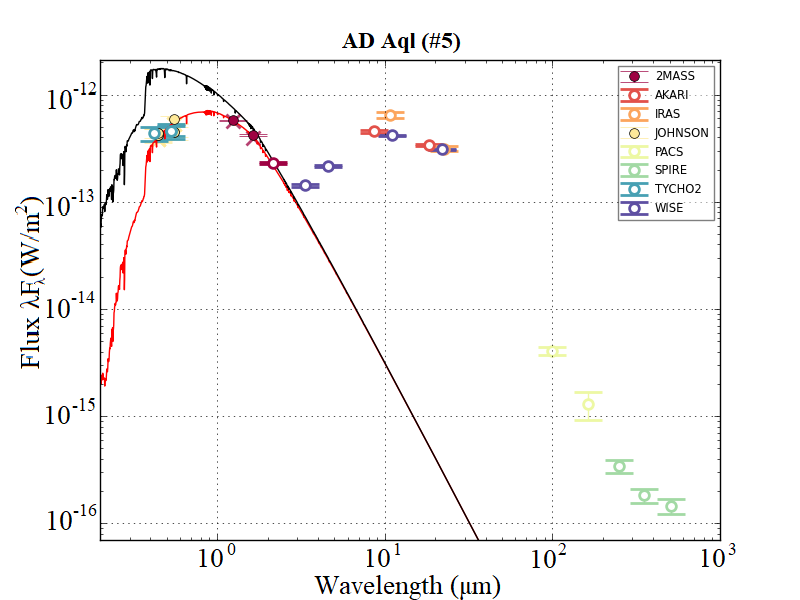}
    \includegraphics[width=.49\linewidth]{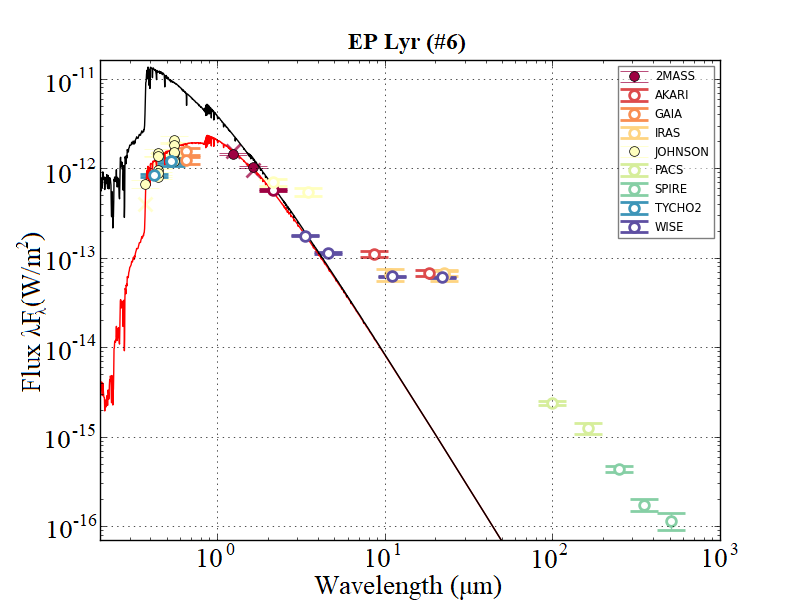}
    \caption{Spectral energy distribution of transition disc stars. The red solid line is the appropriate reddened Kurucz model atmosphere. The black solid line is the de-reddened model scaled to the object. We note that the photometric observations of our sample were obtained with different surveys at different time (pulsation phases). We also note that the IR dust excess in SED plots for AF Crt (\#8) and V1504 Sco (\#10) imply that we see these two targets edge-on. The legend for the symbols and colours used is included within the plot.}\label{figA:allSEDstr}
\end{figure*}

\begin{figure*}
    \centering
    \includegraphics[width=.49\linewidth]{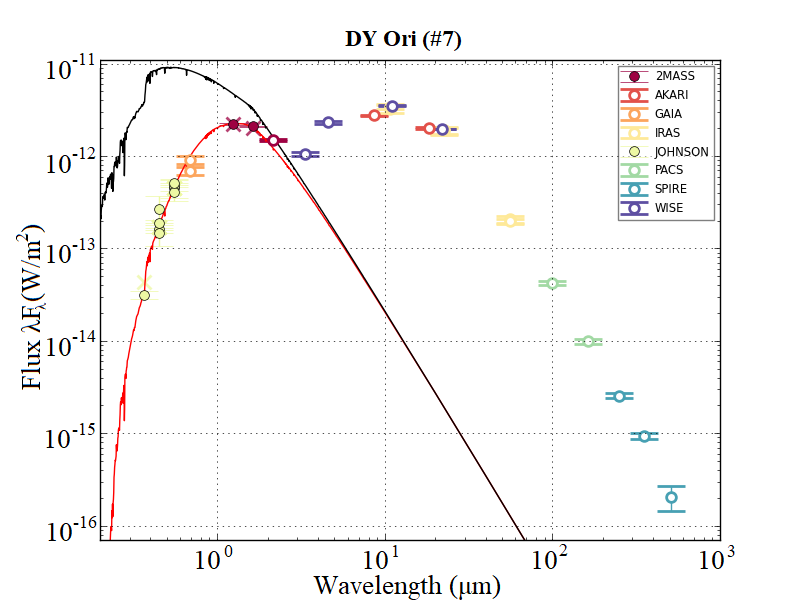}
    \includegraphics[width=.49\linewidth]{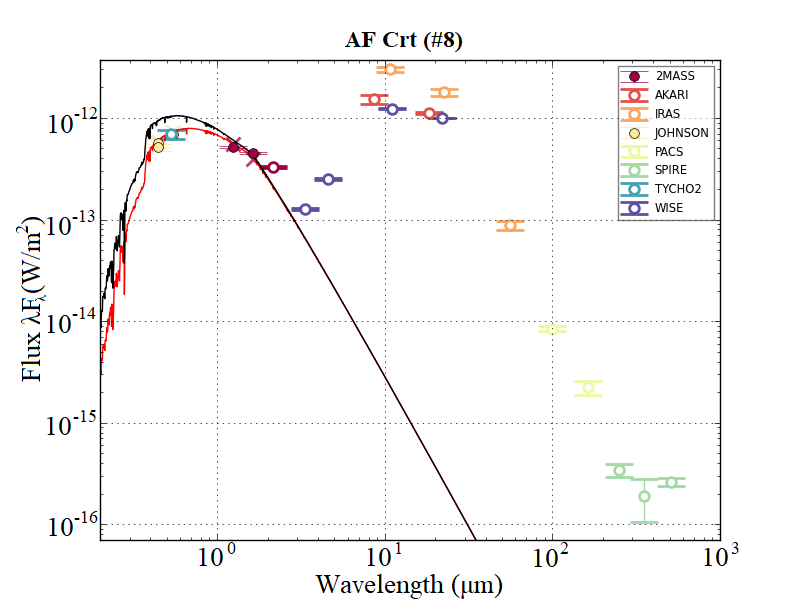}
    \includegraphics[width=.49\linewidth]{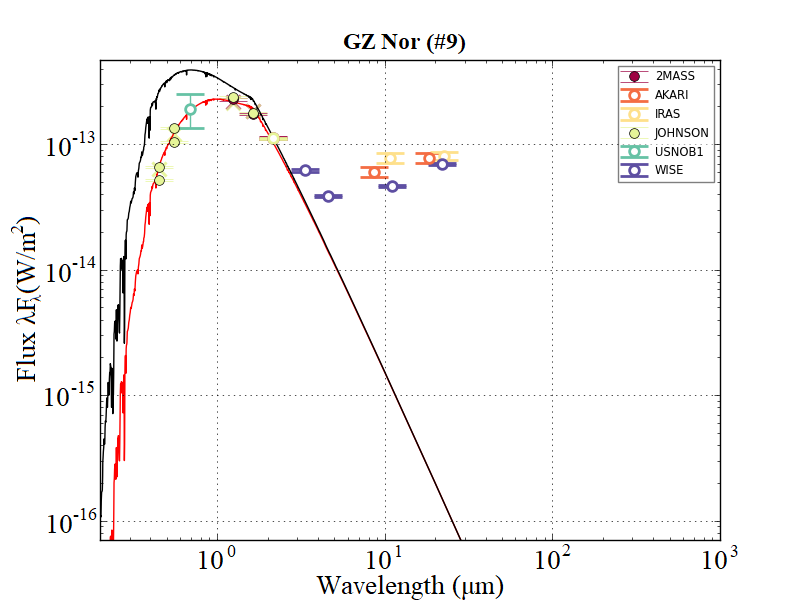}
    \includegraphics[width=.49\linewidth]{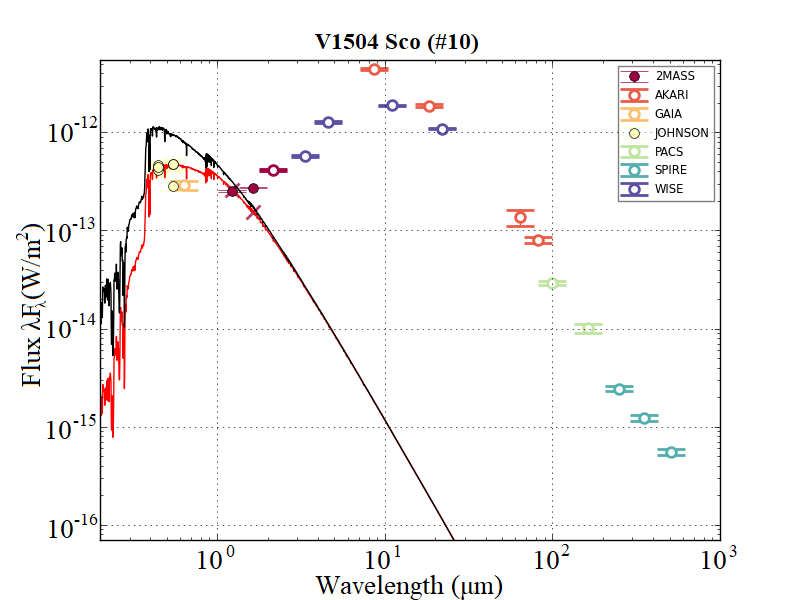}
    \includegraphics[width=.49\linewidth]{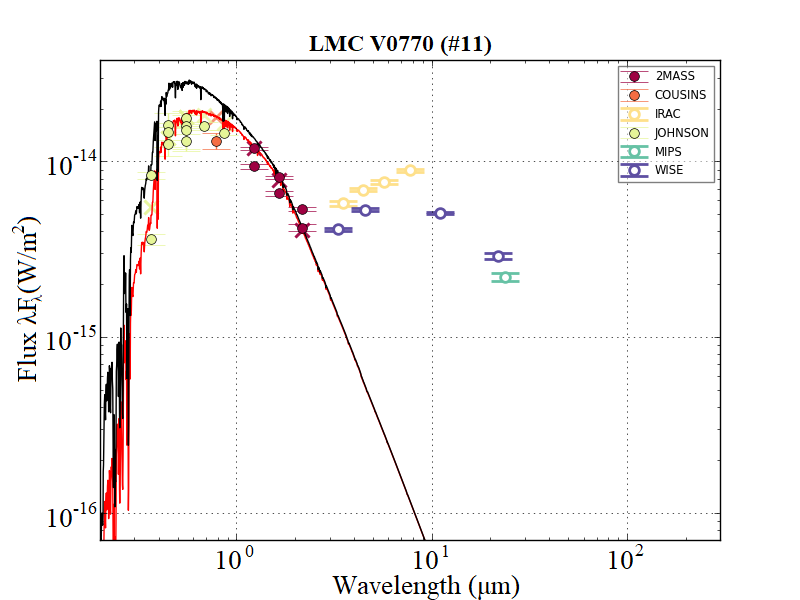}
    \includegraphics[width=.49\linewidth]{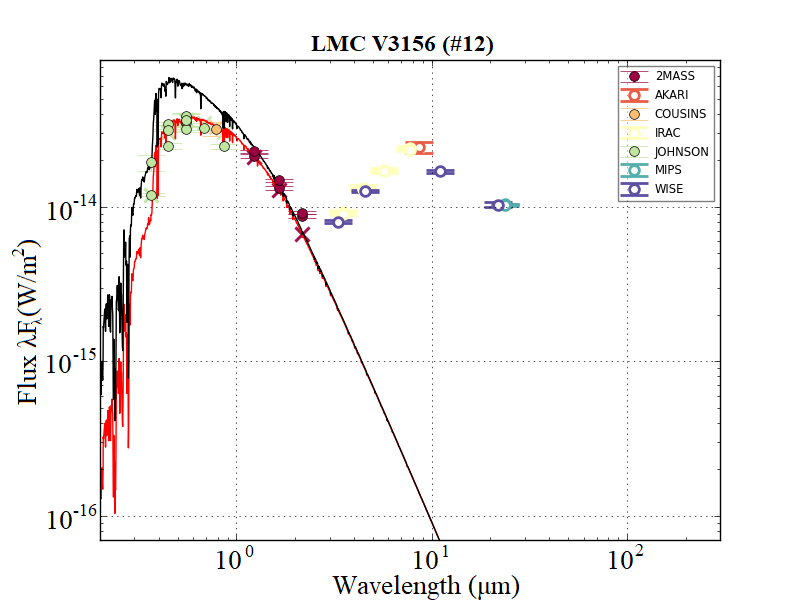}
    \caption{Spectral energy distribution of transition disc candidates. The red solid line is the appropriate reddened Kurucz model atmosphere. The black solid line is the de-reddened model scaled to the object. We note that the photometric observations of our sample were obtained with different surveys at different time (pulsation phases). We also note that the IR dust excess in SED plots for AF Crt (\#8) and V1504 Sco (\#10) imply that we see these two targets edge-on. The legend for the symbols and colours used is included within the plot.}\label{figA:allSEDcnd}
\end{figure*}

In Table~\ref{tabA:phomag}, we present the photometric data collected for SED plots of the target sample. In Fig.~\ref{figA:allSEDstr} and \ref{figA:allSEDcnd}, we provide our SED plots for all transition disc stars and candidates, respectively (see Section~\ref{ssec:dobpht}).

\begin{table*}
    \caption{Photometric data for the target sample (see Section \ref{ssec:dobpht}). For each filter we provide the units and the central wavelengths in $\mu$m. This table is published in its entirety in the electronic edition of the paper. A portion is shown here for guidance regarding its form and content.}\label{tabA:phomag}
    \begin{tabular}{|c@{\hspace{0.001cm}}|c@{\hspace{0.001cm}}|c@{\hspace{0.001cm}}|c@{\hspace{0.001cm}}|c@{\hspace{0.001cm}}|c@{\hspace{0.001cm}}|c@{\hspace{0.001cm}}|c@{\hspace{0.001cm}}|}\hline
        && GENEVA.U (GCPD) & STROMGREN.U (II/215/catalog) & ... & SPIRE.350 (SPIRE350) & SPIRE.500 (SPIRE500) \\
        ID & Name & mag & mag && Jy & Jy \\
        && 0.342 & 0.346 & ... & 348.438 & 500.412 \\ \hline
        1 & CT Ori & -- & -- & ... & 0.018$\pm$0.003 & 0.042$\pm$0.004 \\
        2 & ST Pup & -- & -- & ... & 0.015$\pm$0.003 & 0.014$\pm$0.007 \\
        3 & RU Cen & -- & 12.105$\pm$99.999 & ... & -- & -- \\
        4 & AC Her & 9.419$\pm$0.116 & 11.203$\pm$99.999 & ... & 0.75$\pm$0.04 & 0.35$\pm$0.02 \\
        5 & AD Aql & -- & -- & ... & 0.021$\pm$0.003 & 0.024$\pm$0.004 \\
        6 & EP Lyr & -- & -- & ... & 0.020$\pm$0.003 & 0.019$\pm$0.004 \\
        ... & ... & ... & ... & ... & ... & ... \\ \hline
    \end{tabular}\\
    \textbf{Note:} We list uncertainties which are unavailable as `99.999'.
\end{table*}

\section{Summary of optical spectral visits}\label{app:vis}
In Table~\ref{tabA:obslog}, we list all optical spectral visits considered in this work (HERMES and UVES; see Section~\ref{ssec:dobspc}) for our chemical analysis of the target sample (see Section~\ref{sec:san}).

\begin{table}
    \centering
    \caption{All spectral visits of the target sample. This table is published in its entirety in the electronic edition of the paper. A portion is shown here for guidance regarding its form and content.}\label{tabA:obslog}
    \begin{tabular}{|c|c|c|c|c|c|}
    \hline
        Visit & MJD & Phase & RV & eRV & SNR \\ \hline
        \multicolumn{6}{c}{CT Ori (\#1) (127 visits)} \\ \hline
        00273152 & 55221.91376 & 0.00 & 62.60 & 0.63 & 26.38 \\
        00314017 & 55500.22492 & 0.40 & 60.29 & 0.64 & 24.22 \\
        00314465 & 55507.12069 & 0.61 & 53.92 & 0.97 & 27.18 \\
        00325348 & 55553.04197 & 0.99 & 60.79 & 0.55 & 23.79 \\
        00327525 & 55576.97777 & 0.72 & 51.61 & 1.18 & 22.50 \\
        \multicolumn{6}{c}{\ldots} \\ \hline
    \end{tabular}
\end{table}

\section{Master line list of the target sample}\label{app:lst}
In Table~\ref{tabA:linlst}, we provide the combined optical line list, which we used to derive the atmospheric parameters and elemental abundances of all 12 transition disc targets (see Section~\ref{sec:san}).

\begin{landscape}
\begin{table}
    \centering
    \caption{Combined optical line list of our sample. This table is published in its entirety in the electronic edition of the paper. A portion is shown here for guidance regarding its form and content.}\label{tabA:linlst}
    \begin{tabular}{c@{\hspace{0.1cm}}ccccccccccccccc} \hline
        \multicolumn{4}{c}{Atomic data} & \multicolumn{12}{c}{$W_\lambda$ (m\AA)} \\
        Element & $\lambda$ & $\log gf$ & $\chi$ & CT Ori & ST Pup & RU Cen & AC Her & AD Aql & EP Lyr & DY Ori & AF Crt & GZ Nor & V1504 Sco & LMC V0770 & LMC V3156 \\
        ~ & (nm) & (dex) & (eV) & (\#1) & (\#2) & (\#3) & (\#4) & (\#5) & (\#6) & (\#7) & (\#8) & (\#9) & (\#10) & (\#11) & (\#12) \\ \hline
        \ion{C}{i} & 477.1730 & -1.866 & 7.488 & - & - & 82.6 & - & - & - & - & - & - & 64.6 & - & - \\
        \ion{C}{i} & 493.2049 & -1.658 & 7.685 & 123.6 & 56.7 & 89.8 & 62.8 & 109.0 & 68.0 & 86.5 & - & - & - & 72.3 & 66.3 \\
        \ion{C}{i} & 502.3841 & -2.210 & 7.946 & - & - & - & - & - & - & - & - & - & - & 26.4 & - \\
        \ion{C}{i} & 503.9057 & -1.790 & 7.946 & - & - & - & - & - & - & - & - & - & - & 63.8 & 33.7 \\
        \ion{C}{i} & 505.2144 & -1.303 & 7.685 & - & 94.3 & 123.0 & - & - & - & 134.1 & 124.5 & 92.0 & - & - & 90.4 \\
        \ion{C}{i} & 538.0325 & -1.616 & 7.685 & 120.0 & 52.8 & 98.6 & - & - & 68.3 & 83.8 & - & 57.6 & - & 81.4 & 57.0 \\
        \multicolumn{16}{c}{\ldots} \\ \hline
    \end{tabular}
\end{table}
\end{landscape}

\section{Individual depletion profiles}\label{app:dpl}
\begin{figure*}
    \centering
    \includegraphics[width=.95\linewidth]{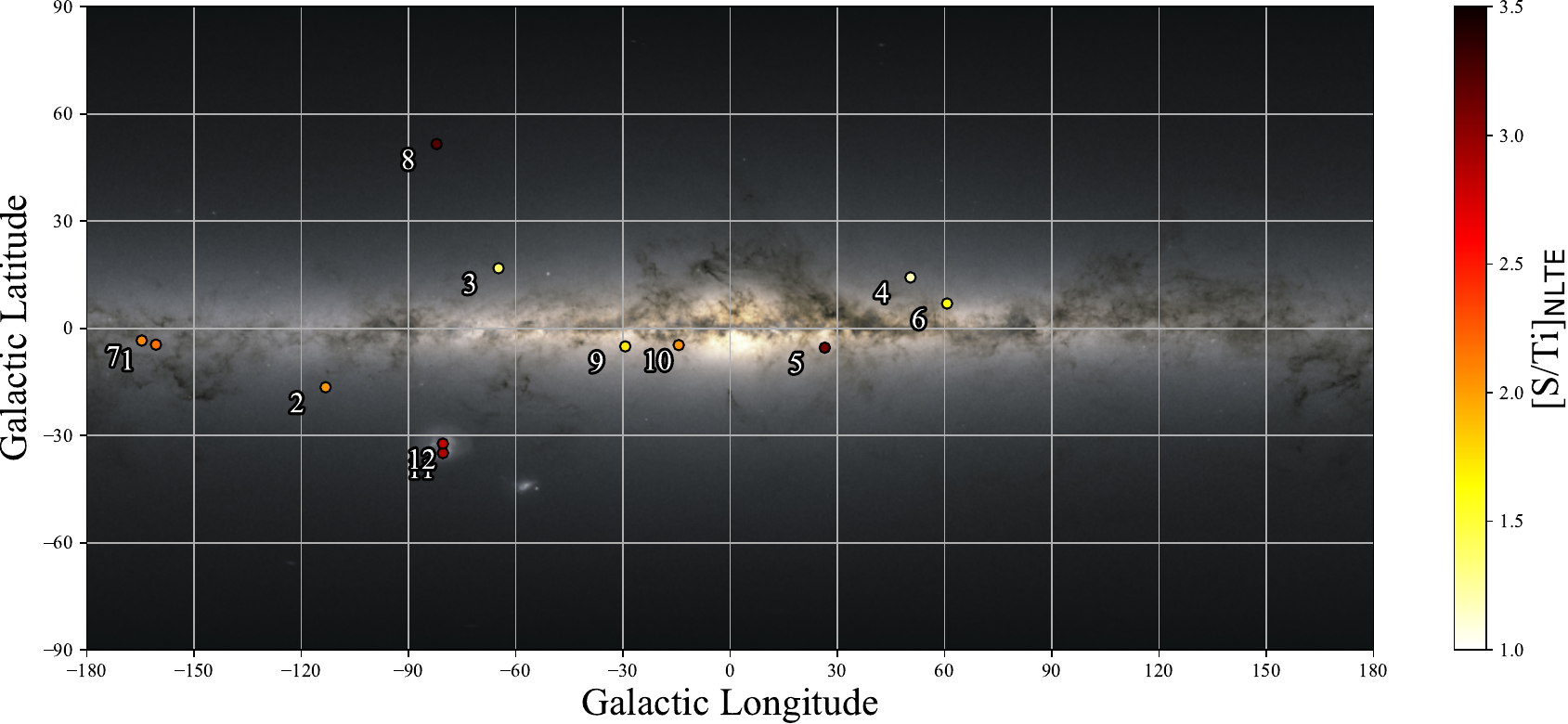}
    \caption{Spatial distribution of the transition disc targets. The targets are labelled by their IDs and coloured by the [S/Ti]$_{\rm NLTE}$ values derived in this study (see Table~\ref{tab:fnlabu}).}\label{figA:allmap}
\end{figure*}

In Table~\ref{tabA:fnlabu}, we provide all elemental abundances derived in this study (see Section~\ref{ssec:sanlte}). In Table \ref{tabA:fnlcor}, we provide the atmospheric parameters and differential NLTE corrections for transition disc targets (see Section~\ref{ssec:sannlte}). In Fig.~\ref{figA:allmap}, we present the spatial distribution of transition disc targets.

\begin{landscape}
\begin{table}
    \centering
    \caption{1D LTE and 1D NLTE elemental abundances of transition disc targets in [X/H] scale.}\label{tabA:fnlabu}
    \resizebox{1.15\textheight}{!}{
    \begin{tabular}{|c|c|c|c|c|c|c|c|c|c|c|c|c|c|} \hline
        \multirow{2}{*}{Ion} & \multirow{2}{*}{$T_{\rm cond}$} & CT Ori & ST Pup & RU Cen & AC Her & AD Aql & EP Lyr & DY Ori & AF Crt & GZ Nor & V1504 Sco & LMC V0770 & LMC V3156 \\
        && (\#1) & (\#2) & (\#3) & (\#4) & (\#5) & (\#6) & (\#7) & (\#8) & (\#9) & (\#10) & (\#11) & (\#12) \\ \hline
        \multicolumn{14}{c}{LTE} \\ \hline
        \ion{C}{i} & 40 & --0.57$\pm$0.09 & --0.50$\pm$0.07 & --0.44$\pm$0.07 & --0.59$\pm$0.06 & --0.10$\pm$0.13 & --0.28$\pm$0.07 & 0.16$\pm$0.08 & --0.41$\pm$0.09 & 0.32$\pm$0.02 & --0.11$\pm$0.07 & --0.75$\pm$0.02 & --0.34$\pm$0.09 \\
        \ion{N}{i} & 123 & --0.78$\pm$0.18 & -- & 0.13$\pm$0.11 & --0.03$\pm$0.11 & --0.03$\pm$0.21 & 0.07$\pm$0.14 & 0.38$\pm$0.19 & --0.54$\pm$0.19 & 0.84$\pm$0.10 & 0.59$\pm$0.13 & -- & -- \\
        \ion{O}{i} & 183 & 0.03$\pm$0.09 & --0.50$\pm$0.05 & --0.01$\pm$0.12 & 0.01$\pm$0.08 & 0.75$\pm$0.17 & 0.14$\pm$0.14 & 0.32$\pm$0.17 & 0.43$\pm$0.17 & 0.50$\pm$0.06 & 0.31$\pm$0.11 & --0.40$\pm$0.10 & 0.78$\pm$0.14 \\
        \ion{Na}{i} & 1035 & --0.33$\pm$0.06 & --0.62$\pm$0.05 & --0.92$\pm$0.05 & --0.55$\pm$0.11 & --0.40$\pm$0.05 & --1.07$\pm$0.14 & 0.32$\pm$0.15 & 0.24$\pm$0.17 & --1.51$\pm$0.10 & 0.48$\pm$0.11 & --0.70$\pm$0.10 & 0.47$\pm$0.14 \\
        \ion{Mg}{i} & 1343 & --1.36$\pm$0.10 & --1.45$\pm$0.08 & --1.68$\pm$0.08 & --1.20$\pm$0.07 & --1.88$\pm$0.10 & --1.91$\pm$0.08 & --1.99$\pm$0.12 & --2.03$\pm$0.12 & --1.65$\pm$0.03 & --1.21$\pm$0.09 & --2.15$\pm$0.07 & --2.21$\pm$0.12 \\
        \ion{Al}{i} & 1652 & --2.40$\pm$0.16 & -- & -- & --1.86$\pm$0.13 & -- & -- & -- & --2.81$\pm$0.16 & -- & -- & --3.83$\pm$0.10 & --2.64$\pm$0.15 \\
        \ion{Si}{i} & 1314 & -- & -- & --1.48$\pm$0.10 & --1.18$\pm$0.03 & -- & --1.66$\pm$0.12 & -- & -- & -- & --0.75$\pm$0.10 & --1.99$\pm$0.10 & -- \\
        \ion{Si}{ii} & 1314 & --1.63$\pm$0.14 & -- & --1.55$\pm$0.12 & --1.20$\pm$0.13 & --1.56$\pm$0.17 & --1.52$\pm$0.08 & --1.47$\pm$0.14 & --1.66$\pm$0.11 & -- & --0.87$\pm$0.13 & -- & --1.47$\pm$0.14 \\
        \ion{S}{i} & 672 & --0.17$\pm$0.06 & --0.34$\pm$0.10 & --0.42$\pm$0.08 & --0.49$\pm$0.03 & 0.03$\pm$0.12 & --0.38$\pm$0.04 & 0.62$\pm$0.08 & --0.07$\pm$0.07 & --0.12$\pm$0.06 & 0.41$\pm$0.10 & --0.36$\pm$0.01 & 0.07$\pm$0.08 \\
        \ion{K}{i} & 993 & --0.47$\pm$0.15 & -- & --0.74$\pm$0.12 & --0.55$\pm$0.07 & --0.09$\pm$0.17 & --1.41$\pm$0.14 & -- & --0.81$\pm$0.15 & --1.31$\pm$0.10 & -- & -- & 0.34$\pm$0.15 \\
        \ion{Ca}{i} & 1535 & --1.64$\pm$0.11 & --2.04$\pm$0.06 & --1.86$\pm$0.05 & --1.45$\pm$0.07 & --2.43$\pm$0.08 & --2.02$\pm$0.11 & --1.76$\pm$0.14 & --2.22$\pm$0.11 & --1.92$\pm$0.03 & --1.41$\pm$0.06 & --2.56$\pm$0.10 & --1.79$\pm$0.12 \\
        \ion{Ca}{ii} & 1535 & --1.81$\pm$0.14 & -- & -- & --1.57$\pm$0.13 & --2.64$\pm$0.16 & -- & --2.04$\pm$0.13 & --2.23$\pm$0.15 & -- & -- & -- & -- \\
        \ion{Sc}{ii} & 1541 & --2.55$\pm$0.10 & --2.48$\pm$0.05 & --2.09$\pm$0.08 & --1.96$\pm$0.07 & -- & --2.29$\pm$0.11 & -- & -- & --2.15$\pm$0.10 & --1.63$\pm$0.06 & --3.74$\pm$0.09 & --2.64$\pm$0.10 \\
        \ion{Ti}{i} & 1565 & -- & --2.53$\pm$0.13 & --1.92$\pm$0.06 & --1.64$\pm$0.10 & -- & -- & -- & -- & -- & -- & -- & -- \\
        \ion{Ti}{ii} & 1565 & --2.48$\pm$0.11 & --2.44$\pm$0.04 & --1.99$\pm$0.07 & --1.85$\pm$0.07 & --3.32$\pm$0.12 & --2.20$\pm$0.10 & --1.77$\pm$0.14 & --3.45$\pm$0.15 & --1.94$\pm$0.03 & --1.81$\pm$0.07 & --3.40$\pm$0.06 & --2.95$\pm$0.09 \\
        \ion{V}{ii} & 1370 & -- & --1.79$\pm$0.02 & --1.81$\pm$0.11 & --1.38$\pm$0.06 & -- & --1.74$\pm$0.13 & -- & -- & --1.78$\pm$0.01 & --1.56$\pm$0.11 & -- & --1.94$\pm$0.13 \\
        \ion{Cr}{i} & 1291 & --1.86$\pm$0.13 & --2.19$\pm$0.07 & --1.97$\pm$0.06 & --1.45$\pm$0.07 & --2.22$\pm$0.10 & --2.19$\pm$0.17 & --2.05$\pm$0.16 & -- & -- & --1.24$\pm$0.08 & --2.79$\pm$0.09 & --2.49$\pm$0.14 \\
        \ion{Cr}{ii} & 1291 & --1.94$\pm$0.11 & --1.95$\pm$0.03 & --1.90$\pm$0.06 & --1.42$\pm$0.07 & --2.16$\pm$0.16 & --2.06$\pm$0.07 & -- & -- & --1.80$\pm$0.04 & --1.27$\pm$0.07 & --2.77$\pm$0.00 & -- \\
        \ion{Mn}{i} & 1123 & --1.62$\pm$0.13 & --1.70$\pm$0.06 & --1.82$\pm$0.05 & --1.23$\pm$0.07 & --1.58$\pm$0.16 & --1.39$\pm$0.14 & -- & --1.60$\pm$0.16 & --2.12$\pm$0.05 & --0.90$\pm$0.09 & --2.22$\pm$0.05 & --1.62$\pm$0.12 \\
        \ion{Mn}{ii} & 1123 & -- & --1.74$\pm$0.10 & -- & -- & -- & --1.35$\pm$0.13 & -- & -- & -- & -- & --2.20$\pm$0.10 & -- \\
        \ion{Fe}{i} & 1338 & --1.85$\pm$0.11 & --1.91$\pm$0.06 & --1.93$\pm$0.05 & --1.47$\pm$0.06 & --2.24$\pm$0.10 & --1.98$\pm$0.10 & --1.97$\pm$0.12 & --2.46$\pm$0.15 & --1.90$\pm$0.03 & --1.06$\pm$0.07 & --2.61$\pm$0.03 & --2.47$\pm$0.11 \\
        \ion{Fe}{ii} & 1338 & --1.89$\pm$0.07 & --1.92$\pm$0.03 & --1.95$\pm$0.07 & --1.48$\pm$0.07 & --2.21$\pm$0.14 & --2.02$\pm$0.08 & --1.98$\pm$0.08 & --2.49$\pm$0.11 & --1.89$\pm$0.04 & --1.03$\pm$0.07 & --2.50$\pm$0.02 & --2.48$\pm$0.09 \\
        \ion{Co}{i} & 1354 & -- & --1.85$\pm$0.13 & --1.84$\pm$0.12 & --1.45$\pm$0.13 & --2.00$\pm$0.14 & --1.91$\pm$0.16 & -- & -- & --1.42$\pm$0.11 & --0.84$\pm$0.14 & --2.55$\pm$0.10 & -- \\
        \ion{Ni}{i} & 1363 & --1.37$\pm$0.16 & --1.92$\pm$0.05 & --1.77$\pm$0.04 & --1.33$\pm$0.06 & --2.14$\pm$0.14 & --2.09$\pm$0.15 & --1.84$\pm$0.15 & -- & --1.62$\pm$0.06 & --1.13$\pm$0.12 & --3.05$\pm$0.12 & --2.00$\pm$0.17 \\
        \ion{Cu}{i} & 1034 & --1.10$\pm$0.17 & --1.46$\pm$0.09 & -- & -- & -- & -- & 0.42$\pm$0.16 & --1.13$\pm$0.18 & -- & -- & -- & --0.08$\pm$0.15 \\
        \ion{Zn}{i} & 704 & --0.57$\pm$0.12 & --0.68$\pm$0.06 & --1.04$\pm$0.04 & --0.71$\pm$0.06 & 0.01$\pm$0.11 & --0.48$\pm$0.12 & 0.08$\pm$0.13 & --0.33$\pm$0.13 & --1.26$\pm$0.08 & --0.16$\pm$0.07 & --0.94$\pm$0.19 & --0.33$\pm$0.09 \\
        \ion{Sr}{ii} & 1548 & -- & -- & -- & -- & -- & -- & -- & --2.51$\pm$0.19 & --1.54$\pm$0.10 & -- & --2.84$\pm$0.10 & -- \\
        \ion{Y}{ii} & 1551 & --2.45$\pm$0.13 & --2.75$\pm$0.05 & --2.22$\pm$0.11 & --1.95$\pm$0.10 & -- & --2.29$\pm$0.14 & --1.64$\pm$0.14 & -- & --2.23$\pm$0.10 & --1.58$\pm$0.07 & -- & --2.91$\pm$0.14 \\
        \ion{Zr}{ii} & 1722 & -- & --2.89$\pm$0.11 & --2.15$\pm$0.12 & -- & -- & -- & -- & -- & --2.05$\pm$0.10 & -- & -- & -- \\
        \ion{Ba}{ii} & 1423 & --1.92$\pm$0.15 & --2.15$\pm$0.12 & --2.01$\pm$0.08 & --1.52$\pm$0.09 & --2.53$\pm$0.16 & --2.14$\pm$0.16 & --1.97$\pm$0.13 & --2.92$\pm$0.17 & -- & --1.16$\pm$0.13 & --3.56$\pm$0.10 & --1.93$\pm$0.13 \\
        \ion{La}{ii} & 1615 & -- & --2.56$\pm$0.05 & -- & -- & -- & -- & -- & -- & --2.07$\pm$0.04 & -- & -- & -- \\
        \ion{Ce}{ii} & 1454 & -- & --2.18$\pm$0.11 & --1.79$\pm$0.12 & --1.41$\pm$0.07 & -- & -- & -- & -- & --2.01$\pm$0.04 & --1.58$\pm$0.12 & -- & -- \\
        \ion{Nd}{ii} & 1630 & -- & -- & -- & -- & -- & -- & -- & -- & --1.97$\pm$0.05 & -- & -- & -- \\
        \ion{Sm}{ii} & 1545 & -- & -- & -- & -- & -- & -- & -- & -- & --1.91$\pm$0.10 & -- & -- & -- \\
        \ion{Eu}{ii} & 1491 & -- & --1.71$\pm$0.11 & -- & -- & -- & -- & -- & -- & -- & -- & -- & -- \\ \hline
        \multicolumn{14}{c}{NLTE} \\ \hline
        \ion{C}{i} & 40 & --0.70$\pm$0.09 & --0.61$\pm$0.07 & --0.57$\pm$0.07 & --0.69$\pm$0.06 & --0.24$\pm$0.13 & --0.41$\pm$0.07 & --0.07$\pm$0.08 & --0.55$\pm$0.09 & 0.09$\pm$0.04 & --0.26$\pm$0.06 & --0.87$\pm$0.01 & --0.52$\pm$0.09 \\
        \ion{N}{i} & 123 & --0.89$\pm$0.18 & -- & --0.01$\pm$0.10 & --0.16$\pm$0.10 & --0.17$\pm$0.21 & --0.12$\pm$0.15 & 0.08$\pm$0.19 & --0.80$\pm$0.19 & 0.70$\pm$0.10 & 0.39$\pm$0.13 & -- & -- \\
        \ion{O}{i} & 183 & --0.02$\pm$0.09 & --0.5$\pm$0.05 & --0.01$\pm$0.12 & 0.01$\pm$0.08 & 0.75$\pm$0.17 & 0.15$\pm$0.14 & 0.23$\pm$0.23 & 0.43$\pm$0.17 & --0.03$\pm$0.06 & 0.31$\pm$0.11 & --0.54$\pm$0.10 & 0.78$\pm$0.14 \\
        \ion{Na}{i} & 1035 & --0.49$\pm$0.06 & --0.74$\pm$0.05 & --1.03$\pm$0.06 & --0.70$\pm$0.11 & --0.55$\pm$0.05 & --1.45$\pm$0.14 & 0.07$\pm$0.12 & --0.53$\pm$0.17 & --1.69$\pm$0.10 & 0.15$\pm$0.11 & --0.82$\pm$0.10 & --0.35$\pm$0.14 \\
        \ion{Mg}{i} & 1343 & --1.33$\pm$0.11 & --1.40$\pm$0.08 & --1.75$\pm$0.04 & --1.20$\pm$0.09 & --1.86$\pm$0.08 & --1.94$\pm$0.07 & --2.04$\pm$0.12 & --2.07$\pm$0.09 & --1.58$\pm$0.05 & --1.29$\pm$0.06 & --2.14$\pm$0.05 & --2.18$\pm$0.12 \\
        \ion{Al}{i} & 1652 & --1.97$\pm$0.16 & -- & -- & --1.45$\pm$0.13 & -- & -- & -- & --2.21$\pm$0.16 & -- & -- & --3.17$\pm$0.10 & --2.06$\pm$0.15 \\
        \ion{Si}{i} & 1314 & -- & -- & --1.62$\pm$0.10 & --1.29$\pm$0.03 & -- & --1.79$\pm$0.12 & -- & -- & -- & --0.84$\pm$0.10 & --1.88$\pm$0.10 & -- \\
        \ion{S}{i} & 672 & --0.31$\pm$0.07 & --0.42$\pm$0.13 & --0.60$\pm$0.08 & --0.66$\pm$0.03 & --0.21$\pm$0.15 & --0.58$\pm$0.04 & 0.31$\pm$0.10 & --0.23$\pm$0.07 & --0.23$\pm$0.07 & 0.23$\pm$0.10 & --0.54$\pm$0.02 & --0.14$\pm$0.06 \\
        \ion{K}{i} & 993 & --0.99$\pm$0.15 & -- & --1.24$\pm$0.12 & --1.03$\pm$0.08 & --0.86$\pm$0.15 & --1.65$\pm$0.14 & -- & --1.1$\pm$0.15 & --1.55$\pm$0.10 & -- & -- & --0.44$\pm$0.15 \\
        \ion{Ca}{i} & 1535 & --1.55$\pm$0.12 & --1.87$\pm$0.05 & --1.75$\pm$0.05 & --1.39$\pm$0.07 & --2.30$\pm$0.08 & --1.92$\pm$0.11 & --1.68$\pm$0.14 & --2.05$\pm$0.11 & --1.79$\pm$0.03 & --1.36$\pm$0.07 & --2.34$\pm$0.10 & --1.67$\pm$0.13 \\
        \ion{Ca}{ii} & 1535 & --2.10$\pm$0.14 & -- & -- & --1.65$\pm$0.13 & --2.75$\pm$0.16 & -- & --2.29$\pm$0.12 & --2.38$\pm$0.15 & -- & -- & -- & -- \\
        \ion{Fe}{i} & 1338 & --1.70$\pm$0.11 & --1.74$\pm$0.06 & --1.76$\pm$0.05 & --1.32$\pm$0.06 & --2.04$\pm$0.10 & --1.81$\pm$0.10 & --1.81$\pm$0.12 & --2.19$\pm$0.15 & --1.79$\pm$0.03 & --1.01$\pm$0.07 & --2.37$\pm$0.03 & --2.14$\pm$0.11 \\
        \ion{Fe}{ii} & 1338 & --1.88$\pm$0.07 & --1.91$\pm$0.03 & --1.94$\pm$0.07 & --1.49$\pm$0.07 & --2.20$\pm$0.14 & --2.01$\pm$0.08 & --1.98$\pm$0.08 & --2.46$\pm$0.11 & --1.89$\pm$0.04 & --1.06$\pm$0.07 & --2.49$\pm$0.02 & --2.44$\pm$0.09 \\ \hline
    \end{tabular}}\\
    \textbf{Notes:} The condensation temperatures for C and N were adopted from \cite{lodders2003CondensationTemperatures}, the condensation temperatures for all other elements were adopted from \cite{wood2019CondensationTemperatures}. Uncertainties below 0.1\,dex are expected for elemental abundances derived from a small (2-4) number of lines displaying individual abundances, which are similar and stable within the uncertainty range of atmospheric parameters.
\end{table}
\end{landscape}
\begin{landscape}
\begin{table}
    \caption{Atmospheric parameters and differential NLTE corrections for studied elemental abundances of transition disc targets.}\label{tabA:fnlcor}
    \begin{tabular}{|c|c|c|c|c|c|c|c|c|c|c|c|c|}
    \hline
        \multirow{2}{*}{Name} & CT Ori & ST Pup & RU Cen & AC Her & AD Aql & EP Lyr & DY Ori & AF Crt & GZ Nor & V1504 Sco & LMC V0770 & LMC V3156 \\
        & (\#1) & (\#2) & (\#3) & (\#4) & (\#5) & (\#6) & (\#7) & (\#8) & (\#9) & (\#10) & (\#11) & (\#12) \\ \hline
        $T_{\rm eff}$ & 5940$\pm$120 & 5340$\pm$80 & 6120$\pm$80 & 6140$\pm$100 & 6200$\pm$170 & 6270$\pm$160 & 6160$\pm$70 & 6110$\pm$110 & 4830$\pm$20 & 5980$\pm$90 & 5750$\pm$100 & 6160$\pm$130 \\
        $\log g$ & 1.01$\pm$0.18 & 0.20$\pm$0.10 & 1.46$\pm$0.15 & 1.27$\pm$0.16 & 1.67$\pm$0.45 & 1.24$\pm$0.18 & 0.88$\pm$0.14 & 0.96$\pm$0.21 & 0.00$\pm$0.18 & 0.98$\pm$0.17 & 0.00$\pm$0.18 & 1.38$\pm$0.20 \\
        $[$Fe/H$]$ & --1.89$\pm$0.11 & --1.92$\pm$0.08 & --1.93$\pm$0.08 & --1.47$\pm$0.08 & --2.20$\pm$0.09 & --2.03$\pm$0.17 & --2.03$\pm$0.04 & --2.47$\pm$0.05 & --1.89$\pm$0.11 & --1.05$\pm$0.07 & --2.61$\pm$0.05 & --2.48$\pm$0.04 \\ \hline
        \ion{C}{i} & --0.13 & --0.11 & --0.13 & --0.10 & --0.14 & --0.13 & --0.23 & --0.14 & --0.23 & --0.15 & --0.12 & --0.18 \\
        \ion{N}{i} & --0.11 & -- & --0.14 & --0.13 & --0.14 & --0.19 & --0.30 & --0.26 & --0.14 & --0.20 & -- & -- \\
        \ion{O}{i} & --0.05 & 0.00 & 0.00 & 0.00 & 0.00 & 0.01 & --0.09 & 0.00 & --0.53 & 0.00 & --0.14 & 0.00 \\
        \ion{Na}{i} & --0.16 & --0.12 & --0.11 & --0.15 & --0.15 & --0.38 & --0.25 & --0.77 & --0.18 & --0.33 & --0.12 & --0.82 \\
        \ion{Mg}{i} & 0.03 & 0.05 & --0.07 & 0.00 & 0.02 & --0.03 & --0.05 & --0.04 & 0.07 & --0.08 & 0.01 & 0.03 \\
        \ion{Al}{i} & 0.43 & -- & -- & 0.41 & -- & -- & -- & 0.60 & -- & -- & 0.66 & 0.58 \\
        \ion{Si}{i} & -- & -- & --0.14 & --0.11 & -- & --0.13 & -- & -- & -- & --0.09 & 0.11 & -- \\
        \ion{S}{i} & --0.14 & --0.08 & --0.18 & --0.17 & --0.24 & --0.20 & --0.31 & --0.16 & --0.11 & --0.18 & --0.18 & --0.21 \\
        \ion{K}{i} & --0.52 & -- & --0.50 & --0.48 & --0.77 & --0.24 & -- & --0.29 & --0.24 & -- & -- & --0.78 \\
        \ion{Ca}{i} & 0.09 & 0.17 & 0.11 & 0.06 & 0.13 & 0.10 & 0.08 & 0.17 & 0.13 & 0.05 & 0.22 & 0.12 \\
        \ion{Ca}{ii} & --0.29 & -- & -- & --0.08 & --0.11 & -- & --0.25 & --0.15 & -- & -- & -- & -- \\
        \ion{Fe}{i} & 0.15 & 0.17 & 0.17 & 0.15 & 0.20 & 0.17 & 0.16 & 0.27 & 0.11 & 0.05 & 0.24 & 0.33 \\
        \ion{Fe}{ii} & 0.01 & 0.01 & 0.01 & --0.01 & 0.01 & 0.01 & 0.00 & 0.03 & 0.00 & --0.03 & 0.01 & 0.04 \\ \hline
    \end{tabular}
\end{table}
\end{landscape}

\section{Individual correlation plots}\label{app:cor}
\begin{figure*}
    \centering
    \includegraphics[width=.49\linewidth]{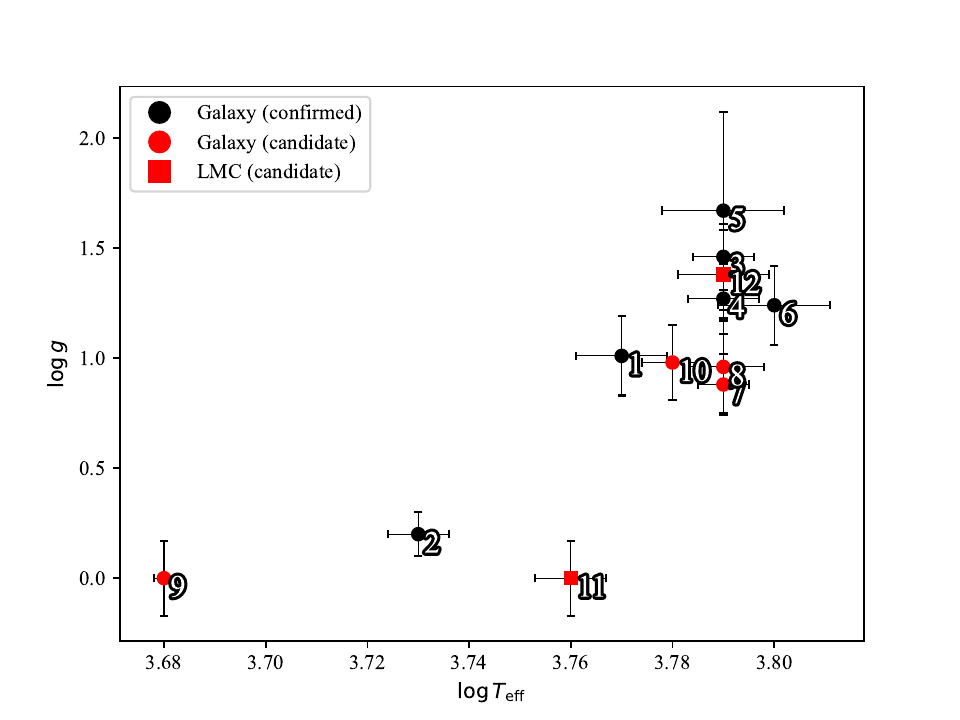}
    \includegraphics[width=.49\linewidth]{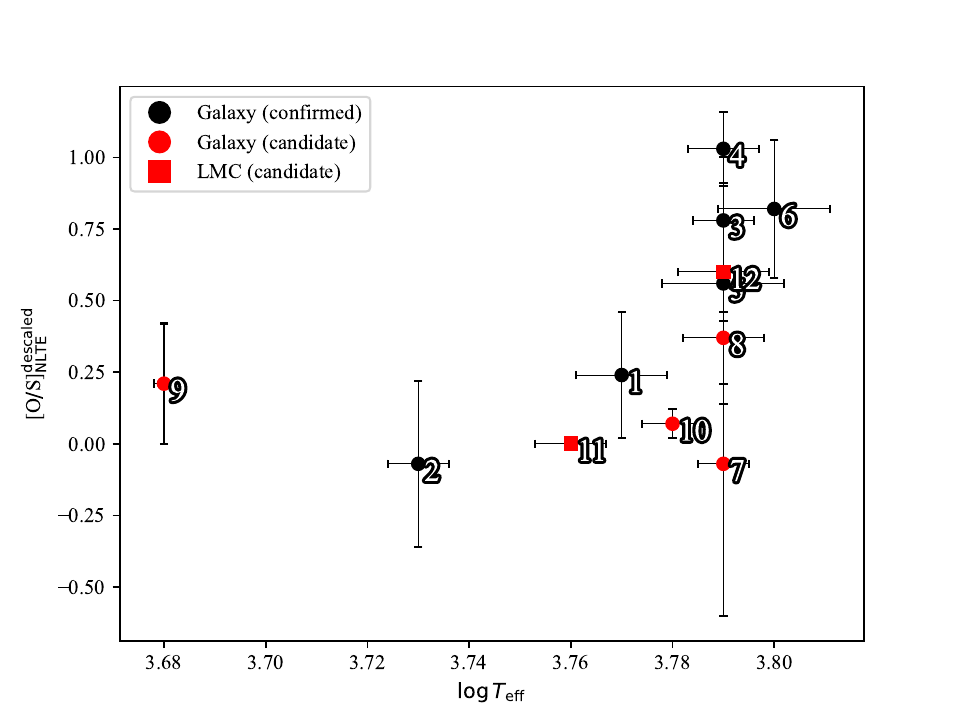}
    \includegraphics[width=.49\linewidth]{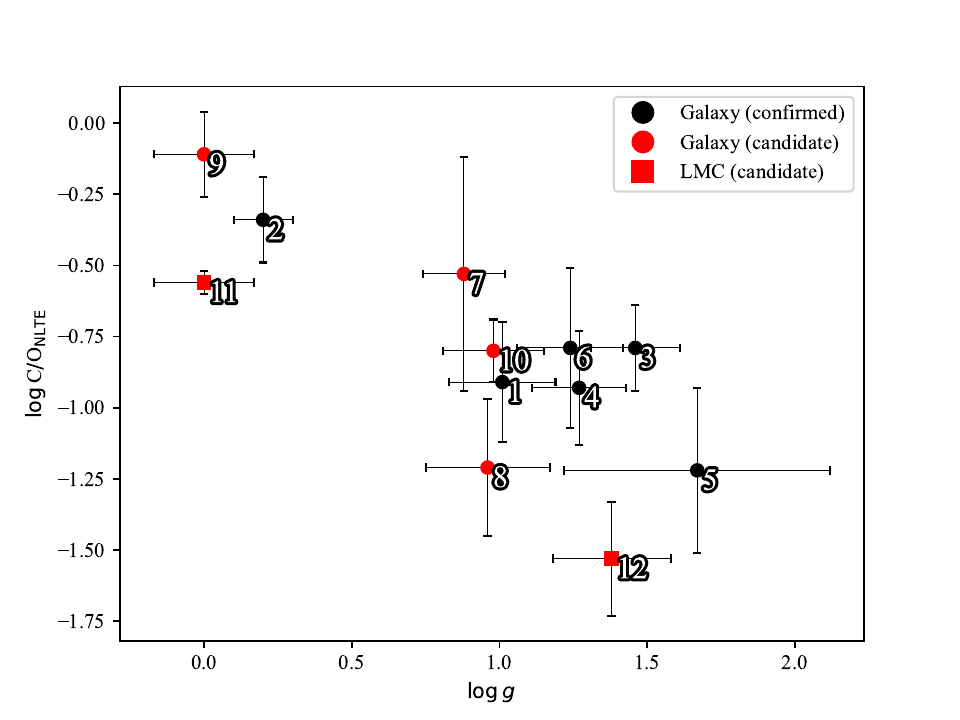}
    \includegraphics[width=.49\linewidth]{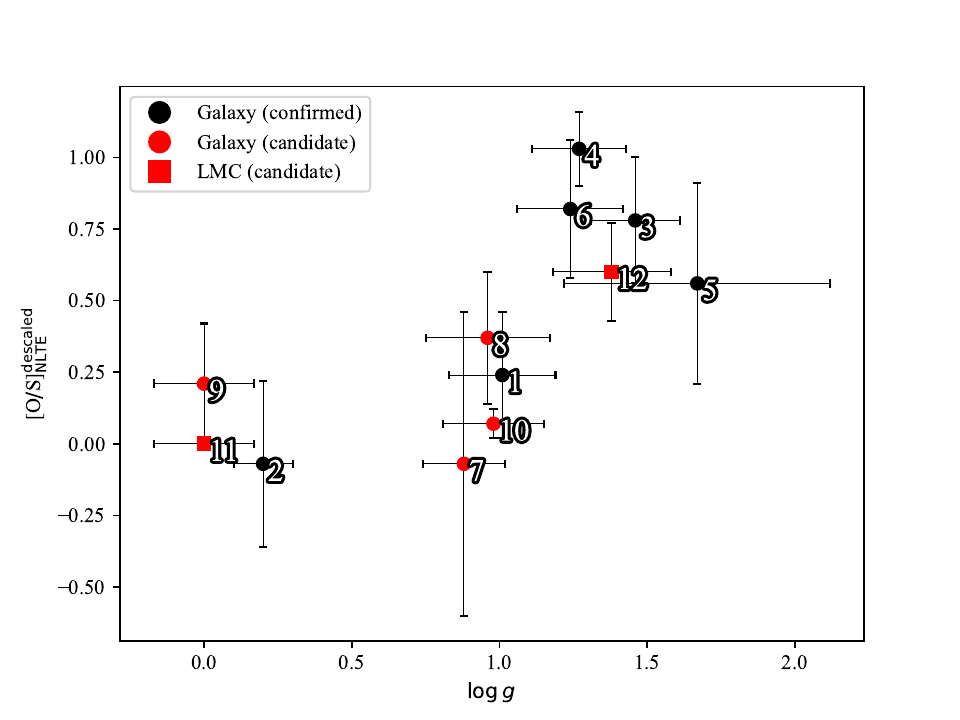}
    \caption{The strongest individual correlations in transition disc targets (see Section~\ref{ssec:dplcor}). The legend for the symbols and colours used is included within the plot. The targets are marked with their IDs.}\label{figA:allcor}
\end{figure*}

In Fig.~\ref{figA:allcor}, we show the strong correlations and anti-correlations (with Spearman's coefficient $|\rho|\geq0.6$) between different parameters of the studied sample (see Section~\ref{ssec:dplcor}). We note that AF Crt (\#8) and V1504 Sco (\#10) were excluded from luminosity plots, since the luminosities of these edge-on targets are less reliable.

\begin{figure*}
    \centering
    \includegraphics[width=.49\linewidth]{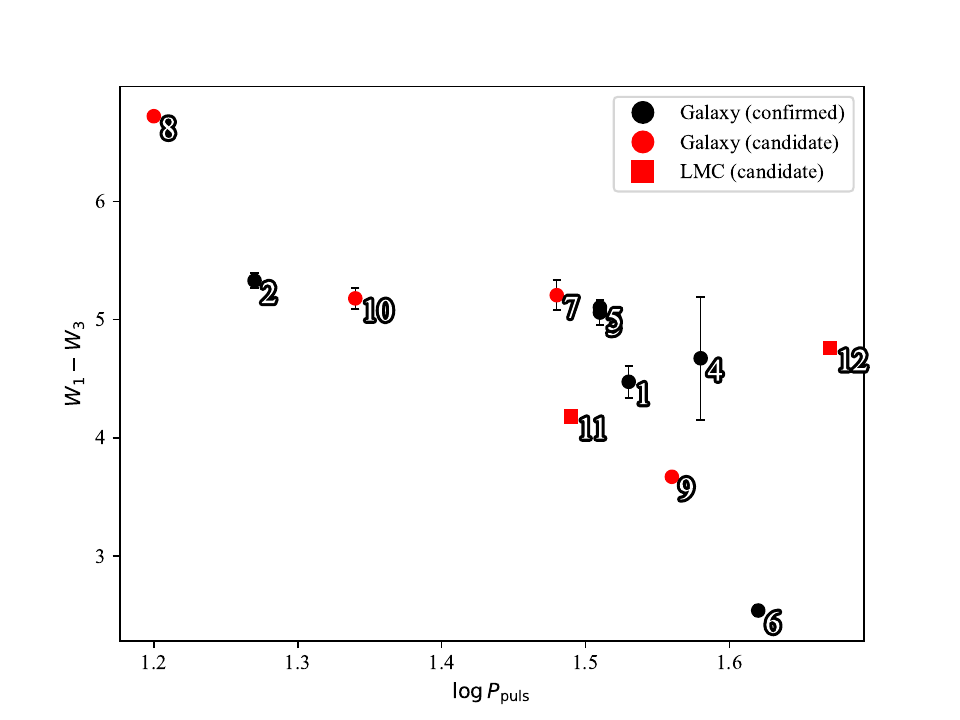}
    \includegraphics[width=.49\linewidth]{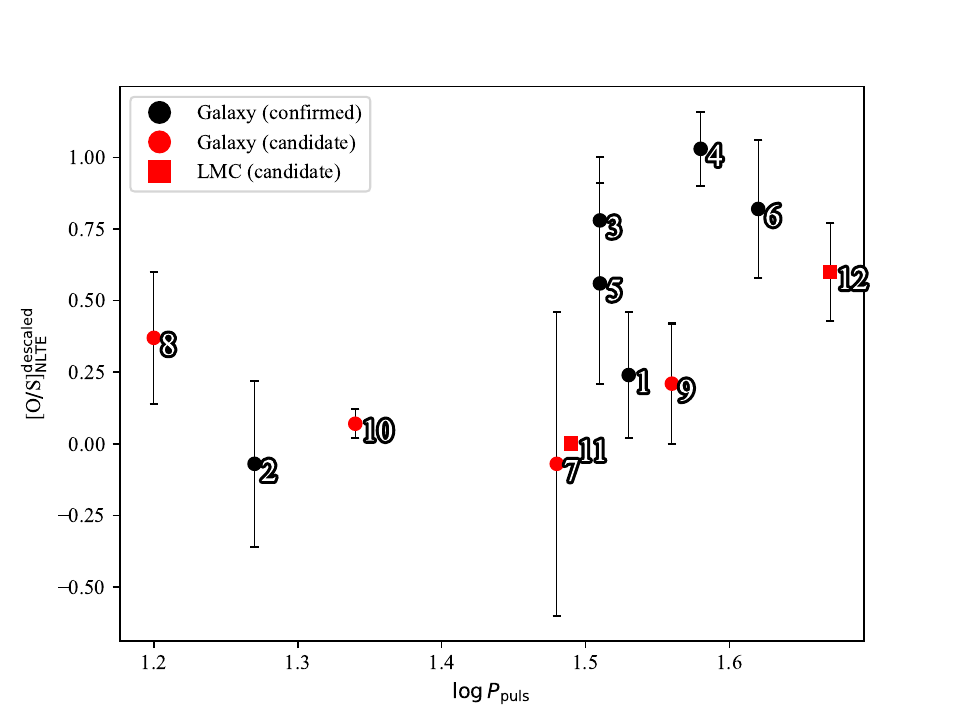}
    \includegraphics[width=.49\linewidth]{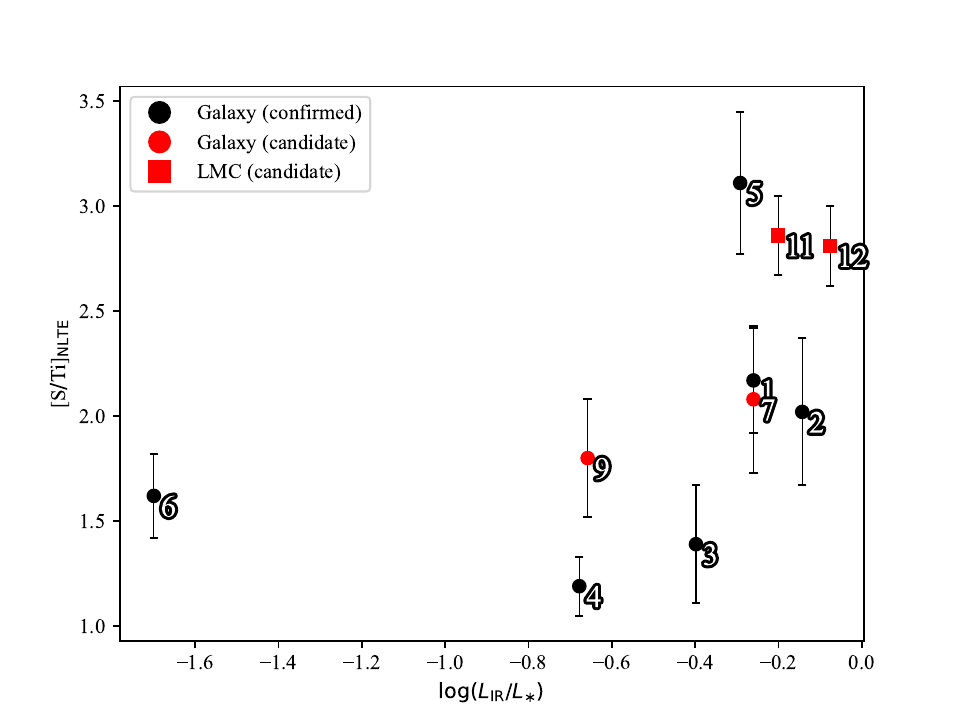}
    \includegraphics[width=.49\linewidth]{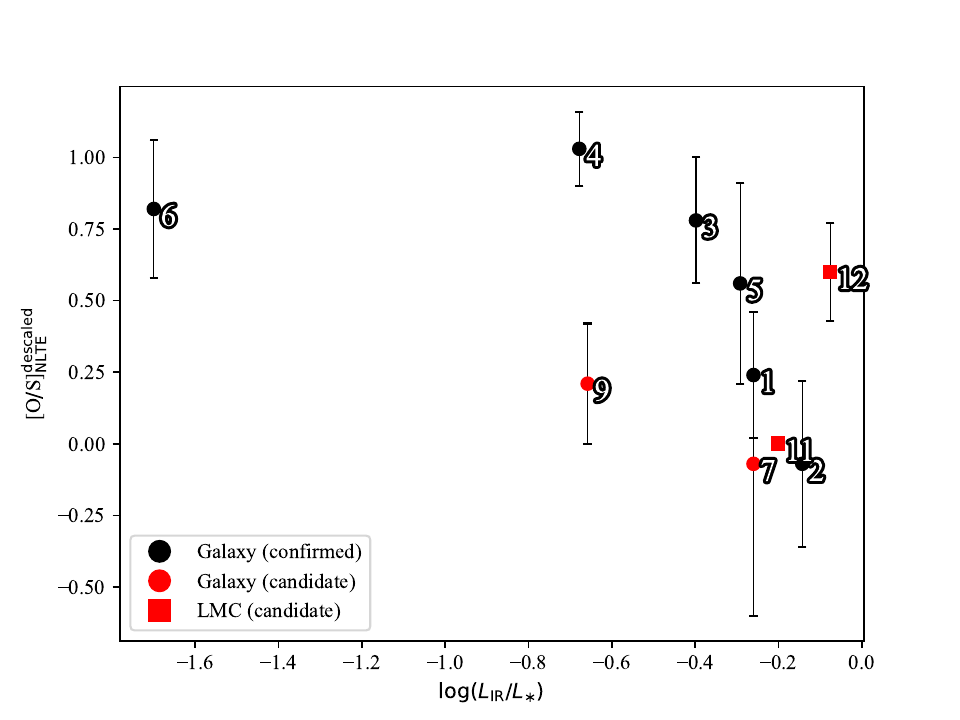}
    \includegraphics[width=.49\linewidth]{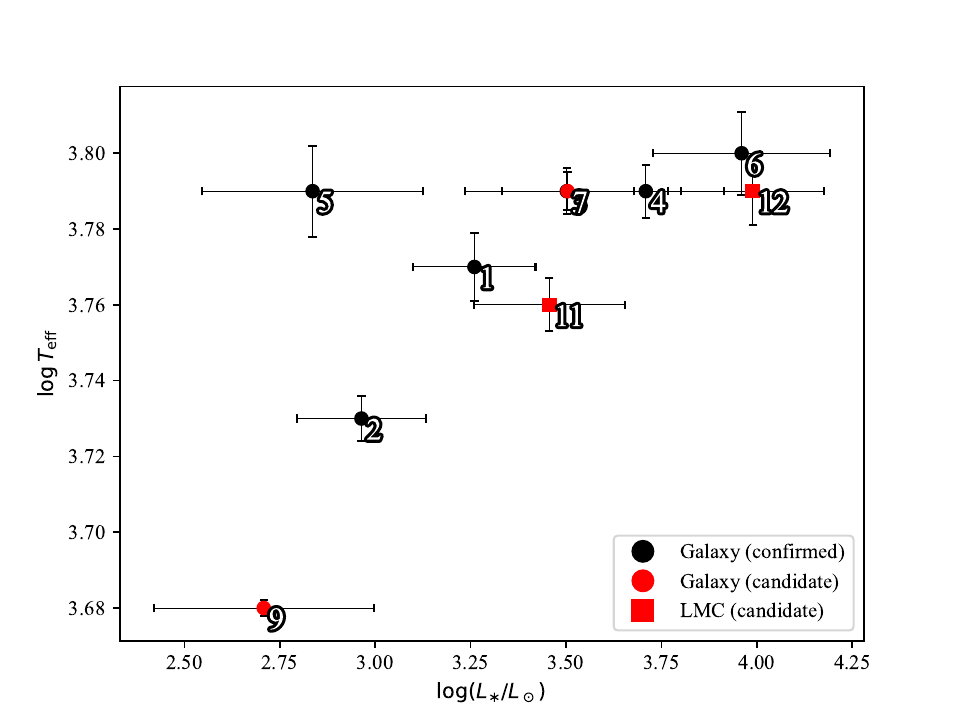}
    \includegraphics[width=.49\linewidth]{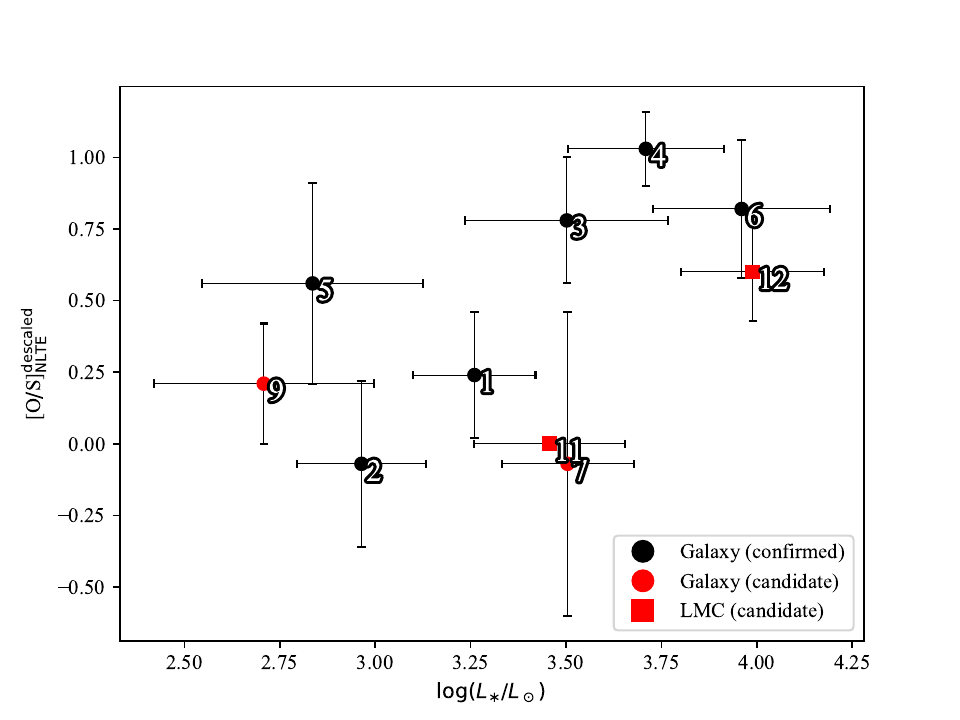}
    \caption{The strongest individual correlations in transition disc targets (see Section~\ref{ssec:dplcor}). The legend for the symbols and colours used is included within the plot. The targets are marked with their IDs.}\label{figA:allcor2}
\end{figure*}

\bsp 
\end{document}